# ANDERSON LOCALIZATION IN LOW-DIMENSIONAL OPTICAL LATTICES

JAN MAJOR

WORK UNDER SUPERVISION OF
PROF. DR HAB. JAKUB ZAKRZEWSKI

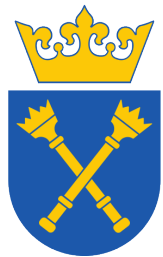

Atomic Optics Department
Marian Smoluchowski Institute of Physics
Jagiellonian University

April 2017

For my sons, Jerzy and Tadeusz

## STRESZCZENIE

Tematem rozprawy doktorskiej jest teoretyczny opis ultrazimnych gazów atomowych w jedno- i dwuwymiarowych sieciach optycznych w obecności nieporządku prowadzącego do lokalizacji Andersona. Nieporządek jest generowany przez oddziaływanie z drugą frakcją losowo rozłożonych i unieruchomionych atomów.

W układach o obniżonej wymiarowości nie występuje przejście z fazy izolatora Andersona do fazy przewodzącej, jednak nie jest wykluczone istnienie dyskretnego zbioru stanów przewodzących. Pierwsza część pracy poświęcona jest własnościom takich stanów, które pojawiają się na skutek istnienia korelacji w rozkładzie nieporządku. W sieciach o skończonych rozmiarach obecność stanów przewodzących powoduje pojawienie się „okien transportu", czyli zakresów energii, dla których długość lokalizacji jest większa od rozmiarów układu. Potencjalnym zastosowaniem takich układów jest tworzenie precyzyjnych, przestrajalnych filtrów energii atomów w sieci optycznej. W pracy opisano metodę pozwalającą analitycznie wyznaczać energie stanów niezlokalizowanych dla korelacji z klasy uogólnionych N-merów. Pokazano również, że nawet w nieskończonym układzie są to stany przewodzące. Następnie zaproponowano metodę tworzenia korelacji tego typu w eksperymencie, a także opisano technikę pozwalającą na generację nieporządku w amplitudach tunelowań, co umożliwia w jeszcze większym stopniu kontrolować ilość stanów przewodzących oraz ich energie.

W drugiej części pracy opisano metodę pozwalającą symulować w dwuwymiarowej sieci optycznej szczególny typ nieporządku: losowe pole magnetyczne. Układy tego typu, stworzone dzięki jednoczesnym modulacjom wysokości sieci i oddziaływań między frakcjami atomów pozwalają badać wiele zjawisk opisanych w ramach fizyki fazy skondensowanej takich jak ułamkowy kwantowy efekt Halla. W pracy zaprezentowano najistotniejsze efekty zaobserwowane podczas numerycznego obliczania długości lokalizacji dla szerokiego zakresu parametrów. W szczególności wytłumaczono anomalnie niską długość lokalizacji występującą dla skorelowanego nieporządku. Przedstawiono także porównanie średniej drogi swobodnej wyznaczonej z teoretycznego związku z długością lokalizacji z wynikami ewolucji w czasie zlokalizowanego pakietu falowego.

W załącznikach zawarto opis metod stosowanych do analizy periodycznie modulowanych układów oraz metod numerycznych służących do wyznaczania długości lokalizacji.

# ABSTRACT

Topic of the thesis is a theoretical description of the ultracold atomic gases in one- and two-dimensional optical lattices in the presence of the disorder leading to the Anderson localization. The disorder is created by interaction of the main fraction of atoms with the second immobilized fraction distributed randomly over the lattice.

In low-dimensional systems there is no transition from the Anderson localized to the conducting phase, although in the presence of correlations a discrete set of extended states can exist. The first part of the thesis is devoted to properties of such states. In the finite size lattices, the presence of those states results in the appearance of 'windows of transport' – energy ranges, in which the localization length is longer than the system size. Potentially, those systems could be used as precise, tunable filters for energies of atoms in the optical lattices. The analytical method of determining the extended states energies for correlations of generalized N-mers type is presented, along with a proof that indeed those states are extended in the infinite system. Subsequently, the way of experimental creation of this type of correlations is proposed, as well as the technique of generation of the disorder in the tunneling amplitudes, which significantly enhances tunability of the proposed energy filters.

The second part of the thesis describes the method which allows to simulate a specific type of the disorder: random magnetic field. Systems of such a class, created in a two dimensional lattice using simultaneous fast periodic modulation of the lattice height and interactions with immobilized species, may allow in future to investigate the range of topics from the condensed matter physics, for example fractional quantum Hall effect at half-filling. In the thesis, the most interesting features observed upon investigation of such systems are presented. Especially, the anomalously low localization length for correlated disorder is explained. The expression linking localization length with the mean free path is tested for the presented model and the results are compared with directly calculated mean free path.

In the appendices the method used to analyze time periodic systems is described together with the numerical techniques which could be used to calculate the localization length.

## PUBLICATIONS

The thesis is based on the following original contributions:

Other published works:

# ACKNOWLEDGEMENTS

I like to thank my supervisor prof. Jakub Zakrzewski for the help with my scientific work and critical reading of a manuscript of this thesis, as well as for his humor and specific attitude, which made the time I spent in the Atomic Optics Department an interesting and enjoyable experience. Also a lot of thanks for all other friends from AOD for contributing to the unique atmosphere.

I am especially grateful to my past and present roommates Andrzej, Ania, Arek, Dmitri, Gosia, Krzysiek, Marcin and Marek for scientific and not so scientific discussions and all the support they gave me. Also thanks to Julka, Romek and (A)Tomek for gossips during the traditional morning tours with coffee (and sorry for all the time I have taken). The special thanks go to Kamil aka Ziemniak for discussions, which somehow justified the 'philosophy' in the PhD title. I am also very grateful to Danusia and Agnieszka for helping with all the frightening administrative stuff.

Last but not least, a lot of thanks to my family for everything, especially to my wife Ewa for inexhaustible patience and to my sons Jerzy and Tadeusz for everyday dose of joy and constant reminding that there are things far more important than my work.

# CONTENTS

# INTRODUCTION

Quantum revolution at the beginning of the 20th century gave us a totally new view on the world and provided answers for many questions unveiling structure of the Universe, along with giving experimental results of accuracy unimaginable before. However, it brought even more unsolved problems and eventually halted its development in contradictory interpretations and technical difficulties. Physics is an experimental science and no matter how beautiful theory is, it has to undergo an experimental validation to be accepted. This became a huge problem not only in fields as a high energy physics, where larger and larger accelerators reaching even higher energies have to be built in order to test fundamental theories, but even in more 'ordinary' topics as a condensed matter physics. Although condensed matter systems could be safely described by well-understood non-relativistic quantum mechanics, due to their great complexity, they show emergent behaviors which can not be easily explained and a lot of unsolved questions persist in this area. Also experiments are problematic despite, comparing with mentioned high energy physics, they are quite easy to be prepared. The macroscopic quantities as conductance, magnetization etc. could be precisely measured but the microscopic informations crucial to understand the systems are mostly inaccessible. This is especially due to a small scale and a very fast pace of the processes[1]. Through the years a range of methods and ideas aimed at solving this problem have been created. Quantum simulators made in systems of ultracold atomic gases in electromagnetic potentials, one of the leading topics of this work are one of them.

## SIMULATING QUANTUM SYSTEMS

Computers had been created thanks to the growing understanding of the quantum-mechanical properties of the solid state systems (especially semiconductors). In a feedback they have a tremendous impact on the physics as they provided an opportunity of fast numerical calculations, which before had been omitted or done with an effort of a long work of whole armies of human computers (as famous 'Pickering's harem' [1]).

---

[1] For example in gold the wavelength of electron at Fermi energy is of the order of picometers and the time between interactions with crystalline lattice is of order of femtoseconds.

Computers quickly became a standard equipment for making numerical simulations of different phenomena from wide variety of disciplines, as for example the simulation of classical nonintegrable, chaotic systems or solving problems with complicated boundary conditions as strains in constructions. It was obvious to use them also to simulate the quantum systems of interest. It was especially important due to the fact that the set of easily solvable models in the quantum mechanics has run out quickly. Alas, it appeared that also conducting the numerical calculations could be a formidable task. In case of the classical physics, a complexity of computation very often scales linearly with a size of the system, thus even when a certain task could not be fulfilled nowadays, we could hope that with quickly growing computer capabilities, it could be done someday. In contrary, for quantum systems, in most of the cases, the size of the Hilbert space grows exponentially fast upon adding degrees of freedom (particles, lattice sites, internal states etc.). Required memory capacity explodes, as well as computational cost of calculating an evolution or finding a ground state of the system. This effectively makes it impossible to find an exact numerical solution, even for a moderately complex quantum system as for example interacting particles on lattice. Number of bits needed to just store the wavefunction of the system quickly exceeds number which could be reached, regardless the technology development[2]. Nowadays, we are limited to around several dozen lattice sites with at most one particle on site, if we allow more particles on site – having not so strongly interacting bosons – length of the lattice drops to less than twenty. There are of course numerous numerical methods which, using various approximations, make calculations for larger system sizes possible, although they usually work well only for a narrow class of systems (for reviews see: Density Matrix Renormalization Group [2], Matrix Product States [3], Quantum Monte Carlo [4]). Nevertheless the general limitation persists.

Brilliant idea of overcoming this exponentially huge problem has been given by Richard Feynman in 1982:

> "And therefore, the problem is, how can we simulate the quantum mechanics? (...) We can give up on our rule about what the computer was, we can say: Let the computer itself be built of quantum mechanical elements which obey quantum mechanical laws."[5]

It was rather a loose concept than a concrete solution. Two main branches grew from it in hope of realizing the *quantum simulator*: digital quantum computing and analog quantum simulation.

Roughly speaking, the digital quantum computing is an attempt to transfer ideas from the classical Turing machines into quantum me-chanical systems. In this manner instead of bits we have qbits (quantum bits) and instead of classical gates we have quantum gates operating on qbits. Array of qbits could be realized for example using a chain of particles with spin $1/2$ (spin up is 1 and spin down 0), we do not have to worry for the memory capacity of those systems as it grows exponentially upon adding new particles to the chain. It have been shown [7] that any unitary operation (as the evolution in a nondissipative system) could be decomposed in terms of the quantum gates – thus the digital quantum simulators seems to be universal. Although it looks promising, and indeed most of the physically important Hamiltonians could be simulated effectively (with at most polynomially growing resources) it is not always easy to find such an effective decomposition into quantum gates [8]. Moreover, as always in digital systems, the decomposition is only approximating the real solution and to obtain higher accuracies more quantum gates are needed, which could be a serious limitation considering present-day experimental capabilities. The first steps in creating the digital quantum computers have been done in systems of nuclear spins controlled by the nuclear magnetic resonance [9], as they gave good results for small systems, albeit are hardly scalable. Nowadays, the best realization of digital quantum simulators are cooled ions trapped in linear harmonic traps. They posses long coherence times and due to strong interactions allow the creation of a high fidelity quantum gates [10, 11].

---

2 For example wavefunction in the Bose-Hubbard model (9) with only 200 sites and mean occupation $1/2$ needs $10^{81}$ bits to be stored, while number of atoms in the visible universe is estimated in range $10^{79}$-$10^{82}$.

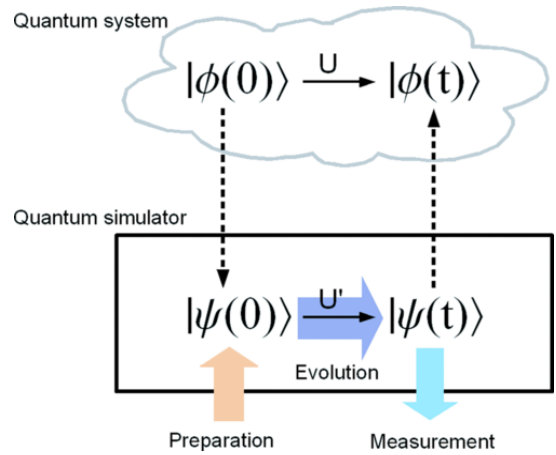

Figure 1: Sketch of the general idea of the quantum simulator. To simulate evolution U of the state $|\phi(t)\rangle$ in considered system we have to find mapping to a state of simulator $|\psi(t)\rangle$ for which initial state could be chosen, evolution could be tailored (to resemble evolution of $|\phi(t)\rangle$) and final state could be precisely measured (source: [6]).

The analog quantum simulation goes the other way – it aims at mimicking the behavior of the specific systems in the simplest possible way. The idea is not new in any way – the analog simulators have been well known long before the era of computers. An interesting example could be astronomical clocks, which are emulating periodic movement of the planets and other astronomical objects by precise clockwork mechanisms. They have been used to obtain informations, such as positions of the planets or possible dates of the eclipses, without need for lengthy calculations. In case of quantum systems the idea is the same: we have to find a system, which is easy to control and study and could be mapped somehow on another system, which is hard to investigate. Depending on the case, this mapping $H_{sim} = U H_{sys} U^{-1}$ could be trivial, as simulating the Hubbard model, which naturally arises in optical lattices [13] (see also Sec. 1.3.3) or very unobvious as for example mapping of a sound waves in Bose Einstein condensate onto some models of cosmological inflation [14]. The analog quantum simulators are not so versatile as their digital counterpart – in most of the cases each new model needs another idea of setup, on the other hand experimental realization is much simpler.

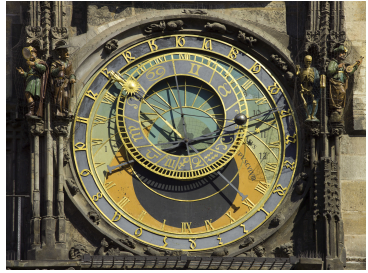

Figure 2: Astronomical clock from Prague, an example of the analog classical simulator (source: [12]).

Quantum simulators attempt to address extremely wide range of systems, the excellent listing could be found in [6]. Historically first and most extensively studied are simulators of condensed matter systems. Especially Hubbard (and Bose-Hubbard) model [13] and spin models [15] are investigated, mostly in order to determine their phase diagrams and characteristics of quantum phase transitions [13, 16]. Also the effects of disorder (Sec. 1.5) and frustration [17] are of the special interest as well as any connections with poorly understood phenomena from condensed matter physics, such as high-$T_c$ superconductivity [18]. Some of the applications appear in the field of the high energy physics and try to simulate particles governed by the Dirac equation [19], also the ideas to simulate effects predicted by cosmological/astrophysical models as the inflation [14], Hawking radiation[20] or Unruh effect [21] exist. Another promising field is simulation of the open quantum systems, topic which in classical computation adds whole new level of complexity, in quantum simulators could be obtained by adding some specific noise or just not insulating system from the environment as well as it is possible [22].

The numerous media are used for the analog quantum simulation, such as arrays of coupled optical waveguides, allowing fine shaping of the potentials but limited mostly to non-interacting particles [23], ions in harmonic traps covering the opposite sector of strongly correlated interacting particles [24] or Rydberg atoms, which due to large distances between atoms makes single site addressing especially simple [25]. Also solid state structures as arrays of Josepshon junctions could be used [26]. Among all this options the most advanced medium existing nowadays are ultracold atoms in electromagnetic potentials, covered in the next section.

### ULTRA-COLD ATOMS

*Overview*

One of the most successful (up to date) realizations of the idea of the quantum simulator are systems of ultracold atoms in electromagnetic potentials. The field dates back to the early 90-thies, time of a great development of techniques of trapping, controlling and precise measurement of the state of ultracold atomic gases by the means of external electric and magnetic fields as well as the laser light.

Two main features have made this field such a good basis for the quantum simulators: scale in which quantum processes appear and the tunability. As I stated before, one of the difficulties in exploring quantum world is usual sub-microscopic scale of quantum mechanical processes and short time scales. However, thermal de Brogile wavelength $\lambda_{dB} = \sqrt{2\pi\hbar^2/mk_BT}$ depends on the temperature and when we consider atoms cooled up to temperatures of the order of nanokelvin it could become quite big[3]. Thus we have systems which exhibit quantum properties but their size is nearly macroscopic and their dynamics are slow enough for electronic devices to easily control and record them. Second property is a great tunability of such systems, as it is facetiously but not so far from reality stated that "everything could be somehow simulated by ultracold atoms". By the appropriate use of the electromagnetic potentials, the atoms could be trapped, cooled, heated, propelled, their internal states could be changed, as well as interactions with other atoms could be tailored. Even complex phases could be added, so the atoms would behave as they were charged particles moving in a magnetic field or even in some non-abelian gauge field.

---

3 For example for rubidium-87 (one of the most widely used species) cooled to $20nK$ the $\lambda_{dB} \approx 1\mu m$.

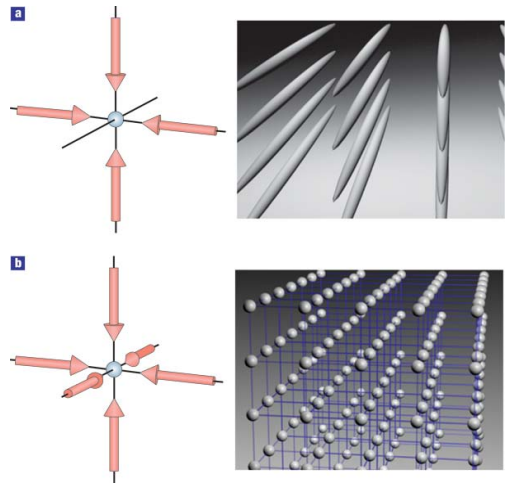

Figure 3: Examples of the optical lattices: (a) two dimensional array of tubes and (b) three dimensional regular lattice. On the left: arrangement of lasers, on the right: visualization of the lattice (source: [28]).

*Optical lattices*

In this context the basic solid state structure – crystal – can be realized in the cold atoms experiments as an optical lattice, a standing wave of light created by two counterpropagating laser beams (or one beam reflected from a mirror). For the first time such a system have been set up in Cohen-Tannoudji group in 1992 [27]. By choosing a frequency of light far detuned from any of the atomic transitions, we could neglect excitations, so we can assume that the atoms move in a purely conservative potential. Then, simple description apply: electric component of the light wave induces a dipole moment in atoms $\mathbf{d} \propto \mathbf{E}(\mathbf{r})$ and interact with those induced dipoles, resulting potential seen by atoms reads $V(\mathbf{r}) = -\mathbf{d} \cdot \mathbf{E}(\mathbf{r}) \propto |\mathbf{E}(\mathbf{r})|^2$ [28]. In the standing wave, the electric component at anti-node fluctuates, but as fluctuations happen at optical frequency which is way higher than any dynamics of the system they could be averaged and effectively we have the potential proportional to intensity of light $V(\mathbf{r}) \propto I(\mathbf{r})$. Usually one of the frequencies of transitions is much closer to the lattice frequency than others and detuning from it governs both strength and the sign of polarizability and in consequence behavior of atoms: for red-detuned laser, atoms are attracted toward maxima of intensity for blue-detuned opposite. The most characteristic features of the optical lattices (which also mark the biggest differences from real crystalline structures) is that they are ideal, in the sense that they do not contain any defects, unless deliberately induced and that they are rigid – atoms are affected by the lattice but the lattice is *not* affected by atoms – there are no phonon excitations (except systems in cavity [29]).

Even for this one specific setup excluded out of the whole field of ultracold atomic gases, there are so many different possibilities and variations I will not try to cover all of them in this introduction. I will concentrate on the model which will be used further – Bose-Hubbard model in the regular lattice, for the whole spectrum of possibilities one could for example check [30, 31] and references therein.

*Bose-Hubbard model*

Standing light wave created by the two counterpropagating beams is seen by atoms as sinusoidal potential $V(x) = A_x \sin(2\pi/b_x x)$ where $b_x$ is half of the wavelength and potential height is given by a formula:

$$A_x = \frac{3\pi^2 \Gamma I_0}{2\omega_0^3 \delta},\tag{1}$$

where $I_0$ is a maximal intensity of light (at antinode), $\delta$ is a detuning from the closest atomic transition frequency ($\omega_0$) and $\Gamma$ is an inverse time of decay of state excited by frequency $\omega_0$. Usage of pairs of perpendicular beams allows creation of two- or three-dimensional regular lattices (see Fig 3). In the case of optical lattice created only in one or two dimensions[4], without any additional potentials the dynamics of free particles in remaining dimensions should be considered, and in experimental conditions the atoms could 'leak out'. Usually, to prevent this deep harmonic potential is used $V(z) = 1/2\Omega^2 m z^2$ where $\Omega$ is angular frequency of trap and $m$ is atom mass. For definiteness I will describe the system with the optical lattice in one direction ($x$) and the harmonic confinement in others ($y$ and $z$). The one particle Hamiltonian reads:

$$h(\mathbf{r}) = -\frac{\hbar^2}{2m}\nabla^2 - A\sin\left(\frac{2\pi}{a}x\right) - \Omega m \left(y^2 + z^2\right).\tag{2}$$

Convenient property of our regular lattice potential is its separability. We could treat wavefunction in different directions separately and whole wavefunction of a system is just $\Psi(\mathbf{r}) = \psi(x)\psi(y)\psi(z)$ where each of components could be determined by solving one dimensional problem with Hamiltonian:

$$h(x_i) = -\frac{\hbar^2}{2m}\partial_{x_i}^2 - V_i(x_i).\tag{3}$$

In the directions of the harmonic confinement we have eigenstates of harmonic oscillator, if $\Omega$ is high enough we could neglect any excitations in those directions as their energy is much higher than any excitations in lattice. The wavefunction is then $\Psi(\mathbf{x}) = \psi(x)\phi_h(y)\phi_h(z)$ (where $\phi_h$ is a ground state of the harmonic oscillator) and we could focus on dynamics in the $x$ direction. As the lattice is spatially periodic, it allows use of the Bloch's theorem [32]. We could write $\psi(x)$ as $e^{ikx}\phi_k^\alpha(x)$ where $\phi_k^\alpha(x)$ are periodic Bloch waves, index $k$ enumerates quasi-momenta and $\alpha$ enumerates energy bands. If the lattice is high enough that there exists a gap between ground and first excited band, instead of Bloch waves we can use a basis of Wannier functions which are localized on sites of lattice [33, 34]. One could go from Bloch to Wannier basis by making the Fourier transform:

$$w_n^\alpha(x) = \sum_k e^{ib_x nk}\phi_k^\alpha(x). \qquad (4)$$

---

4 Those systems are usually called one- or two- dimensional, although more precise will be quasi one- or two- dimensional as all internal properties of particles (as statistics) remain three-dimensional, only their movement is confined to line/plane.

The function $w_n^\alpha(x)$ describes particle in the $\alpha$ band localized on the $n$-th site of the lattice. As phases of the different Bloch waves could be changed independently we are still left with vast choice of form of the Wannier functions. Standard option is to choose real functions with maximum at chosen site for symmetric states (ground and even excited) and with maximum first derivative at chosen site for asymmetric states (odd excited). The wavefunction could be expressed in the basis of the Wannier functions:

$$\psi(x) = \sum_n \psi_n^\alpha w_n^\alpha(x), \qquad (5)$$

where $\psi_i$ is occupation of $i$-th site. We could also rewrite Hamiltonian (2) in this basis as:

$$H_{TB} = \sum_{ij\alpha\beta} |w_i^\alpha\rangle\langle w_i^\alpha|h|w_j^\beta\rangle\langle w_j^\beta| = \sum_{ij\alpha\beta} [h]_{ij}^{\alpha\beta}|w_i\rangle\langle w_j|, \qquad (6)$$

where $[h]_{ij}^{\alpha\beta} = \langle w_i^\alpha|h|w_j^\beta\rangle$ are matrix elements of the Hamiltonian (2) in chosen basis[5]. As the Bloch functions for different bands are orthogonal, also the Wannier functions are, so in one-particle Hamiltonian bands are separated. From now on I will assume that system is in the lowest band and drop band indices. Furthermore, we could distinguish the on-site energies $\epsilon_i = [h]_{ii}$ which denotes energy of particle placed in some specific lattice site from tunneling rates $t_{ij} = [h]_{ij}$ for $i \neq j$. As the tunneling rates fall exponentially with distance, in most of the cases we consider only terms $t_{ii\pm1} \equiv t_i$. In this way we reach the standard tight binding model with the nearest neighbor tunneling:

$$H_{TB} = \sum_i \epsilon_i|w_i\rangle\langle w_i| - t_i|w_i\rangle\langle w_{i+1}| - t_{i-1}|w_i\rangle\langle w_{i-1}|. \qquad (7)$$

---

5 Of course we should use full wave-function $\Psi(\mathbf{r})$ but as we have assumed that in all other directions we have ground state of harmonic oscillator $\psi(y)$ and $\psi(z)$ just integrates out to 1.

Values of the on-site energy $\epsilon_i$ and the tunneling rates $t_i$ are given by action of Hamiltonian (2) onto Wannier functions and for regular lattice should be homogeneous but anticipating various possibilities of inducing inhomogeneities in the system I have written Hamiltonian with site dependent parameters.

Further, $H_{TB}$ could be extended by means of the second quantization (for derivation and description see for example [35]) to describe systems of many interacting particles. If we assume that only the $s$-wave scattering is present (which is reasonable for dilute ultracold atomic gases), it could be modeled by a contact interaction pseudopotential :

$$U(\mathbf{r}, \mathbf{r'}) = \frac{4\pi\hbar^2\sigma}{m}\delta^{(3)}(\mathbf{r} - \mathbf{r'}), \tag{8}$$

where $\sigma$ is scattering length. Single particle tight binding Hamiltonian (7) with the contact two-particle interactions can be written as:

$$H_{BH} = \sum_i \left( \epsilon_i n_i - t_i a_i^\dagger a_{i+1} - t_i a_{i+1}^\dagger a_i + \frac{U_i}{2} n_i(n_i - 1) \right), \tag{9}$$

where $i$ enumerates lattice sites, $a_i(a_i^\dagger)$ are bosonic ($[a_i, a_i^\dagger] = 1$) operators of annihilation(creation) of particle localized at site $i$ (the Wannier function $w_i(x)$), $n_i = a_i^\dagger a_i$ is particle number operator, $U_i$ is energy of contact interactions given by

$$U_i = \frac{4\pi\hbar^2\sigma}{m}\int dx|w_i(x)|^4 \int dy|\phi_0^h(y)|^4 \int dz|\phi_0^h(z)|^4, \tag{10}$$

in this case the integrands of ground states of harmonic oscillator could not be simply omitted as now the higher powers are integrated in result one gets:

$$U_i = 2\hbar\sigma\Omega\int dx|w_i(x)|^4. \tag{11}$$

Another issue which should be remembered is that integral (11) gives nonzero results also for different sites and different bands, thus in principle there are long range interactions and band mixing and all elements $U_{ijkl}^{\alpha\beta\gamma\delta}$ should be calculated and placed in the Hamiltonian. The long range interactions can be neglected for deep enough lattices. Coupling to higher bands also could be discarder if only the energy of excitations is big enough (or density of particles is small enough)[6].

Having done all the assumptions above we arrive at the simple model (9) with three parameters (or sets of parameters): $\{\epsilon_i\}$, $\{t_i\}$ and $\{U_i\}$. All of them could be somehow modified:

---

6 As an indicator the ratio of energy of excited band to interaction energy could be used – if it is smaller than one it is energetically favorable for particles to occupy higher bands.

ON-SITE ENERGY in most of the experimental setups is modified by addition of a shallow harmonic trap which prevents leaking out of atoms at the edges of the lattice, it may be made stronger to create significant inhomogeneity. Another method of breaking the homogeneity of on-site energies is imposing another optical lattice – if only it is weak enough we could assume it changes only the on-site energies of the main lattice. In this way one could get periodic variation of the on-site energies (superlattice) or quasi-periodic Aubry-Andre model [36]. Disordered on-site energies could be created by means of speckle potentials or second species of atoms immobilized in the lattice (described further in section 1.6).

TUNNELING RATES could be simply changed by changing the lattice height. Inhomogeneity of tunnelings appears naturally if on-site energies are changed (by one of the methods described above), but usually it is small effect (estimation could be found in [37]). The effective method of making larger changes of tunnelings (amplitudes as well as complex phases) is using fast periodic modulation of the lattice parameters, examples are given in sections 2.2.2 and 3.5.

INTERACTIONS value (precisely the scattering length $\sigma$ in equation (8)) could be changed or even set to zero using external magnetic field in vicinity of the Feshbach resonances [38]. Inhomogeneous interaction strengths could be obtained in the optical lattices placed above an atom chip, where magnetic field can vary on small length scales [39].

FLOQUET THEOREM

Usually, when considering ultracold atomic gases we assume that Hamiltonian of the system is time independent during the evolution (the elements of system as fields, lattice height etc. do not change considerably). Time dependencies appear at a transitions between different phases of experiment and are either adiabatically slow to preserve ground state or very quick to provide fast readout, also resonant driving is used to place system in some excited state. Yet another notable form of time dependency, which has entered the toolbox of ultracold atomic gases around ten years ago is a fast off-resonant periodic modulation of the system parameters. The theory describing the class of periodically driven systems has been created in the end of 19th century by Gaston Floquet [40], originally to deal with linear differential equations with periodic time dependence. He made two main statements:

1. for every differential equation which could be written as:

$$\dot{x}(t) = A(t)x(t), \text{ where} : A(t+T) = A(t) \tag{12}$$

there exist matrix B for which: $x(t) = e^{itB}x_p(t)$ where $x_p(t)$ has the same period as $A(t)$.

2. there exist a coordinate change $y(t) = Q^{-1}(t)x(t)$ which results in:

$$\dot{y}(t) = Ry(t), \tag{13}$$

where R is time independent.

Especially first claim looks familiar – it resembles the Bloch theorem for the case of the space periodicity. In classical systems the Floquet theory have been used for example to solve the Hill equation determining stability of lunar motion [41], or to deal with other problems considering movement of astronomical objects [42]. First application to quantum systems has been done by Shirley in 1965 [43] for two level system with time dependent coupling (two level atom in a varying magnetic field).

Its importance for the ultracold atomic systems came from the fact that upon applying a periodic modulation to a system, its behavior could be changed in a way impossible to obtain by other means. In this way the Floquet engineering (as it was dubbed due to ample opportunities and ease of changing Hamiltonians on the will) greatly enhances ultracold atomic gases as quantum simulators. I will be interested mostly in the case of fast modulations, which allows to describe system dynamics for a long times (much longer than a modulation period) using the time independent effective Hamiltonian.

Application to the field of ultracold atoms have been first proposed in the 'hot' era just after condensate had been created in 1997 [44], the fast development has started around ten years later by the idea of suppressing the tunneling in the optical lattice or even reversing its sign by the periodic shaking of the lattice [45]. Also the transition from a superfluid to a Mott insulator have been induced in this way [46, 47]. Yet another significant application is using periodic modulations to create artificial gauge fields [17] allowing simulation of the orbital magnetism using neutral atoms.

The statements of the Floquet theorem look deceptively simple, but in most of interesting situations, finding appropriate change of the coordinates is impossible. Usually one have to appeal to approximate solutions. More precise description of methods of finding such effective time independent Hamiltonians is given in appendix A.

## ANDERSON LOCALIZATION

*Overview*

Anderson localization (known also as a strong localization) is a phenomenon appearing in a disordered quantum systems. It presents a stark distinction from transport in classical disordered systems. If we consider the classical particle traveling in disordered potential landscape it will move diffusively[7]. Thus the transport is slowed down but not stopped. The particle could be also localized but only if it has started in some 'well' in potential disconnected from the rest of the system and does not have enough energy to escape it.

On the other hand, for the quantum particles (or in general any coherently propagating waves) the situation is qualitatively different. One thing is that particle will not remain trapped, if the well is not infinitely deep. Especially, if the wells create periodic structure (for example the optical lattice), the Bloch theorem applies [32] and we are assured that the eigenstates of the Hamiltonian are Bloch waves spanning across the whole system. Upon considering the quantum particle traveling in a considerably weak random potential (such that, in classical case it will only slightly affect the particle movement), even more striking difference appears – the wavefunction of the particle appears to become localized with exponentially decaying tails (Fig. 4(a)). The wavefunction gets form:

$$\psi(x) \sim e^{-|x-x_0|/\lambda}. \tag{14}$$

Position on which function is localized ($x_0$), depends on the disorder realization but the localization length $\lambda$ is a nonrandom quantity – it depends only on statistical properties of the disorder. This effect is a famous Anderson localization, first described nearly sixty years ago by Philip Anderson [49]. It is an interference effect (Fig. 4(b)). The wavefunction of a particle gets scattered on numerous scatterers and all those scattered waves interfere destructively almost everywhere. In the effect the particle localizes – the probability of finding it falls with distance exponentially. In low dimensional systems the effect appears for nearly all energies and disorder types[8]. In three dimensions, for growing energies the system undergoes quantum phase transition from the Anderson insulator (all states localized) into the metallic phase (all states extended) – the transition point is called the mobility edge.

---

7 At sufficiently large scale, if the scale will be too short the movement will be almost ballistic.

8 The exceptions include for example some specific correlations [50, 51] or two-dimensional systems with spin-orbit coupling [52].

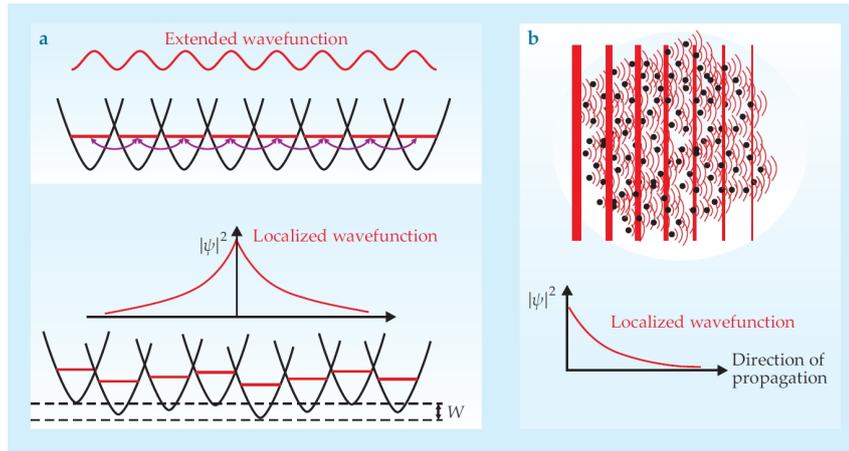

Figure 4: Two facets of the Anderson localization phenomenon. (a) In the lattice system eigenfunctions are Bloch waves, but after introducing the disorder, eigenfunctions become localized with exponentially falling envelope. (b) Coherent wave is traveling through randomly placed scatterers, scattered waves interfere destructively causing exponential decay of the conductance with the distance (source: [48]).

*Weak localization*

Weak localization is often depicted as predecessor of strong (Anderson) localization. It is an interference effect appearing for coherent waves traveling in the disordered media – same as Anderson localization, however weak localization correction only reduces the diffusion constant, not sets it to zero.

To describe a movement of a quantum particle traveling through the system between points $\mathbf{r_0}$ and $\mathbf{r}$ we have to take into account all possible paths it could take. Probability of reaching the goal $P(\mathbf{r_0}, \mathbf{r})$ is the squared absolute value of sum of all amplitudes of probability ($a_i$):

$$P(\mathbf{r_0}, \mathbf{r}) = \left| \sum_{i \in \text{paths}} a_i \right|^2 = \sum_{i \in \text{paths}} |a_i|^2 + \underline{\text{interference terms}}$$
$$= P_{cl}(\mathbf{r_0}, \mathbf{r}). \tag{15}$$

In the disordered system in most of the cases the interference term averages to zero when summed over all possible paths and we are left with the classical sum of probabilities. One important exception is when we consider particle returning to its origin ($\mathbf{r} = \mathbf{r_0}$). Then, for all paths we could find another one which is going on exactly the same contour but in opposite direction (as at the picture 5). Those

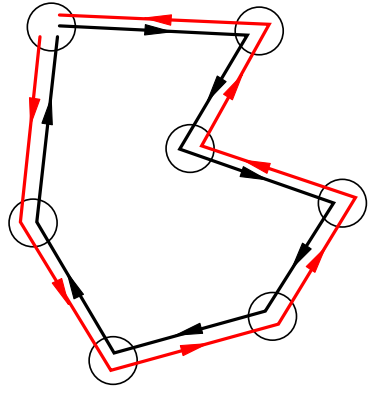

Figure 5: When considering two reversed paths their phases are the same so they add constructively doubling the probability of returning to the origin for a quantum particle (if the transport is coherent).

two paths (as long as the time reversal symmetry holds) have the same phase factor, so we could write:

$$P(\mathbf{r_0}, \mathbf{r_0}) = \left| \sum_{i \in \text{paths}/2} 2a_i \right|^2 = 4 \sum_{i \in \text{paths}/2} |a_i|^2 + \underline{\text{interference terms}}$$
$$= 2P_{cl}(\mathbf{r_0}, \mathbf{r}),$$

(16)

so for a coherently propagating quantum particle the probability of returning to the origin is twice higher than for classical one.

*Scaling theory of the localization*

One of the first attempts to better understand localization processes and the apparent dependence on the dimensionality has been done by means of the scaling theory of localization [53]. General question in case of the transport and localization could be framed as: we want to know how the conductance of the system is changing upon changing the scale of the system. The conjecture made by Thouless [54] was that the conductance ($g$) of $d$ dimensional block of size $2L$ is given solely by the conductance of $2^d$ blocks of size $L$ of which it is composed. In other words, when we determine the conductance of some part of the system, the conductance of the part elongated in one direction by some factor $b$ will be given by function depending only on $b$ and conductance itself:

$$g(bL) = f(b, g(L)),$$

(17)

(obviously to satisfy this relation the added part has to be big enough to have representative realization of the disorder). Further we assume that the above equation could be written in continuous form:

$$\frac{d \log g(L)}{d \log L} = \beta(g(L)). \tag{18}$$

If we are able to find a form of the $\beta(g)$ function we could use it to calculate the conductance for any system size starting from just one sample. In order deduce the shape of $\beta(g)$ we need to know its limiting behaviors:

- For very large conductance we could assume that the classical theory of transport is correct and use *classical* dimensionless conductance:

$$\tilde{g}(L) = \frac{2k\ell}{d}(kL)^{d-2}, \tag{19}$$

  where $k$ is momentum and $\ell$ is a mean free path (a distance which particle can travel before its movement becomes randomized). In the considered limit scaling function is constant:

$$\beta(g) \underset{g \to \infty}{\longrightarrow} d - 2. \tag{20}$$

- For the opposite case, the exponential localization is present[9], so $g \sim e^{-\alpha L}$ and the scaling function has form:

$$\beta(g) \underset{g \to 0}{\longrightarrow} \log\left(\frac{g}{g_0}\right), \tag{21}$$

  where $g_0$ is of order of one.

Having those limiting cases we could determine the behavior for systems of different dimensionalities.

ONE DIMENSION    In one dimensional systems the classical conductance falls linearly with the system size (the Ohm's law) so the $\beta(g)$ is $-1$. As it is negative, for growing system size $g(L)$ will fall, $\beta(g)$ will be even smaller so the fall of $g$ will accelerate eventually reaching the regime of strong localization where $\log g(L)$ starts to fall linearly so the decay of conductance becomes exponential (as presented in the Fig. 6).

TWO DIMENSIONS    In two dimensions the classical conductance is independent on the size of the system. It left us with rather ambiguous situation, as we are around zero and do not know in which direction the renormalization flow will take us. In this situation the details

---

9 If we imagine d-dimensional system as $(d-1)$-dimensional array of wires, the exponential decay of conductance in one channel will win with polynomially growing number of channels.

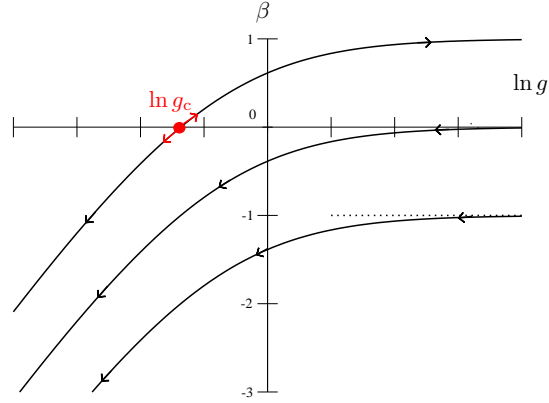

Figure 6: The picture showing the scaling curves $\beta(g)$ in function of $\log g$ for different dimensionalities, arrows mark the direction of flow (source: [56]).

of the model becomes relevant. We could write the limiting $\beta(g)$ with the corrections of order $g^{-1}$:

$$\beta(g) = \frac{c_2}{g} + O(g^{-2}). \tag{22}$$

The correction obtained from scaling theory: $-1/2$ is wrong but as long as the system preserve time reversal and spin change symmetries (stays in *orthogonal* class) the weak localization correction applies and the $\beta(g)$ is always negative. Exact value of coefficient $c_2$ could be determined, using self consistent theory of localization [55], to be $-1/\pi$. In the result the system is always localized but in contrary to the one-dimensional case the localization length grow exponentially for larger energies (momentum) or weakening disorder. The approximate expression connecting localization length with the mean free path reads:

$$\lambda = \ell \exp\left(\frac{\pi}{2}k\ell\right). \tag{23}$$

three dimensions   In three dimensions the situation is qualitatively different. The classical conductance grows with growing system size, so the $\beta(g)$ function is positive for large $g$. If we start with the dimensionless conductance in this region, for larger system sizes we will get even larger $g$ and the flow will take us to $g \to \infty$ – the system will be conducting. On the other hand starting at the other side will again lead us to exponentially decaying $g(L)$. Point $g_c$ for which $\beta(g_c) = 0$ is an unstable fixed point and marks the phase transition between localized and extended states, the mobility edge.

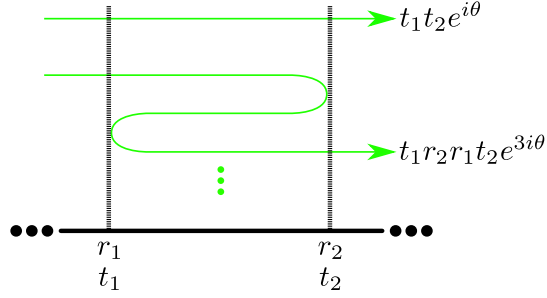

Figure 7: Wave passing through two scatterers could undergo multiple reflections between them. To get transmission probability all those paths have to be summed.

*Example: localization in one dimension*

The case of one dimensional systems could be treated more rigorously to show that exponential localization indeed appears. The behavior of wave interacting with scattering potential is typically given by scattering matrix connecting left and right outgoing waves. For our purposes more convenient is the transfer matrix[10] which links waves on the left side of the obstacle with those on the right side. It could be written as:

$$M = \begin{bmatrix} 1/t^* & -r^*/t^* \\ -r/t & 1/t \end{bmatrix},$$ (24)

where $r$ and $t$ are reflection and transmission coefficients (I assume that scatterer is symmetric). The advantage of the transfer matrix over the scattering matrix is that it could be chained so the transfer through two scatterers, having individual transfer matrices $M_1$ and $M_2$ is given just by $M_{12} = M_1 M_2$. The probability of transfer through them $T_{12} = |t_{12}|^2$ could be calculated by summing all possible internal reflections (as shoved in the picture 7). The result is given by:

$$T_{12} = \frac{T_1 T_2}{|1 - \sqrt{R_1 R_2} e^{i\theta}|^2},$$ (25)

where $\theta$ is a phase the wave acquires traveling between scatterers. To investigate the localization we want to know how the transmission behaves upon adding subsequent scatterers. Alas, the relation (25) could not be simply extended, as then we will have to take into account all possible paths between all scatterers. The best option is to find a function of $T$ which is additive. Solution is a logarithm of transmission probability as:

$$\log T_{12} = \log T_1 + \log T_2 - 2 \log \left| 1 - \sqrt{R_1 R_2} e^{i\theta} \right|,$$ (26)

---

10 Not to be confused with transfer matrix T introduced in 2.4. They could be connected but are different objects.

and upon averaging over disordered distance between scatterers: $\langle \ldots \rangle_\theta = \int_0^{2\pi} d\theta/2\pi$ the last term averages to zero:

$$\langle \log(T_{12}) \rangle = \log T_1 + \log T_2. \tag{27}$$

Thus, if we have wire of length $L$ with density $n$ of randomly placed identical scatterers (with transmission probability $T$), then the logarithm of transmission probability trough the system will be $\log\langle T_{sys} \rangle = nL \log T$. As could be shown (for example in [56]) the logarithm of transmission is a self-averaging quantity so it could be used as a typical value of transmission upon taking the average over different realizations of disorder. Following this we could write the typical transmission as:

$$T_{typ} = \exp\langle \log T_{sys} \rangle = \exp\left(-Ln \log T_1\right), \tag{28}$$

where $(n \log T_1)^{-1}$ is defined as the localization length $\lambda$.

*Many body localization*

Although whole of this work is concentrated on Anderson localization – single particle effect, it would be unfair not to mention vividly developing field of Many Body Localization. At the first glance it seems that interactions between particles being source of strong decoherence should destroy the localization. It was dominant opinion for a long time and the localization has been investigated in systems with negligible interactions. Recently, it was shown that for sufficiently strong disorder, interacting systems could also show signatures of localization. It could be interpreted as localization in the Hilbert space rather than in the real space. Such a localization stops the thermalization of the states. In most of the quantum systems (even if they are isolated from the environment), only a finite number of local observables are conserved quantities, so upon the evolution the system loses most of informations of its initial state, apart from the very basic ones (energy, magnetization etc.) – it thermalizes. On the contrary in many body localized systems there is an extensive number of conserved local observables. The most studied example is probably the charged density wave order which, being the highly excited state, quickly diminishes in ergodic systems while persist for long times in many body localized cases. Review of the field could be find in [57, 58].

DISORDERED SETUPS FOR THE ULTRACOLD ATOMS

Although the Anderson localization have been devised for the electrons in condensed matter systems it is extremely hard to observe it in this class of media due to a strong decoherence caused by the electron-electron and the electron-phonon interactions. Instead, it have been experimentally studied in the ultracold atomic gases, which provide highly tuneable setups, without phonons (due to the rigidity of optical lattices) and with the interactions that could be set to zero by means of the Feshbach resonances[11].

There exist basically three main methods of creating disorder in ultracold atoms experiments. Each of them provides different disorder distributions and gave different possibilities of tailoring correlations, disorder strength etc. I will shortly review all three of them.

*Speckle potential*

Localization in the speckle potential have been the first one to be shown in the experiment (half of century after Anderson published his theory) [62]. Speckle potential could be (and in most of the cases is) applied to the ultracold atomic system without the underlying optical lattice. It could be created in the setup presented in Fig. 8: laser beam is illuminating a diffusive plate, the scattered light is focused by the lens and the pattern of light appearing on the focal plane is the speckle pattern. Disorder is introduced by the scattering on a diffusive plate as phases of the scattered waves are randomized, but as the time coherence is preserved they interfere. Because of its origin, speckle potential posses some very specific correlations which, apart from yielding simple theoretical description, could be easily modified in the experiment by modification of the aperture of the diffusive plate. For example, by adding some shutter in the midst of it one could create correlations resulting in the appearance of an effective mobility edge – a range of the energies for which the localization length grows by several orders of magnitude [63]. This method could be taken even further into so called holographic microtraps allowing creation of a wide range of potential shapes by precisely preparing the diffusive plates [64]. However the possibilities of changing the aperture are limited and due to the fact that its size is restricted by the laser beam width and the geometry of experimental setup (which may not allow to let in arbitrary wide beam) the variability of the speckle potential (i.e. width of one speckle) is small comparing for example with the optical lattice. Due to this fact, methods based on high variability of potential as ones presented in the next chapter (2.2.2) are much less useful in case of the speckle.

*Frozen particles*

In this class of methods (proposed in [66, 67]) we use two fractions of the particles (which could be different species of atoms as good as different internal states of atoms of the same element). One species is *frozen* in random positions in the lattice and acts as a static disorder (all parameters and operators related to it are denoted by f superscript) while the other are *mobile* particles which dynamics we consider. The general idea of creating such system is as follows: We start from the two fractions of bosons in the optical lattice, described by the Hamiltonian:

$$H_{TF} = \sum_{\langle ij \rangle} (t a_i^\dagger a_j + t^f a_i^{f\dagger} a_j^f)$$
$$+ \sum_i \frac{U}{2} n_i(n_i - 1) + \frac{U^f}{2} n_i^f(n_i^f - 1) + V n_i^f n_i, \quad (29)$$

where $\langle ij \rangle$ denotes nearest neighbors, $U$ and $V$ are intraspecies and interspecies interactions respectively. At the beginning we assume that we have only the (not yet)*frozen* particles in the system. We could easily obtain two different random distributions of disorder: binary and Poisson. For the first we should use small density of strongly repulsing particles ((a) $\rho_f < 1$ and $U^f \gg t^f$) while for the latter higher density of *frozen* atoms in a deep superfluid regime ((b) $\rho_f > 1$ and $t^f \gtrsim U^f$). After letting *frozen* particles to evolve for some time we proceed with a fast quench of $t^f$ for example by the rapid increase of a lattice height. In the (a) case we get in this way a binary disorder – on each site of the lattice there is either 0 or 1 frozen particles. If we used option (b), after quenching a superfluid the distribution of the *frozen* particles should be given by the Poisson distribution with the mean at the $\rho_f$. Into system prepared in such a way we put *mobile* particles. As now the $t^f = 0$ we could treat $n_i^f$ just as a number and element $(V n_i^f)$ of the Hamiltonian as the on-site energy term. The question could rise how one species could be immobilized while second has non-vanishing tunnelings? It could be done for example by using two elements which have different detuning from the lattice frequency or by using two hyperfine states susceptible only to one circular polarization of light and then creating two overlapping lattices with different polarizations and independently controlled height[12]. To sum up we get Hamiltonian:

$$H_D = \sum_{\langle ij \rangle} t a_i^\dagger a_j + \sum_i \epsilon_i n_i, \tag{30}$$

where $\epsilon_i = V n_i^f$ is random on site potential.

---

11 Another systems allowing experimental observation of the localization are for example photons in optical waveguides [59], ultrasound in elastic network [60] or laser light on simple set of randomly placed transparencies [61].

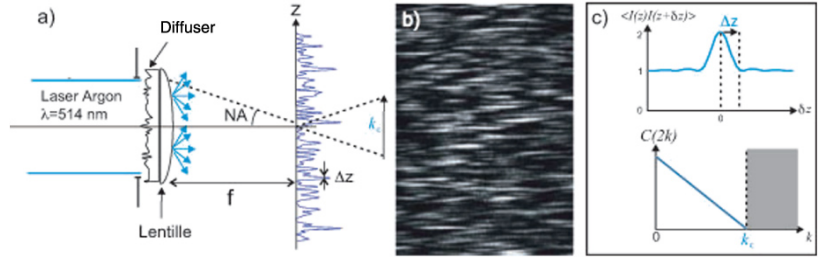

Figure 8: (a) Sketch of a speckle potential setup: laser light is scattered by a diffusive plate, and focused by a lens – on the focal plane the speckle pattern appears. (b) Anisotropic speckle pattern. (c) The autocorrelation of the intensity (upper plot) and power spectrum (lower plot) both showing the lower limit on the size of one speckle – given by the width of autocorrelation and by the upper boundary of power spectrum (source: [65]).

*Incommensurate lattice*

This class is quite specific as it do not has any disorder in the common sense. The setup is composed of two overlapping optical lattices created by light with different wavelengths (so called bichromatic lattice). If the lattice constants are commensurate we get superlattice – still a periodic structure but with changed period. More interesting for us is a case when the lattice constants are incommensurate. Then the spatial periodicity is lost so the Bloch theorem could not be used anymore. The obtained potential obviously is not random, but it could be treated as sort of quasi-disorder and it appears that to some extent it gives similar results as really disordered potentials.

To simplify the description we could assume that one lattice (main) is much higher and the second is only a small perturbation. Then Bloch theorem is still approximately satisfied and we could find the Wannier functions for the main lattice, while the second lattice effectively changes only the on-site energies[13]. In this way we eventually approach the Hamiltonian of the Aubry-Andre model [36]:

$$H_{AA} = \sum_{\langle ij \rangle} t a_i^\dagger a_j + \sum_i V_{AA} \cos(\beta i) \psi_i, \tag{31}$$

where $V_{AA}$ is height of incommensurate potential and $\beta$ measure of incommensurability. It have been shown that such a system has a transition between the localized and the extended states at $V_{AA} = 2t$. For lower potential all states are extended while for higher all are exponentially localized.

Is this interesting region $V_{AA} \approx 2t$ accessible by proposed setup? Quite standard optical lattices have heights around $500t$ so the second lattice with height of order $2t$ certainly could be treated as a small perturbation. Indeed, experiment showing exponential localization in bichromatic systems had been done [69].

---

12 This method is more suitable for my work as it allows creation of lattices with different lattice constants, which is used in section 2.3.3.

13 It could change also the tunneling rates for bigger but still applicable lattice heights (calculations for this case are presented in [68]).

# OFF-DIAGONAL DISORDER AND CORRELATIONS

---

## INTRODUCTION

In this chapter I am going to concentrate on one-dimensional systems. Two main topics interwoven in the course of this part are the influence of correlations in disorder distribution onto localization properties and the possibility of creating systems with disordered tunneling amplitudes in experiments with ultracold atoms in optical lattices.

My leading goal will be to design experimentally realizable systems, which have both localized and extended states. Such systems could be possibly used as tunable filters for the energies of the particles. Benefits could be two-fold: They could be used as a mean of making a precise measurement of the energies of the particles coming from an another experiment, by selectively trapping and letting through atoms with different energies. On the other hand, they could be used as a source of the particles with precisely chosen energy, which can be further used, for example in an atom interferometry [70]. Another promising application for the setups with several delocalized states is shaping of a wavefunction leaving the system in the manner similar to shaping the light waves by adding several pulses with different wavelengths [71].

Chapter is structured as follows: At the beginning I will describe some specific properties of the systems with the disorder in the tunneling amplitudes (off-diagonal disorder), then I will present procedure of obtaining such type of a disorder using fast periodic modulation of the model parameters. Further I will switch to a description of the correlations in the disorder. I will present some general remarks as well as a method of treating one type of correlations (generalized N-mers) [51], which is conceptually simple and gives a good insight into the reasons of the appearance of the extended states. Further, I will discuss the methods allowing creation of the correlations in the optical lattices and show results for several cases, which either have interesting analytical properties or are realizable in the ultracold atoms experiments.

## OFF-DIAGONAL DISORDER

Apart from a dimensionality of the problem (covered in a section 1.5.3) and specific type of disorder distribution and its correlations (described further in Sec. 2.3) another important factor that has impact on the localization properties is *what* is really disordered, or more precisely: in which part of a Hamiltonian parameters are taken from random distribution. For the Bose-Hubbard model (9) we could roughly sketch three classes: disordered on-site energies, tunneling amplitudes or interaction strengths. First is the most extensively studied case – one originally proposed by Anderson. Last one of course could be present only in interacting systems, taking us into scope of many body localization [57, 72]. Lets concentrate here on the remaining one – model in which tunneling amplitudes are taken from a random distribution. Systems of this class are often called the systems with off-diagonal disorder (jargon comes from a matrix representation of the Hamiltonian in the tight-binding approximation in which disorder is present in off-diagonal terms) in contrary to the diagonal disorder (disordered on-site energies).

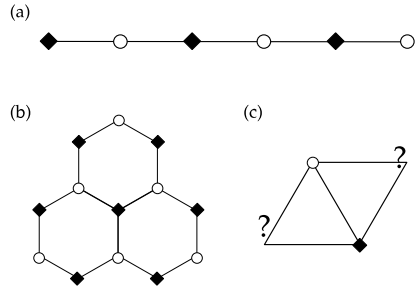

Figure 9: Examples of the lattices which are bipartite for $E = 0$: (a) and (b), as well as one which is not (c). White disks and black squares denotes two decoupled sub-lattices.

Models with a purely off-diagonal disorder have been investigated nearly from the beginning of the Anderson localization theory. An interesting property pointed out first in year 1975 by Weißmann and Cohan [73] is that the states with the energies around a center of the band (i.e., $E = 0$) seem to be extended even in one-dimension. It had been a rather exciting possibility of the existence of the mobility edge in the one-dimensional systems, thus the properties of the states for $E = 0$ have been extensively studied. A lot of arguments for and against the extendedness of this states have been undertaken [74–76]. The discussion substantially have been cut in 1981 by a work [77], which showed that for $E = 0$ the transmission coefficient for infinite system is zero (the wave-function envelope scales as $\exp(-\gamma\sqrt{L})$, with L being the system size). The localization persist, but its form is different than for the standard Anderson localization (other works on this topic are [78, 79]). The further investigation of this class [80], showed that anomalous transport properties for the states in the band center are not strictly connected with the sole off-diagonality of the disorder, but rather it appears when lattice becomes bipartite (when it could be decomposed into two independent lattices). This indeed takes place for one-dimensional lattice with the nearest neighbor hoppings, but also for example for a square lattice or a honeycomb lattice, but not for a triangular lattice (Fig. 9). In bipartite case we could describe behavior of the particle as a random walk [80]. For the one-dimensional lattice it is straightforward to observe that, if there is no inhomogeneity in the on-site energies, we could set $\epsilon_i = 0$ and for $E = 0$ the Schrödinger equation becomes:

$$0 = -t^*_{i-1}\psi_{i-1} - t_i\psi_{i+1}, \tag{32}$$

thus odd sites are decoupled from even ones.

For the one-dimensional systems with both the diagonal and the off-diagonal uncorrelated disorder all states are Anderson localized, however when the diagonal disorder becomes small compared to the off-diagonal one, the residue of this anomaly could be observed as a significant increase of localization length in the band center. Furthermore, as we will see, presence of the off-diagonal disorder could significantly modify the properties of the systems with correlated disorder, changing energies of extended states or even causing their appearance in the otherwise localized systems [37, 81].

*Flores method*

For certain distributions of the off-diagonal disorder, there exist a method (devised by Flores [82]), which allows to write a system in a form with only on-site disorder. Then it is possible to treat systems with techniques applicable only for the diagonal disorder (as Izrailev's Hamiltonian map approach [83] described further in the section 2.3.1). We start from the time-independent Schrödinger equation with the disordered tunnelings (possibly also with the disordered onsite energies):

$$E\psi_i = \epsilon_i\psi_i - t^*_{i-1}\psi_{i-1} - t_i\psi_{i+1}. \tag{33}$$

We can make a transformation $\psi_i = \eta_i\phi_i$, where $\eta_i$ is defined by a recursion formula:

$$\eta_i\eta_{i-1} = \frac{1}{t_i} \tag{34}$$

or, if expressed in a more convenient way:

$$\eta_i = \frac{t_{i-1}}{t_i} \dots \frac{t_1}{t_2}\eta_0. \tag{35}$$

Point in which we start recurrence and the starting value ($\eta_0$) can be chosen arbitrary. Applying (34) to (33) we get:

$$0 = |\eta_1|^2(\epsilon_i - E)\phi_i - \phi_{i-1} - \phi_{i+1}, \tag{36}$$

so all of tunnelings have been set to one. Unfortunately, there is one issue rendering this method inapplicable in many situations. Even if $\{t_i\}$ are distributed in a way that the mean value of $\eta_i$ is finite (which in fact is usually the case), in the absence of correlations, the variance of the distribution grows fast. As an example we could use binary off-diagonal disorder. For uncorrelated disorder the fluctuations of $\phi_i$ grows exponentially fast while if we have correlated dual random dimer model (discussed further in Sec. 2.5.3) $\eta_i$ takes only three possible values.

*Periodic modulation*

Effects of a fast periodic modulation on the quantum systems are described in the Sec. 1.4 as well as in the App. A. Here I am going to present only two specific cases which allow to transfer the diagonal disorder into the off-diagonal terms in the tight-binding models.

PERIODIC MODULATION OF THE ON-SITE ENERGY. It was proposed for superlattices [84], but it could be immediately adopted to systems with on-site disorder (or any system with inhomogeneous on-site energies). Hamiltonian (7) with periodically modulated on site energy $\epsilon_i \rightarrow \epsilon_i^0 + \epsilon_i^1 f_\omega(\tau)$, where $f_\omega(\tau)$ is a periodic function with frequency $\omega$ and period $T = 2\pi/\omega$, has a form[1]:

$$H(\tau) = \sum_i (\epsilon_i^0 + \epsilon_i^1 f_\omega(\tau)) n_i - t(a_i^\dagger a_{i+1} + \text{h.c.}). \qquad (37)$$

Following the standard procedure I am making transformation:

$$\mathcal{U} = \exp\left(iF_\omega(\tau) \sum \epsilon_i^1 n_i\right), \qquad (38)$$

to go to the rotating frame ($F_\omega(\tau)$ is the antiderivative of $f_\omega(\tau)$). Transformed Hamiltonian reads:

$$H_{\text{rot}}(\tau) = \sum_i \epsilon_i^0 n_i - t \exp\left(i(\epsilon_{i+1}^1 - \epsilon_i^1)F_\omega(\tau)\right)\left(a_i^\dagger a_{i+1} + \text{h.c.}\right). \qquad (39)$$

If only the tunneling between sites takes much longer time than the modulation period $t \ll \omega$ (from now on we set $\hbar = 1$), we could rely on the high frequency approximation (zeroth order of the Magnus expansion). Effective Hamiltonian is then just a time average of (39):

$$H_{\text{eff}} = \sum_i \epsilon_i^0 n_i - t_i^{\text{eff}}[f_\omega](a_i^\dagger a_{i+1} + \text{h.c.}). \qquad (40)$$

For the most widely used harmonic modulation $f_\omega(\tau) = \sin(\omega\tau)$, the effective tunneling is given by:

$$t_i^{\text{eff}}[\sin \omega\tau] = t\mathcal{J}_0\left(\frac{\epsilon_{i+1} - \epsilon_i}{\omega}\right), \qquad (41)$$

---

[1] I have used the 2nd quantized form of the Hamiltonian as it is more convenient for calculations and it is easier to generalize it further to interacting many-body case.

with $\mathcal{J}_0(x)$ – zeroth order Bessel function. Effective Hamiltonian (40) with $t_i^{\text{eff}}[\sin \omega \tau]$ further denoted just as $t_i^{\text{eff}}$ is a basis for most of the results in this chapter.

Throughout this work, the used modulation of the on-site energies is *de facto* modulation of interspecies interaction between *mobile* and *frozen* particles. Interactions between two species could be tuned using the Feshbach resonances similarly to interactions among one type of atoms. By periodically varying the magnetic field in vicinity of the Feshbach resonance (as in [85]) the fast periodic modulation of interactions could be generated. The problem appears when we want to set interactions for *mobile* particles to zero ($U = 0$), the magnetic Feshbach resonance cannot be used anymore, instead microwave or optical Feshbach resonances could be employed [38].

MODULATION OF TUNNELING RATES.    Another simple scheme of the periodic modulation is changing the tunneling amplitudes. Actually, this effect usually appears also when we modulate on-site energies but is very small. Better method of varying tunneling amplitude is modulation of the lattice height:

$$H(\tau) = \sum_i \left( \epsilon_i n_i - (t_0 + t_1 \sin(\omega \tau)) a_i^\dagger a_{i+1} + \text{h.c.} \right), \qquad (42)$$

where the mean value of tunneling is $t_0$, the amplitude of modulation $t_1$ and frequency $\omega$. As we are modulating a nonlocal operator $\sum_{\langle ij \rangle} a_i^\dagger a_j$, upon making transformation to a rotating frame it will generate long-range tunnelings. However, their amplitudes are vanishing quickly with the distance (approximately as $2^{3/2j}/(j!!)^2$ for tunneling range $j$), thus the method can not be used to effectively create a system with tunnelings further than next-neighbors. The effective Hamiltonian reads:

$$H_{\text{eff}} = - \sum_i t_0 a_i^\dagger a_{i+1}$$
$$+ \sum_i \sum_{j=0}^\infty \sum_{k=j}^\infty \mathcal{K}_k^j \left( \frac{t_1}{\omega} \right)^{2k} \delta^{2k}[\epsilon_i] a_{i-j}^\dagger a_{i+j} + \text{h.c.}, \qquad (43)$$

where

$$\mathcal{K}_k^j = \frac{(-1)^j (2k-1)!!}{(k+j)!(k-j)!k!} \qquad (44)$$

and

$$\delta^n[\epsilon_i] = \sum_{j=0}^n (-1)^j \binom{n}{j} \epsilon_{i+n/2-j} \qquad (45)$$

is the $n$-th order central finite difference of the on-site energies on site $i$. In expansion to first nontrivial order we get:

$$H_{\text{eff}} \approx \sum_i \left( \epsilon_i - 2 \left( \frac{t_1}{\omega} \right)^2 (\epsilon_{i+1} - 2\epsilon_i + \epsilon_{i-1}) \right) n_i - t a_i^\dagger a_{i+1}$$

$$+ \left( \frac{t_1}{\omega} \right)^2 (\epsilon_{i+1} - 2\epsilon_i + \epsilon_{i-1}) \, a_i^\dagger a_{i+1} + \text{h.c.} \tag{46}$$

As a modulation frequency have to be big compared to tunneling rates we do not expect very strong effects. However, even a weak coupling to next-nearest neighbors could sometimes significantly alter the system properties (as for example shown in section 2.5.3).

### CORRELATIONS

The original theory of the localization have been created for an uncorrelated disorder [49, 53, 56]. It is an interesting question, if and how correlations can change properties of disordered systems. Especially, would they allow appearance of the mobility edge in otherwise fully localized systems? First attempts to address this problem have been done for random dimer model [86] described in the next section. Indeed, even in the one-dimensional systems it have been shown that adding correlations could result in appearance of the extended states – which however do not form a dense set – we do not have mobility edge. Still the possibility of getting those states is intriguing, they could also be of practical importance in the finite systems, as localization length in their vicinity grows to large values, thus finite systems could become transparent in a window of energies around the extended state.

There is no general description of effects of correlations for all cases. Methods described in further parts concentrate on specific 'block' type of correlations, although they work for arbitrary values of disorder amplitude. Another, very interesting approach have been carried out by Izrailev and Krokhin [50] as they gave formula for localization length with the arbitrary correlations but only in the weak disorder limit. I believe it is a method worth knowing so I will briefly resume here the main points:

*The Hamiltonian map approach [50]*

Lets start again with the time independent Schrödinger equation:

$$0 = \psi_{i+1} + \psi_{i-1} + \mathcal{E}_i(E)\psi_i, \tag{47}$$

where $\mathcal{E}_i$ is general expression for local part of the Schrödinger equation. In case of standard tight binding model it is just $(\epsilon_i - E)$, but in principle it can also have other forms (for example obtained using

Flores method described in Sec. 2.2.1). We could write equation (47) as a two-dimensional Hamiltonian map with a position and a momentum defined as $x_i = \phi_i$, $p_i = (\phi_i \cos(k) + \phi_{i-1})/\sin(k)$, where $k$ is quasimomentum[2]:

$$\begin{cases} x_{i+1} &= -(p_i + A_i x_i)\sin(k) + x_i \cos(k) \\ p_{i+1} &= (p_i + A_i x_i)\cos(k) + x_i \sin(k), \end{cases} \tag{48}$$

where $A_i$ depends on diagonal term $\mathcal{E}_i$ (for $\mathcal{E}_i = \epsilon_i - E$, $A_i = -\epsilon_i/\sin k$). The equation (48) describes the dynamics of a classical kicked rotor – harmonic oscillator undergoing periodic delta kicks with (random) amplitude $A_i$. After transforming the map into action-angle variables ($x = r\sin(\theta)$, $p = r\cos(\theta)$) we get:

$$\begin{cases} \sin(\theta_{i+1}) = D_i^{-1}\left(\sin(\theta_i - k) - A_i \sin(\theta_i)\sin(k)\right) \\ \cos(\theta_{i+1}) = D_i^{-1}\left(\cos(\theta_i - k) + A_i \sin(\theta_i)\cos(k)\right) \end{cases}, \tag{49}$$

where

$$D_i \equiv \frac{r_{i+1}}{r_i} = \sqrt{1 + 2A_i \sin(2\theta_i) + A_i^2 \sin^2(\theta_i)}. \tag{50}$$

The Anderson localization length do not depends on angle, only on the radius $r$ and is given by the expression:

$$\lambda^{-1} = \lim_{N \to \infty} \frac{1}{N} \sum_{i=1}^{N} \ln D_i. \tag{51}$$

If the disorder is small (i.e. $A_i \ll 1$) we could expand the logarithm to the second order in $A_i$:

$$\lambda^{-1} = \frac{1}{8}\langle A_i^2 \rangle + \frac{1}{2}\langle A_i \sin(2\theta_i) \rangle, \tag{52}$$

where $\langle ... \rangle$ is an average over $i$. What is left to calculate is the 'kick-angle' correlator $\langle A_i \sin(2\theta_i) \rangle$. To do this we expand the map (49) to the second order in $\theta_i$:

$$\theta_i = \theta_{i-1} - k + A_{i-1}\frac{\sin^2(\theta_{i-1})}{\sin(k)} \tag{53}$$

and recursively write $\langle A_i \sin(2\theta_i) \rangle$ using $\langle A_i \sin(2\theta_{i-1}) \rangle$ 'kick-angle' correlator. Defining correlation of the angle $\theta_i$ with the kick's strength $A_i$:

$$a_n = -\frac{2i}{\langle A_i^2 \rangle} e^{2ik} \langle A_i e^{2i\theta_{i-n}} \rangle, \tag{54}$$

---

2 As we are assuming weak disorder limit we could use dispersion relation for ordered one-dimensional lattice: $E = -2\cos k$.

(where $\langle A_i^2 \rangle$ is variance of the $A_i$) and applying (53) several times on $a_n$ give us the relation:

$$a_{n-1} = e^{-2ik} a_n + q_n, \tag{55}$$

with $q_n = \langle A_i A_{i-n} \rangle / \langle A_i^2 \rangle$ an autocorrelation of $A_i$. From definition of $a_n$ we can notice that

$$\langle A_i \sin(2\theta_i) \rangle = \mathrm{Re}\left( \frac{\langle A_i^2 \rangle}{2} e^{-2ik} a_0 \right), \tag{56}$$

from (55) we obtain $a_0 = \sum_{n=1}^{\infty} q_n e^{-2ik(n-1)}$.

Finally the inverse Anderson localization length $\lambda^{-1}$ is given by [83]:

$$\lambda^{-1} = \frac{\langle A_i^2 \rangle}{8} \left( 1 + 2 \sum_{n=1}^{\infty} q_n(k) \cos(2kn) \right). \tag{57}$$

All information about correlations in the system is stored in $q_n$ – two point correlators of $A_n$. If the correlations part (in parenthesis) is zero for some quasimomentum $k_e$ also $\lambda^{-1}(k_e)$ is zero and we get extended state.

*Random N-mer model*

Specific type of the correlations, which could appear in the systems with a binary disorder are random N-mers. We can assume that a certain value of the on-site energy always comes in the rows of length N. Such system has been proposed in [86] in its simplest form: a dimer model, where possible on-site energies $\{\epsilon_a, \epsilon_b\}$ always comes in pairs. An interesting property, pointed out by the authors, was fact that although it seems that we could map this system onto an effective lattice with the uncorrelated disorder (as in Fig. 10), the localization properties change significantly and two extended states with energies $\epsilon_a$ and $\epsilon_b$ appear (as in Fig. 11). Model has been further generalized to a N-mer model, where the rows with one (or both) energies have length N. It have been shown using several methods, that such systems have multiple extended states [83, 87–91] (if one energy comes in the rows of length $N_a$ and second of length $N_b$ there are $N_a + N_b - 2$ extended states) given by the formula:

$$E_R \in \left\{ \epsilon_a + 2\cos\left(\frac{\pi}{N_a} i\right) \quad \text{for} : i \in \{1, N_a - 1\} \right\} \cup$$
$$\left\{ \epsilon_b + 2\cos\left(\frac{\pi}{N_b} j\right) \quad \text{for} : j \in \{1, N_b - 1\} \right\}. \tag{58}$$

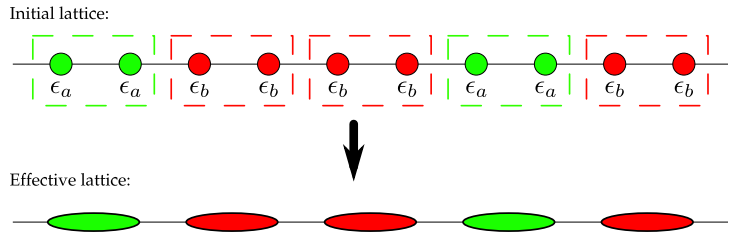

Figure 10: Random dimer model in the one-dimensional lattice: energies $\epsilon_a$ and $\epsilon_b$ always comes in pairs. Lattice could be mapped onto effective lattice without correlations but its transport properties differ significantly from the original one.

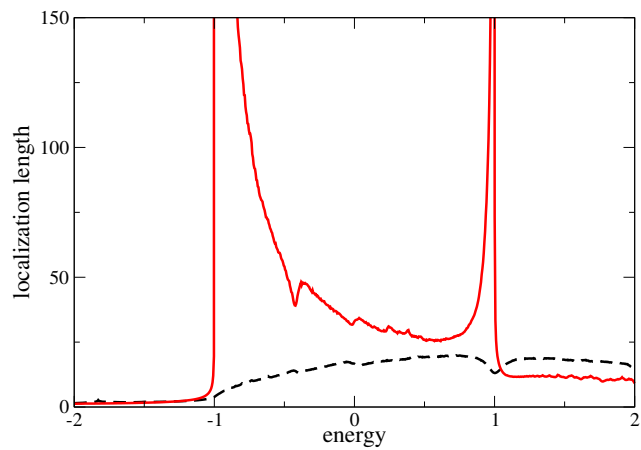

Figure 11: The localization length (in units of lattice constant) as a function of the state energy (in units of the tunneling rate) for the binary disorder with $\{\epsilon_a, \epsilon_b\} = \{-1, 1\}$. Black dashed line – uncorrelated binary disorder, red solid – dimer model. The delocalization resonances are clearly visible for $E_R = 1$ and $-1$, localization length as a function of energy for both models differs significantly, even far from the resonant energies.

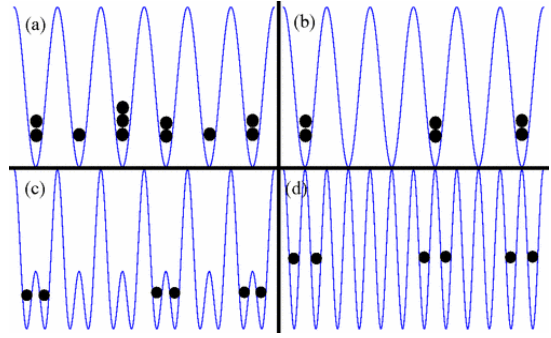

Figure 12: Procedure of creating a random dimer model. (a) Loading the *frozen* particles into the auxiliary lattice, (b) removing particles from all sites but those with occupation 2, (c) adiabatic turning on the main lattice, (d) turning off the auxiliary lattice (source: [92]).

*Experimental realization*

Using the standard setup for the binary disorder described in section 1.6 we will get uncorrelated disorder. I will describe two methods of adding correlations into distribution of frozen particles in the experiment.

TAILORING CORRELATIONS.    One of methods allowing creation of a system with N-mer correlations is to use the same lattice as for uncorrelated disorder but modify preparation procedure. Instead of letting *frozen* particles to distribute randomly over lattice, one needs to add additional steps to obtain desired correlations.

In the case of the random dimer model, such a procedure is described in [92]. As presented in Fig. 12: first, one lets *frozen* atoms to distribute randomly in the auxiliary lattice which lattice constant is twice of the planned one. Further the sites with the occupation number different than 2 are purified by a selective removal of the atoms (using techniques described for example in [93, 94]). After the purification there should be only sites with zero or double occupancies left. Next step is an adiabatic splitting of wells, done by slowly turning on main lattice – this should result in atoms from the doubly occupied states falling into both of the new sites. Then auxiliary lattice is turned off and resulting system has correlations of type of random dimer model. A dual random dimer model also could be created by even simpler procedure (without atom removal phase) described in [37]. In principle N-mer with arbitrary length could be created using more complicated versions of this procedure but it increases risk of errors (randomly placed impurities in N-mer structure) which are quickly consuming the most interesting feature of those systems – appearance of extended states for certain energies.

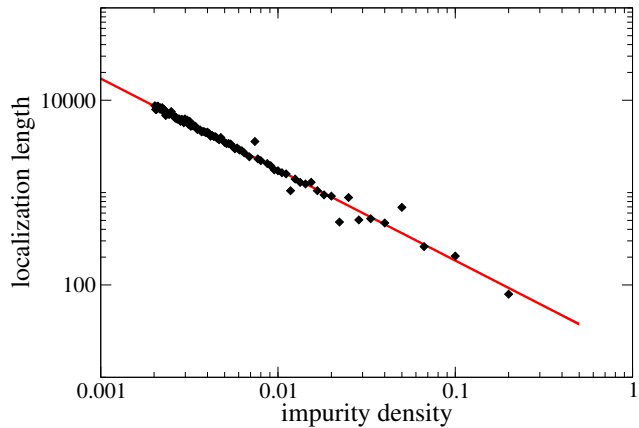

Figure 13: Localization length (in units of lattice constant) as a function of density of impurities in 6-mer. Results for $E = 0$ and $\epsilon = 1$ shows power-law decay of localization length with number of imperfectly created N-mers (with exponent close to one).

It could be visualized on a simple toy model where I assume that there is certain probability that particles in step (c) do not distribute evenly. Results showing power-law decay of localization length for the 6-mer with on site energies $0$ or $1$ for the energy $E = 0$ are shown in the figure 13.

BICHROMATIC LATTICE.    Another possible way of creating the correlations in the optical lattices is using a bichromatic lattice [51]. We could use two overlapping lattices with lattice constants $b$ and $b^f$ and opposite circular polarizations. I consider *mobile* and *frozen* particles to be two hyperfine states of the same element – each of them 'feeling' only one polarization. In this way the *mobile* particles move in lattice with constant $b$ (this one is nicknamed main and all further numbering of sites refers to it) while *frozen* ones are hold in the second.

When both lattice constants $b$ and $b_f$ are equal we have back again binary disorder, but if we take $^{b_f}/b = l$ where $l \in \mathbb{N}$ we have system presented in the figure 14. When there is no frozen particle on some site we have $l$ sites of main lattice with effective on-site energy zero but when there is *frozen* particle, on-site energies for mobile particles could be changed on several sites in a manner depending on the range of interactions, the length of blocks etc.

Using two optical lattices with lattice constants differing by a large factor (in further examples $l = 7$) may look as the most problematic request for the experimental realization. However it could be met, for example with the use of $CO_2$ laser (with long wavelength $10.6\mu m$) [95] and lasers from the telecommunication band (1260-1675nm) or similar [96].

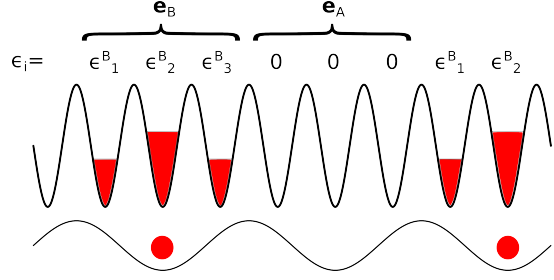

Figure 14: Experimental setup for the generalized N-mer (with length $l = b^f/b = 3$): two optical lattices overlap (shifted vertically on the picture). One lattice affects the *mobile* particles, the other *frozen* ones. The system could be decomposed into two types of randomly placed blocks of length $l$ (if only the interactions between the species falls with the distance fast enough). When the *frozen* particle is present (depicted as red dot) it effectively changes on-site energies for *mobile* particles, whereas when it is absent we have the row with length $l$ and zero energy.

### generalized random N-mer model

The correlations which could be created using setup described above reaches beyond the N-mer class as they are inhomogeneous blocks. The new wider class could be defined and called a generalized N-mer model. In this case we also have blocks of fixed length but now they are inhomogeneous – they could have internal structure. Also number of different structures, which for N-mers is taken to be 2, could be larger (in principle it can be even infinite countable). In other words, we have a system which could be divided into finite number of different blocks, which have fixed internal structure and came in random order. It could be also presented as the effective uncorrelated lattice as in Fig. 15. I denote different block types by capitals $\{A, B \ldots\}$, each block is defined by its length $l_A$, set of the on-site energies $\mathbf{e}_A = \{\epsilon_1^A \ldots \epsilon_{l_A}^A\}$ and possibly also the inhomogeneous tunneling rates $\mathbf{t}_A = \{1, t_1^A \ldots t_{l_A-1}^A, 1\}$. One could notice that tunneling rates are set to constant value at the edges (here chosen to be the energy scale) – it is required for consistency of the presented method. If the tunnelings for different blocks differ, it is impossible to treat block distribution as uncorrelated because the edge tunnelings are shared among two blocks and appearance of some tunneling value at the end of one block creates constraint for possible next blocks. Further I will show that the condition of homogeneity of the edge tunnelings is physically reasonable at least for most of the presented models.

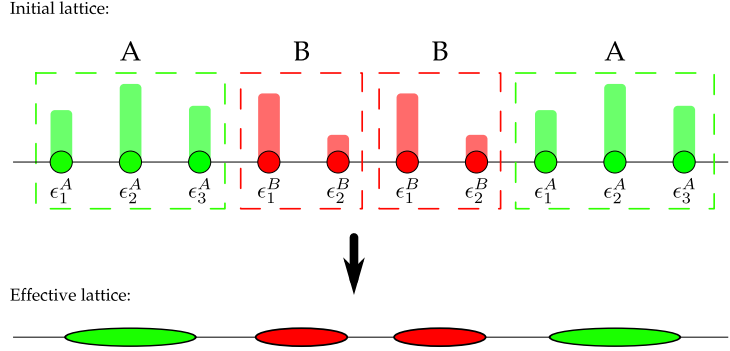

Figure 15: Generalized N-mer in the one-dimensional lattice. Different blocks are denoted by boxes, for each block type set of on-site energies and tunnelings is always the same. One can find transfer matrix $T_X(E)$ describing transport through any type of the block.

*Method of finding the extended states*

To analyze transport properties of this class of systems we could use analytical properties of transfer matrices. For simplicity we will limit ourselves to one dimensional tight-binding with nearest neighbor tunnelings (although method could be used for more complicated systems). In this case the time independent Schrödinger equation:

$$E\psi_i = \epsilon_i\psi_i - t_i\psi_{i+1} - t_{i-1}^*\psi_{i-1}, \qquad (59)$$

could be rewritten into a matrix form:

$$\begin{bmatrix} \psi_{i+1} \\ \psi_i \end{bmatrix} = \begin{bmatrix} \frac{\epsilon_i - E}{t_i} & -\frac{t_{i-1}}{t_i} \\ 1 & 0 \end{bmatrix} \begin{bmatrix} \psi_i \\ \psi_{i-1} \end{bmatrix} \equiv T_i \begin{bmatrix} \psi_i \\ \psi_{i-1} \end{bmatrix}, \qquad (60)$$

where $T_i$ is called transfer matrix through the site $i$. Transfer matrix between any two sites $i$ and $j$ could be easily calculated by chaining:

$$\begin{bmatrix} \psi_{j+1} \\ \psi_j \end{bmatrix} = T_j \cdot \ldots \cdot T_i \begin{bmatrix} \psi_i \\ \psi_{i-1} \end{bmatrix} \equiv T_i^j \begin{bmatrix} \psi_i \\ \psi_{i-1} \end{bmatrix}. \qquad (61)$$

Matrix $T_i^j$ describes transport through given part of the system including effects of internal scattering, interference of scattered waves etc. In the presented method we will not calculate transfer matrices for specific parts of the system, but rather for different block types $\{A, \ldots\}$:

$$\mathbb{T} = \{T_A(E), T_B(E) \ldots\} \qquad (62)$$

where:

$$T_A(E) = T(\mathbf{e}_A, \mathbf{t}_A, E) = \prod_{i=1}^{l_A} \begin{bmatrix} \frac{\epsilon_i^A - E}{t_i^A} & -\frac{t_{i-1}^A}{t_i^A} \\ 1 & 0 \end{bmatrix}. \qquad (63)$$

We could now write transfer matrix of the whole system:

$$\mathcal{T}(E) = \prod_i T_i(E), \tag{64}$$

as the product of transfer matrices of blocks:

$$\mathcal{T}(E) = \prod_X T_X(E), \quad \text{where } T_X(E) \in \mathbb{T}. \tag{65}$$

In fact we have just rewritten our model into a form of the effective lattice as in Fig. 15, the question is how it will affect the transport properties? To answer it, we have to check commutation properties of our transfer matrices. We search for an energy (or energies) $E_R$ for which transfer matrices for all of blocks will commute pairwise:

$$[T_X(E_R), T_Y(E_R)] = 0 \quad \forall T_X(E_R), T_Y(E_R) \in \mathbb{T}. \tag{66}$$

State with the energy which fulfills above condition will not be Anderson localized. It is so, as from commutation follows that we could find one matrix $C$ diagonalizing all transfer matrices in $\mathbb{T}$. Thus we could write (by $D_X$ denoting diagonal form of $T_X$)

$$\mathcal{T}(E_R) = C \left( C^{-1} \mathcal{T}(E_R) C \right) C^{-1} = C \left( \prod_X C^{-1} T_X(E_R) C \right) C^{-1}$$

$$= C \left( \prod_X D_X(E_R) \right) C^{-1}. \tag{67}$$

Now the transport through the whole system is given by the product of diagonal matrices. The change of state amplitude is given solely by product of eigenvalues of those transfer matrices regardless their ordered/disordered arrangement. To check whether our state is extended we have to determine amplitudes of those eigenvalues. In the considered cases transfer matrix always has a determinant equal one[3] and there are two possible forms of its eigenvalues $\{\alpha_1, \alpha_2\}$: they could be either real and opposite ($\alpha_1 = 1/\alpha_2 \in \text{Re}$) or complex with amplitude one and opposite phases ($|\alpha_1| = |\alpha_2| = 1$ and $\alpha_1 = \alpha_2^*$). For infinite system composed of one block type (ordered system) those two cases correspond to 'out of band' and 'in band' energies and I will use this terminology here, although for only one block it could be a bit misleading. If $E_R$ lies within the band for all types of the blocks, then every diagonal transfer matrix $D_X$ is only adding some complex phase to initial state $(\psi_1, \psi_0)$, thus state amplitude is unchanged and we get just a plane wave propagation. On the other hand, if $E_R$ falls out of the bandwidth of even one block, each time when such a block occurs our initial state amplitude will be changed, after passing many of them the wavefunction will drop exponentially, in this case it is not the multiple scattering effect as for Anderson localization but just the evanescent decay of the wave as we are making calculations out of the band.

---

3 Determinants of elementary transfer matrices (60) depends on tunneling amplitudes but one tunneling amplitude $t_i$ appears always in two transfer matrices $T_i$ and $T_{i+1}$ – once in denominator and once in numerator of determinant. As long as tunnelings on the edges of considered structure are homogeneous determinant of it is 1.

In the further analysis I will usually restrict myself to the case of two structures $T_A(E)$ and $T_B(E)$. Then there is only one commutator and we search for $E_R$ for which:

$$[T_A(E_R), T_B(E_R)] = 0. \tag{68}$$

*Few specific cases*

Before utilizing theory developed above I will make some comments to point out interesting properties and possibilities given by presented method.

SYMMETRIC STRUCTURES    I will only consider structures with symmetric on site energies and tunnelings as such are created in the experimental setup sketched above (Sec. 2.3.3). In such a case it is straightforward to check that the off-diagonal elements of transfer matrix are opposite: $[T]_{12} = -[T]_{21}$ for all $E$ (it somewhat simplifies calculations f.e. allows obtaining formula (93)).

When the structure is not symmetric the ratio $[T]_{12}/[T]_{21}$ usually depends on $E$, however for the relation (66) to hold, ratios for all structures at energy $E_R$ must be the same.

TWO *vs.* MANY STRUCTURES    Choice of two structures is dictated by proposed experimental setup (Sec. 2.3.3), also in this case the extended states should be more robust. Even with some parameters changed slightly, if only a perturbation is weak enough, $E_R$ could change but the resonance will not vanish. On the other hand, although, it is in principle possible to construct an arbitrary large set of $2 \times 2$ matrices commuting pairwise, it will be much harder to observe the resonance then. In an ideal case the resonant energy will be the same for every commutator $E_r^{(j)} = E_R$ for all $j$, but in the perturbed system those energies could change differently: $E_r^{(j)\prime} = E_R + \varepsilon_j$ and the resonance will vanish. This can be possibly method of making some very precise measurements, as we get strong signal only for fine tuned parameters, but I will not continue this thread here.

EXTENDED STATES: IN, OUT AND AT THE EDGE OF A BAND    Two interesting properties could be obtained by checking the condition for commutation in conjunction with the condition for positions of the band edges. We have two arbitrary transfer matrices $T_A(E)$ and $T_B(E)$ of symmetric structures ($[T_X]_{12} = -[T_X]_{21}$). If we want to check what energies are in band for some structure we have to find eigenvalues of its transfer matrix. They read:

$$\alpha^X_{+/-}(E) = \frac{1}{2}\big([T_X(E)]_{11} + [T_X(E)]_{22}$$
$$\pm \sqrt{([T_X(E)]_{11} - [T_X(E)]_{22})^2 - 4[T_X(E)]^2_{21}}\,\big). \qquad (69)$$

The most important is the sign of the part under the square root, which we could denote:

$$\sigma_X(E) = ([T_X(E)]_{11} - [T_X(E)]_{22})^2 - 4[T_X(E)]^2_{21}. \qquad (70)$$

If $\sigma_X(E)$ is negative we are in band. On band edges it reaches zero (it could also be zero inside the band, although it is rare). On the other hand the condition for commutation of $T_A$ and $T_B$ is:

$$\frac{[T_A(E_R)]_{11} - [T_A(E_R)]_{22}}{[T_A(E_R)]_{21}} = \frac{[T_B(E_R)]_{11} - [T_B(E_R)]_{22}}{[T_B(E_R)]_{21}}. \qquad (71)$$

From this equations we could get two facts:

1. If we assume that (71) is satisfied, by putting (70) into it we can easily get:

$$\frac{\sigma_A(E_R)}{[T_A(E_R)]^2_{21}} = \frac{\sigma_B(E_R)}{[T_B(E_R)]^2_{21}}. \qquad (72)$$

    Thus, if only $[T_X(E_R)]_{21}$ is nonzero, we know that the resonant energy is in the band or out of the band for both structures simultaneously (this condition do not apply for example for N-mer resonances for which off-diagonal elements of transfer matrix are zero for $E_R$).

2. If the band-edges of two structures coincide for some energy, so that $\sigma_A(E_b) = \sigma_B(E_b) = 0$ it is also straightforward to check that (71) is satisfied up to a sign. Due to the sign not all such coinciding edges will mark extended state but as the chances are even such points are worth checking. Resonances of this type could be especially interesting, as due to being at the edge of the band they are usually very narrow.

BEYOND ONE-DIMENSION    I am concentrating on the structures, which could be created in the optical lattices but it is worth emphasizing that general condition (66) is applicable for any system for which the transfer matrix could be found. Especially, higher dimensional systems with single in and out leads could be interesting. If we have a *device* described by an arbitrary Hamiltonian: $H_D$ with the only assumption that its first and last sites are connected to the one-dimensional leads, in most of the cases it is possible to find transfer matrix describing such system. We start from a block Hamiltonian:

$$H = \left[ \begin{array}{c|c|c} H_L & \mathcal{C}_0 & \\ \hline \mathcal{C}_0^\dagger & H_D & \mathcal{C}_0 \\ \hline & \mathcal{C}_0^\dagger & H_L \end{array} \right] \tag{73}$$

where $H_L$ is a Hamiltonian of an one-dimensional ideal lead:

$$H_L = \sum_i -a_i^\dagger a_{i+1} + h.c. \tag{74}$$

and $\mathcal{C}_0$ is connection between the leads and the *device*:

$$\mathcal{C}_0 = \left[ \begin{array}{ccc} \vdots & & \iddots \\ 0 & 0 & \\ 1 & 0 & \dots \end{array} \right]. \tag{75}$$

If we apply a transformation:

$$\mathcal{U} = \left[ \begin{array}{c|c|c} \mathbb{I} & & \\ \hline & u & \\ \hline & & \mathbb{I} \end{array} \right] \tag{76}$$

where $u$ is matrix diagonalizing $H_D$, we get:

$$H' = \left[ \begin{array}{c|c|c} H_L & \mathcal{C} & \\ \hline \mathcal{C}^\dagger & D_D & \mathcal{C}' \\ \hline & \mathcal{C}'^\dagger & H_L \end{array} \right]. \tag{77}$$

Above $D_D$ is diagonal form of $H_D$ and new coupling matrices $\mathcal{C}$ and $\mathcal{C}'$ reads:

$$\mathcal{C} = \left[ \begin{array}{ccc} \vdots & & \iddots \\ 0 & 0 & \\ u_{11} & u_{12} & \dots \end{array} \right], \quad \mathcal{C}' = \left[ \begin{array}{ccc} u_{n_D 1} & 0 & \dots \\ u_{n_D 2} & 0 & \\ \vdots & & \ddots \end{array} \right], \tag{78}$$

Thus they are filled by *first* and *last* elements of eigenvectors of $H_D$. We denote $J_\alpha = u_{\alpha 1}$ and $J'_\alpha = u_{\alpha n_D}$. Now we could write set of equations:

$$E\psi_i = -\psi_{i-1} - \sum_\alpha J_\alpha \psi_\alpha, \tag{79}$$

$$E\psi_j = -\psi_{j+1} - \sum_\alpha J'_\alpha \psi_\alpha, \tag{80}$$

$$E\psi_\alpha = \varepsilon_\alpha \psi_\alpha - J_\alpha \psi_i - J'_\alpha \psi_j \quad \text{for}: \alpha = 1 \dots n_D, \tag{81}$$

where $i$ is last site of *in* lead and $j$ the first site of *out* lead, $\psi_\alpha$ are occupations of the channels (eigenmodes) of the $H_D$ and $\varepsilon_\alpha$ are eigenvalues. If $E \notin \{\varepsilon_\alpha\}$ we could apply set of equations (81) to (79) and (80) obtain:

$$E\psi_i = -\psi_{i-1} - \tilde{t}\psi_j + \tilde{\varepsilon}_i\psi_i, \tag{82}$$

$$E\psi_j = -\psi_{j+1} - \tilde{t}\psi_i + \tilde{\varepsilon}_j\psi_i, \tag{83}$$

where I have denoted:

$$\tilde{\varepsilon}_i = \sum_\alpha \frac{1}{E - \varepsilon_\alpha} J_\alpha^2, \tag{84}$$

$$\tilde{\varepsilon}_j = \sum_\alpha \frac{1}{E - \varepsilon_\alpha} J_\alpha'^2, \tag{85}$$

$$\tilde{t} = \sum_\alpha \frac{1}{E - \varepsilon_\alpha} J_\alpha J_\alpha'. \tag{86}$$

Further it is straightforward to calculate transfer matrix describing transport through the *device*:

$$\begin{bmatrix} \psi_{j+1} \\ \psi_j \end{bmatrix} = \begin{bmatrix} \frac{(\varepsilon_i - E)(\varepsilon_j - E)}{t} - t & -\frac{(\varepsilon_j - E)}{t} \\ \frac{(\varepsilon_i - E)}{t} & -\frac{1}{t} \end{bmatrix} \begin{bmatrix} \psi_i \\ \psi_{i-1} \end{bmatrix}. \tag{87}$$

This transfer matrix could be treated on the same grounds as normal transfer matrices for generalized N-mers. Only special case in when $E \in \{\varepsilon_\alpha\}$ then, the system could not be simply mapped onto one-dimensional lattice.

## EXAMPLES

Below I will present exemplary applications of the method described in preceding section. In most cases I am using the unchanged tunneling amplitude as an energy scale ($t = 1$).

### Binary disorder

A simple binary disorder could be treated with the method. In this case we have $\mathbf{e}_A = \{\varepsilon_a\}$ and $\mathbf{e}_B = \{\varepsilon_b\}$. Then, it is straightforward to check that commutator

$$[T_A(E), T_B(E)] = \begin{bmatrix} 0 & \epsilon_B - \epsilon_A \\ \epsilon_B - \epsilon_A & 0 \end{bmatrix} \tag{88}$$

will be zero only for a trivial case when $\epsilon_A = \epsilon_B$, as we can expect.

### N-*mers*

As the resonance energies for N-mers are well known, this example will allow us to check the method. Also it will show some interesting properties which will help in the further analysis of the generalized N-mers.

Let us denote transfer matrix through N-mer with length $l$ and height $\epsilon$ by: $T_\epsilon^l(E)$. By subsequent multiplying transfer matrices one could check that elements $[T_\epsilon^l(E)]_{ij}$ of $T_\epsilon^l(E)$ have to satisfy recurrence relation:

$$
\begin{aligned}
[T_\epsilon^l(E)]_{11} &= (\epsilon - E)[T_\epsilon^l(E)]_{21} + [T_\epsilon^l(E)]_{22}, \\
&= (\epsilon - E)[T_\epsilon^{l-1}(E)]_{11} - [T_\epsilon^{l-2}(E)]_{11}
\end{aligned}
\tag{89}
$$

which is nearly the recurrence relation satisfied by the Chebyshev polynomials of the second kind ($U_{l+1}(x) = 2xU_l(x) - U_{l-1}(x)$). What follows:

$$
T_\epsilon^l(E) = \begin{bmatrix} U_l(\frac{\epsilon-E}{2}) & -U_{l-1}(\frac{\epsilon-E}{2}) \\ U_{l-1}(\frac{\epsilon-E}{2}) & -U_{l-2}(\frac{\epsilon-E}{2}) \end{bmatrix}.
\tag{90}
$$

If we have in the system rows with constant energy $\epsilon$ ($T_A = T_\epsilon^l(E)$) and empty spaces of arbitrary length ($T_B = T_0^1(E)$), it is straightforward to check that $[T_\epsilon^l, T_0^1] = 0$ only if $U_{l-1} = 0$. Using well known expression for the zeros of the Chebyshev polynomials:

$$
U_l(x_k) = 0, \quad \text{for} : x_k = \cos\left(\frac{2}{l+1}\pi\right), \quad k = 1, \dots, n,
\tag{91}
$$

we retrieve expression (58) for resonant energies (for case of blocks with lengths $l$ and 1):

$$
E_R = \epsilon + 2\cos\left(\frac{\pi}{l}i\right), \quad \text{for } i \in \{1, \dots, l-1\}.
\tag{92}
$$

More important for further analysis are two properties present if one of two-structure generalized N-mer is a 'standard' N-mer:

1. For resonant energies $T_\epsilon^l(E_R) = \mathbb{I}$. As identity matrix commutes with everything, we will always get set of extended states given by (92) as long as they do not fall out of band of the second block.

2. Using recurrence relation (89) we could simplify equation for resonant energies (68) getting formula:

$$
[T_B(E)]_{11} = (\epsilon - E)[T_B(E)]_{21} + [T_B(E)]_{22}.
\tag{93}
$$

*Dual Random Dimer Model (DRDM)*

This model is as old as the dimer model, as it was introduced in the same article [86]. I will present it separately, as it provides several simple yet interesting results. It is also good example, showing how things get complicated when we break the assumption that tunnelings at the edges of the structures always have the same value.

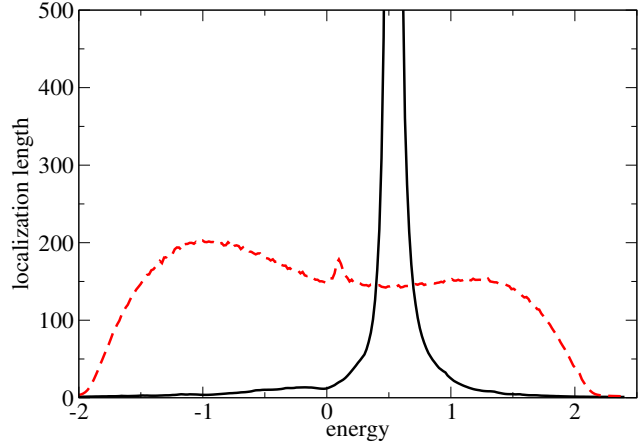

Figure 16: Localization length (in units of lattice constant) plotted as a function of energy (in units of tunneling amplitude) for DRDM with $V_0 = 0.4$. Red dashed line is for $t' = 1$ (unchanged tunnelings), while black solid line for $t' = 0.5$ [91].

In DRDM we assume that the *frozen* particles could not occupy two adjacent sites. In generalized N-mer terminology it will mean that we have $\mathbf{e}_A = \{0, V_0\}$ and $\mathbf{e}_B = \{0\}$. It is straightforward to check that, there are no extended states in this case, which could be confirmed by a numerical calculation shown in Fig. 16. However, we can alter tunnelings by means of fast periodic modulation (as in Sec. 2.2), which in this case means that the tunnelings are changed to $t' = \mathcal{J}_0(\frac{V_1}{\omega})$ around each frozen particle. It explains the name of the model, as changed tunnelings $t'$ always comes in pairs. In this model, for certain values of $t'$ and $V_0$ there exist one extended state as could be seen in numerical results (Fig. 16). Resonant energy is given by expression:

$$E_R = \frac{V_0}{1 - t'^2},\tag{94}$$

which is derived for example in [37] with use of the Green functions. The extended state will be observable only if $E_R$ lies within the band for empty lattice (the relation: $V_0 \leqslant 2|1 - t'^2|$ has to be fulfilled). How could we reproduce this result? It could be immediately seen that it is impossible to use the same definitions of blocks as above, because the tunnelings at block edges will depend on the adjacent blocks and the whole reasoning made in section 2.4 fails. It is tempting to just use $\mathbf{e}_A = \{0, V_0, 0\}$ and $\mathbf{t}_A = \{1, t', t', 1\}$ (which, in fact gives the proper result), but in this way we change correlations, assuming that particles are separated by at least *two* empty sites. The proper way of address-ing this problem is a bit more complicated. We define infinite number of block types with on-site energies and tunnelings given by:

$$\mathbf{e}^{(0)} = \{0\}, \qquad\qquad\qquad \mathbf{t}^{(0)} = \{1, 1\}$$
$$\mathbf{e}^{(1)} = \{0, V_0, 0\}, \qquad\qquad \mathbf{t}^{(1)} = \{1, t', t', 1\}$$
$$\mathbf{e}^{(2)} = \{0, V_0, 0, V_0, 0\}, \qquad \mathbf{t}^{(2)} = \{1, t', t', t', t', 1\}. \qquad (95)$$
$$\vdots \qquad\qquad\qquad\qquad \vdots$$

It allows us to catch all possible structures and gives homogeneous tunnelings at the edges of the blocks. By defining auxiliary matrices:

$$W(E) = \begin{bmatrix} -E & -t' \\ 1 & 0 \end{bmatrix} \cdot \begin{bmatrix} \frac{V_0 - E}{t'} & -1 \\ 1 & 0 \end{bmatrix} \cdot \begin{bmatrix} \frac{1}{t'} & 0 \\ 0 & 1 \end{bmatrix}, \qquad (96)$$

$$B(E) = \begin{bmatrix} -E & -t' \\ 1 & 0 \end{bmatrix} \cdot \begin{bmatrix} \frac{V_0 - E}{t'} & -1 \\ 1 & 0 \end{bmatrix} \cdot \begin{bmatrix} -\frac{E}{t'} & -\frac{1}{t'} \\ 1 & 0 \end{bmatrix}, \qquad (97)$$

we could write:

$$T^{(0)}(E) = \begin{bmatrix} -E & -1 \\ 1 & 0 \end{bmatrix}, \qquad (98)$$

$$T^{(n)}(E) = W(E)^{n-1} \cdot B(E). \qquad (99)$$

One could check easily that for $E_R$ given by (94) (and only for it):

$$[T^{(0)}(E_R), T^{(1)}(E_R)] = [T^{(0)}(E_R), B(E_R)] = 0, \qquad (100)$$

$$[T^{(0)}(E_R), W(E_R)] = 0, \qquad (101)$$

$$[T^{(1)}(E_R), W(E_R)] = [B(E_r), W(E_R)] = 0. \qquad (102)$$

As the commutator $[T^{(0)}(E), T^{(n)}]$ for any $n$ could be decomposed into sum of commutators (100,101) as well as $[T^{(n)}(E), T^{(k)}(E)]$ could be decomposed into sum of commutators (102) it is clear that all blocks will commute for $E_R$ given by (94)[4]. Before moving on I will present several possible extensions of DRDM:

LONG RANGE TUNNELINGS    Up to now we have used the simplest version of tight-binding, with only nearest neighbors considered. In optical lattices with typical depths of the order of $10-20E_{rec}$(recoil energy) it is usually justified as the next-nearest neighbor tunnelings are at least two orders of magnitude weaker than the nearest neighbor ones. However, considering shallower lattices ($\sim 5E_{rec}$) long-range tunnelings have to be taken into account. We could expect that appearance of such long range coupling could destroy delocalization, so it is good idea to check strength of this effect. If we assume that we have only nearest and next-nearest coupling, Hamiltonian reads

$$H^{(nn)} = \sum_i \epsilon_i n_i + (a_i^\dagger a_{i+1} + \tilde{t} a_i^\dagger a_{i+2} + \text{h.c.}). \qquad (103)$$

---

4 One could point out that for larger $n$ there will be even more energies for which those commutators vanish but it is not the point here as $E_R$ is the only energy for which *all* commutators vanish. Those rejected energies give $E_R$ for so called dual N-mers (t' coming in rows of length N).

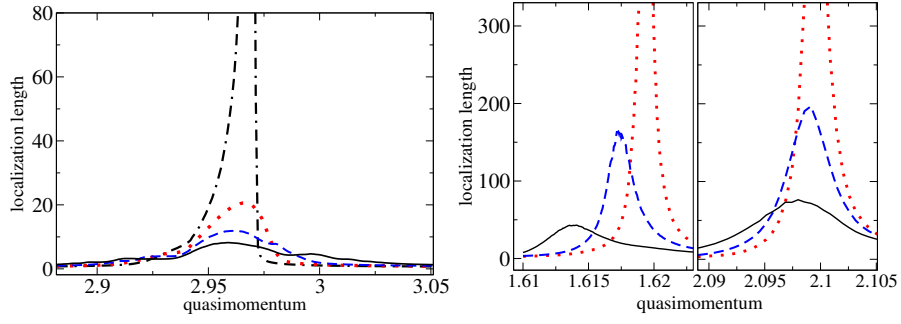

Figure 17: The localization length (in units of lattice constant) as a function of quasimomentum for: left panel $V_0 = 1.95$; right panel $V_0 = 1$; middle panel $V_0 = 0.1$, in all cases $t' = 0.1$. Black dashed-dotted curve denotes results for $\tilde{t}' = 0.0005$, red dotted for $\tilde{t}' = 0.0025$; blue dashed for $\tilde{t}' = 0.005$ and black solid for $\tilde{t}' = 0.01$ [81].

Upon the fast periodic modulation of on site energies, both tunnelings are changed so we get system governed by an effective Hamiltonian:

$$H_{eff}^{(nn)} = \sum_i \epsilon_i n_i + t_i^{eff}(a_i^\dagger a_{i+1} + \tilde{t}_i^{eff} a_i^\dagger a_{i+2} + \text{h.c.}) \qquad (104)$$

where $t_i^{eff} = \mathcal{J}_0((\epsilon_{i+1} - \epsilon_i)/\omega)$ and $\tilde{t}_i^{eff} = \tilde{t}\mathcal{J}_0((\epsilon_{i+2} - \epsilon_i)/\omega)$. Again we have tunnelings changed only around the *frozen* particles, we denote $\tilde{t}' = \tilde{t}\mathcal{J}_0(V_1/\omega)$. Localization length in such a system could be calculated using technique of winding lattice into form of quasi one-dimensional stripe and calculating the localization length in it (described in appendix B.1.2). Results (Fig. 17) show that indeed resonance is quenched by appearance of the next-nearest neighbor tunneling. As we can see at the left panel, in situation when resonant state is very close to disappearing ($V_0 \lesssim 2(1 - t'^2)$) even very weak next-neighbor tunneling destroys extended state. For other parameters, deeper in the regime with delocalized state, resonance persist longer but is also substantially unobservable for $\tilde{t}' \approx 0.01$ (right panels).

N-MERS WITH PERIODIC MODULATION. In a certain sense we could consider DRDM as a 1-mer with tunnelings changed around the *frozen* particle and with additional correlation that no two frozen particles could occupy two adjacent sites. In the same way we could consider N-mer, with tunnelings modified at edges of blocks and the 'no-touch' correlation. The same reasoning as at the beginning of this section allow us to analyze this system. We need to check only the resonant energies satisfying (93) for the N-mer block with empty sites on its edges. As we know explicit expression for N-mer transfer matrix we could calculate:

$$\bar{T}(V_0, l, t')(E) = \begin{bmatrix} -E & -t' \\ 1 & 0 \end{bmatrix} \cdot \begin{bmatrix} \frac{V_0 - E}{t'} & -\frac{1}{t'} \\ 1 & 0 \end{bmatrix}$$

$$\cdot T_{V_0}^{N-2}(E) \cdot \begin{bmatrix} \epsilon - E & -t' \\ 1 & 0 \end{bmatrix} \cdot \begin{bmatrix} -\frac{E}{t'} & -\frac{1}{t'} \\ 1 & 0 \end{bmatrix}$$

$$= T_0^1(E) \cdot \begin{bmatrix} \frac{1}{t'^2} U_l(\bar{V}) & -U_{l-1}(\bar{V}) \\ U_{l-1}(\bar{V}) & -t'^2 U_{l-2}(\bar{V}) \end{bmatrix} \cdot T_0^1(E), \quad (105)$$

where $\bar{V} = {}^{(V_0 - E)}/{}_2$. Putting it into equation (93) we get:

$$(E(t'^2 - 1) + V_0)U_{l-1}(\bar{V}) + (t'^4 - 1)U_{l-2}(\bar{V}) = 0, \quad (106)$$

which could not be easily solved analytically. One simple and interesting property appear for limiting values of $t'$. For $t' = 1$ we get $U_{l-1}(\bar{V}) = 0$ so equation for energies of the extended states in standard N-mer with length $l$ as we should expect. For $t' \to \pm\infty$ we get $U_{l-2}(\bar{V}) = 0$, thus resonant energies fit those of N-mer with length $l - 1$ (I remind that using presented scheme of modulation to obtain $t'$ we could not go into this regime as $t' \leqslant 1$). Going the other way with $t'$ approaching $0$ we get trivial case of fragmented chain but the extended states energies converge to solutions of $U_l(\bar{V}) = 0$ – resonant energies for N-mer with length $l + 1$.

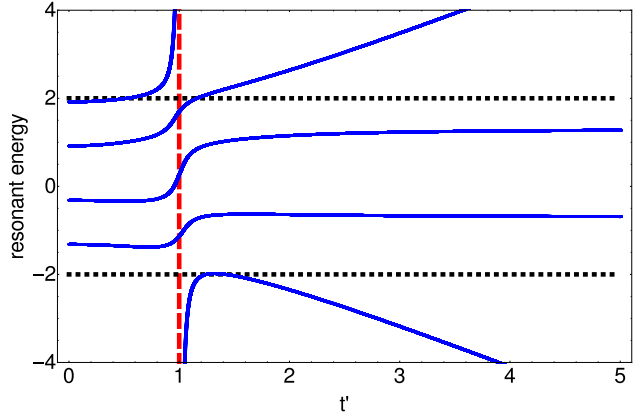

Figure 18: Energy of extended states in the N-mer with length 4, for $V_0 = 0.3$ in function of value of changed tunneling $t'$ (solutions of (106)). Dotted lines marks band edges ($E = \pm 2$) while dashed line denotes the limiting case $t' = 1$ (N-mer model) [91].

Behavior of the extended states energies could be seen in Fig. 18.

*Generalized* N-*mers*

Having the setup as described in section 2.3.3 with $b^f = lb$ where $l \in \mathbb{N}$ we have correlations of type of generalized N-mers with two structures given by (as in Fig. 14):

- empty rows ($T_A(E) = T_0^l(E)$),

- structures with inhomogeneous set of energies $\mathbf{e}_B$ and/or tunnelings $\mathbf{t}_B$ ($T_B(E)$).

As $T_A(E)$ describes an empty row – special case of the N-mer – it will provide $l-1$ resonances (92) (for some of them the localization length could be finite as they are out of band for B block type).

Whether resonances provided by (92) will exist and any additional ones appears is governed by the specific structure of the block B. We will be interested in examples which could be generated in the optical lattices. We could safely assume that an interaction with a *frozen* atom is strongest in the central site of the block and it gets weaker with a distance. Furthermore, we have to assume that the interactions become negligible for distances bigger than the width of one block.

SHORT RANGED INTERACTIONS    The simplest example is case of short ranged interactions (sr). Interaction between the *mobile* and the *frozen* particles does not vanish only in the center of the block. Thus we have $\mathbf{e}_{sr}$ defined by $\epsilon_0^{sr} = V$ and $\epsilon_i^{sr} = 0$ for $i \neq 0$[5]. Transfer matrix could be written in the form:

$$T_{sr}(V, l, E) = T_0^l(E) + T_0^{(l-1)/2}(E) \cdot \begin{bmatrix} V & 0 \\ 0 & 0 \end{bmatrix} \cdot T_0^{(l-1)/2}(E). \quad (107)$$

Putting elements of $T_{sr}(V, l, E)$ into equation (93), we could check if the system has more resonant energies apart from $l-1$ provided by $T_0^l(E)$. For the first part of $T_{sr}$ equation is satisfied trivially for any E, while the second part reads:

$$0 = V \left( U_{\frac{l-1}{2}}^2 (^-E/2) + E U_{\frac{l-1}{2}} (^-E/2) U_{\frac{l-3}{2}} (^-E/2) + U_{\frac{l-3}{2}}^2 (^-E/2) \right). \quad (108)$$

It is matter of short calculation using trigonometric form of Chebyshev polynomials:

$$U_n(\cos \phi) = \frac{\sin(n+1)\phi}{\sin \phi}, \quad (109)$$

and substitution $E = -2 \cos \phi$ to show that in equation (108) whole part in parenthesis is identically 1, thus equation is satisfied only for $V = 0$ and we do not have any additional resonances. Looking at numerical results (Fig. 19) we could observe that all resonances expected for $T_A$ are present but they are very narrow and asymmetric. This happens usually, if the resonant energy is at the band edge. Indeed, by a simple calculation we could determine that it is not a coincidence for chosen parameters but property of generalized N-mers with short ranged interactions. If we find eigenvalues of $T_{sr}$, the part under the square root:

$$\sigma(E) = \left[ (E - V) \left( U_{\frac{l-1}{2}}^2 \left( {}^{-E}\!/2 \right) - U_{\frac{l-3}{2}}^2 \left( {}^{-E}\!/2 \right) \right) + 2U_{\frac{l-1}{2}} \left( {}^{-E}\!/2 \right) U_{\frac{l-3}{2}} \left( {}^{-E}\!/2 \right) \right.$$
$$\left. -2U_{\frac{l-3}{2}} \left( {}^{-E}\!/2 \right) U_{\frac{l-5}{2}} \left( {}^{-E}\!/2 \right) \right]^2 - 4 \left[ U_{\frac{l-3}{2}}^2 \left( {}^{-E}\!/2 \right) - U_{\frac{l-1}{2}} \left( {}^{-E}\!/2 \right) U_{\frac{l-5}{2}} \left( {}^{-E}\!/2 \right) \right]^2 \quad (110)$$

can be taken as an indicator of band edges positions as $\sigma(E_b) = 0$ is necessary (but not sufficient) condition for $E_b$ to be at the band edge. We know that $U_{l-1} \left( {}^{-E}\!/2 \right) = U_{\frac{l-1}{2}}^2 \left( {}^{-E}\!/2 \right) - U_{\frac{l-3}{2}}^2 \left( {}^{-E}\!/2 \right)$ and assume that $E = E_R$ (resonant energy for empty row with length $l$) for which $U_{l-1} \left( {}^{-E_R}\!/2 \right) = 0$. It is easy to further check that $\sigma(E_R) = 0$. In this case indeed $E_R$ seems to be always at band edge (I omit here quite complicated checking whether is it possible in this systems to meet rare situation in which $\sigma(E)$ is zero but inside a band).

---

5 For the sake of clearness, from now on I will use only odd $l$ and adopt intuitive numbering with $i = 0$ in central site of the block ranging $i \in [-(l-1)/2, (l-1)/2]$.

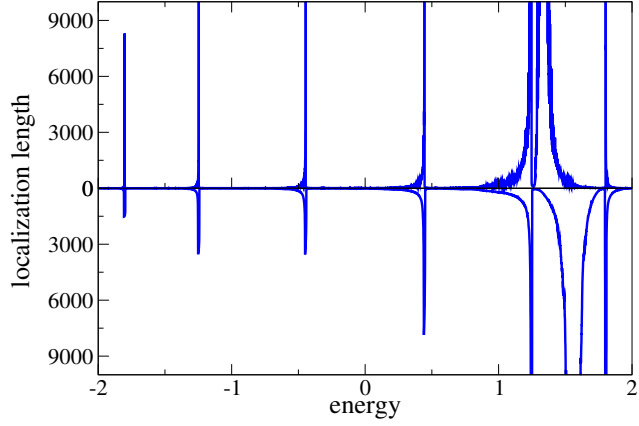

Figure 19: The localization length (in units of lattice constant) in function of the energy (in units of tunneling amplitude) calculated numerically by the transfer matrix method for generalized N-mer with $T_A = T_0^1(E)$ and $T_B = T_{sr}(V, t', l, E)$ for $l = 7$, $V = 1$. Upper pane shows results for $t' = 0.5$, lower for $t' = 0.6$. Resonances are very narrow and asymmetric as most of them lies at the edges of gaps in spectrum of $T_{sr}$, one resonance changes its position due to change of $t'$.

SHORT RANGED INTERACTIONS WITH PERIODIC MODULATION
In the above setup if the tunnelings around central site are changed (again using periodic modulation of interaction energy) we get model described by transfer matrix $T_{sr'}(V_0, t', l, E)$ where $t_i$ are changed only around central site $i = 0$ (namely $t_0^{sr} = t_{-1}^{sr} = t'$, where $t' = \mathcal{J}_0(V_1/\omega)$). Upon making this change we get model similar to DRDM. Indeed the new resonance appears with exactly the same energy as in the DRDM. It could be derived even easier as we have homogeneous tunnelings at edges (this in fact coincides with this 'naive' method of solving DRDM mentioned before). In this case one interesting thing is that by changing parameters of shaking (thus changing $t'$) we could change new resonant energy, leaving others intact so we have simple, possible to experimental realization model in which we could change relative positions of extended modes. The results are shown in Fig. 19 for $T_B = T_{sr}(V, t', l, E)$ with $V = 1$, $l = 7$ and for two different $t' \in \{0.5, 0.6\}$ (upper/lower pane). All predicted extended modes are present, one of them (created by the presence of $t' \neq 1$) can be moved relatively to others.

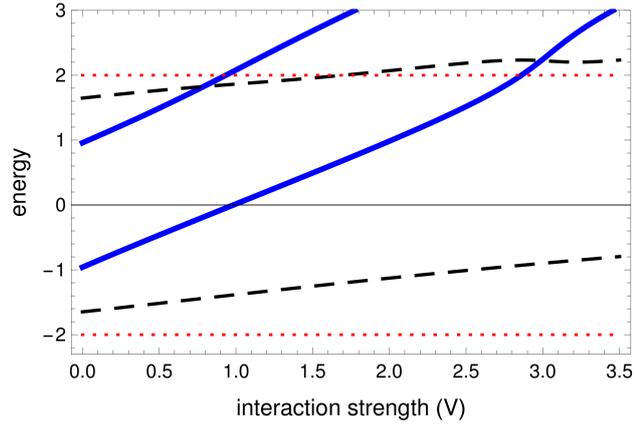

Figure 20: Solutions of equation (93) for $T_B = T_{dim}(V, d, l, E)$ for $l = 7$ and $d = 0.9$ as a function of energy and interaction strength $(V)$ (both in units of tunneling amplitude). Blue solid lines represents real solutions – for those parameters extended states exists, black dashed lines are complex solutions, in vicinity of them localization length could be increased (especially if imaginary component is small) however is finite, red dotted lines are band edges for $T^A = T_0^l$.

LONG RANGE INTERACTIONS    In case of slower decay of interaction strength with a distance, for example exponential or polynomial one, it is impossible to give analytical solutions[6]. We have to resort to numerical solving the problem for given $T_B(E)$. Results were qualitatively similar for most of investigated models so I will present just one. We assume that interaction strength is falling with the third power of distance, namely $\epsilon_0^B = V$, $\epsilon_i^B = \frac{V_d}{|i|^3}$ for $i \neq 0$. Solutions of (93) for $T_B = T_{dim}(V, V_d, l, E)$ ($l = 7$ and $V_d/V = 0.9$) are plotted in Fig. 20, solid lines are resonant energies (real solutions), while dashed denotes complex solutions. Localization length for this model calculated for $V = 1$ is shown in Fig. 21 (black line). Most of the resonances appear on predicted positions (marked by green dotted lines). Three of them are absent as they are lying out of bands. By changing $V$ or $V_d$ it is possible to move the resonance for $E \approx 0.02$ relatively to the rest.

---

6 In fact it is possible to write closed form of transfer matrix for linearly falling interaction strength, but as it is impossible to further solve it to get resonant energies I omit this example.

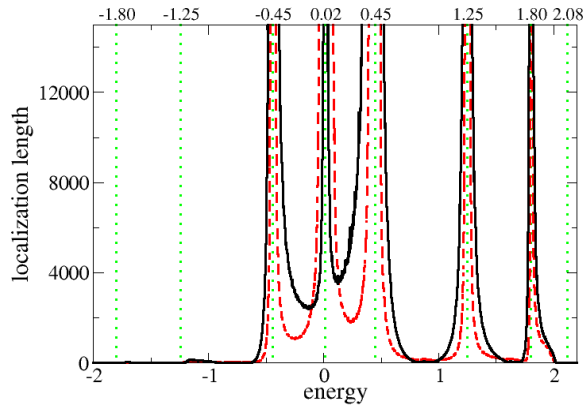

Figure 21: The localization length (in units of lattice constant) as a function of the energy (in units of tunneling amplitude). Black line is the result for the disorder in form of generalized N-mer with $T_A = T_0^z(E)$ and $T_B = T_{dim}(V, d, l, E)$ for $l = 7$, $d = 0.9$ and $V = 1$. Red dashed line is the result for the same system but without interaction range cutoff assumed in the generalized N-mer case. Green dotted vertical lines represent predicted resonances, some of them are out of band.

To close this section I will comment on the assumption of block separability. While the situation is clear for short ranged interactions for case presented in this paragraph it is not so obvious. Moreover we have seen in section 2.5.3 that even very weak breaking of separability of the blocks by the long-range tunnelings could quickly destroy delocalized states.

For the case presented above, red dashed line in the Fig. 21 shows results for calculation with the same potential $\mathbf{e}_{dim}$ and parameters but without cutoff on block edges[7]. Obviously some features differ but qualitatively the plots are the same. Thus, at least in some cases, interactions which reach further than block size do not cause immediate disappearance of the observed effects.

At the end I will shortly comment on, assumed before, homogeneity of tunnelings at blocks' edges. Due to their physical origin, con-sidered structures have the highest on-site energies in the center. We assume that at the block edges they are, if not negligible, considerably smaller than in the center. As the tunnelings are changed due to difference of on site energies (see Sec. 2.2.2), in the proposed experimental setup the tunnelings at the block edges should be homogeneous with a good approximation.

---

7 In calculations cutoff was set for interaction strength smaller than $2 \cdot 10^{-5}$.

# LOCALIZATION IN RANDOM ARTIFICIAL GAUGE FIELD

In this chapter I will focus on a specific subclass of the two-dimensional systems with both diagonal and off-diagonal disorder. Their main feature is that the complex phases of tunnelings are disordered. Although, it may seem a small difference, it constitutes a totally new class of systems, as the dynamics of a neutral particle (i.e. atom) in such a disorder resemble dynamics of a charged particle moving in a randomly varying magnetic field. Thus we could call this type of disorder a random artificial gauge field or just a random magnetic field[1].

I will discuss the two-dimensional lattices, as in one-dimensional systems with nearest neighbor coupling the changed phases of tunnelings could be always taken care of by changing the phases of Wannier functions (which we always could do). Also in two-dimensional systems effect of introducing the disordered magnetic field is much stronger than for example in quasi one-dimensional stripes. Possibility of simulating this class of systems in the optical lattices may allow to check various interesting topics from condensed matter physics. For example the fractional quantum Hall effect at half-filling could be mapped onto particles moving in randomly varying gauge field [97, 98], also some models of high-$T_c$ superconductors could be tested, as they predict strong suppression of a critical temperature due to random field fluctuations [99]. The case with both on-site and magnetic field disorder presents an interesting interplay between two effects, as upon the appearance of the magnetic field the localization length should be increased due to breaking of the time reversal symmetry (see next section) but also the new source of disorder appears, which in principle could decrease the localization length.

I will start with a short review of the theory of disordered systems in the magnetic field. Further I will describe specific case when the magnetic field itself is a random variable, especially a history of studies in case of solely magnetic disorder (in correspondence to case of purely off-diagonal disorder in section 2.2). In the next part I am going to describe various methods which could be used to obtain the artificial gauge fields in the ultracold atomic gases. After this lengthy introduction I will present main model considered in this chapter and a specific setup which could be used to create it in the experiment. Further, I will show results of numerical calculations and explain observed phenomena, particularly the unexpectedly strong localization appearing for the systems with correlated disorder. I will also present results showing the concurrence between strengthening and weakening of the localization by the random fluxes. Finally, I will check for presented model the validity of expressions (23,111) connecting the mean free path with the localization length.

---

[1] In order to not overuse long 'artificial gauge field' I will often use 'magnetic field' in this chapter, though we have to remember its meaning in this scope.

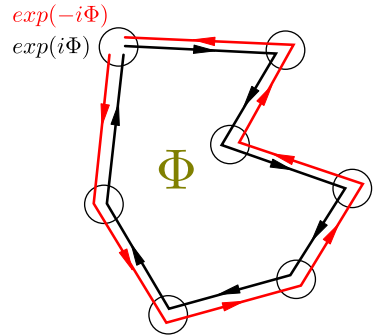

Figure 22: When considering two reversed paths their phases are changed by presence of the external magnetic field – probability of returning to the origin is not increased (weak localization correction vanish).

### DISORDERED SYSTEMS IN EXTERNAL MAGNETIC FIELD

The problem of the Anderson localization in the external magnetic field has not been fully solved yet. I will present several remarks and known results.

In the systems with broken time-reversal symmetry the weak localization correction (see Sec. 1.5.2) vanishes as two counterpropagating paths acquire different phases (for the case of magnetic field, the phase difference is twice the flux through the contour) and in the effect do not add constructively (Fig. 22). The effect on the Anderson localization is most significant in two-dimensional systems, where the correction which makes the scaling function $\beta(g)$ always negative and therefore assures localization of all states is due to the weak localization effect. However, in contrary to systems with spin-orbit coupling (*symplectic* ensemble), the delocalization term do not appear. To determine behavior of those systems further corrections have to be calculated. It has been done by Wegner using the nonlinear σ-model [100]. He showed that the correction to $\beta(g \to \infty)$ of order $1/g^2$ do not vanish and moreover is negative, so again all states in two dimensions are localized. However, due to higher order of first non-vanishing correction the connection between mean free path and localization length should have form:

$$\lambda = \ell e^{\zeta(k\ell)^2}.$$  (111)

Similar but weaker effect appears for the quasi one-dimensional stripes, for which it was shown, using random matrix theory, that upon breaking the time reversal symmetry the localization length will be simply doubled [101].

### RANDOM MAGNETIC FLUX MODELS

Special class of the disordered systems in the external magnetic field consists of systems in which the magnetic field itself is disordered. Historically first – classical not quantum – attempt to describe such a situation had been done by Fermi long before the Anderson localization was even proposed [102]. The model of classical charged particles moving in randomly varying magnetic field provided a surprising result, that particles will accelerate. It was proposed as a model for the origin of a high energy cosmic radiation.

First work considering random fluxes in scope of quantum mechanics and the localization phenomena have been published in year 1981 [103], it have shown that when one add such a magnetic field to system with the standard on-site disorder it appears that the localization length grows. It seemed that weakening of the localization due to breaking of the time reversal symmetry is stronger effect than localization on appearing disorder. This topic has been further investigated in [104, 105].

Most of the effort in the following years has been focused on the systems with solely flux disorder with mean magnetic flux equal to zero, such systems are conceptually simple and could be connected with some other interesting situations as nonstandard superconductivity models [99, 106] or spin liquids [107]. It have been also shown that random magnetic field alters significantly critical temperature of the Bose-Einstein condensation [108].

There have been a long lasting controversy whether such systems have the mobility edge. In principle they should belong to the *unitary* class and all the states should be localized, but many opposed arguing that in this case σ-model [100, 109] does not apply and the mobility edge will appear. This statement have been supported by numerical numerical calculations [110–114]. On the other hand different authors have shown numerical results confirming localization of all states [115–117]. Those studies could not be conclusive (especially in early 90-thies) as the localization length in two-dimensional systems grows quickly with increasing energy, moreover due to purely off-diagonal character of the disorder it posses the same 'extended-like' state in the band center as described in section 2.2. Finally it have been proven analytically – at least in the limit of long range correlated disorder – that indeed all states are localized [118, 119].

Out of other interesting results in this field one could pick for example work showing effects of screening on transport properties [120] or surprising lack of influence of long range tunnelings on the localization length [110].

### CREATION OF ARTIFICIAL GAUGE FIELDS

The serious limitation in usage of the ultracold atomic gases as quantum simulators had been inability to simulate effects of the orbital magnetism. Luckily, this obstacle has been overcome and nowadays there exist a range of methods allowing creation of artificial gauge fields (not necessarily only U(1) but even non-abelian gauge fields).

The charged quantum particle moving in the external magnetic field, after completing closed path acquires some complex phase which could not be removed by gauge transformation and which is *geometric* in the sense that it depends on the shape of the path not for example time it has taken to complete it. It is also strictly connected with the flux of magnetic field surrounded by the contour. If only it is possible to create conditions in which a neutral particle also acquires such a phase, then it will effectively behave as the charged particle in the magnetic field. This idea of giving to neutral particle a *geometric* phase underlies the majority of methods used to create the artificial gauge fields.

### Rotation of the system

Rotating the system with constant angular velocity seems to be the simplest idea to create desired effects. Indeed, in the frame rotating with some angular velocity $\Omega$ the particle feels Coriolis force equivalent to the Lorenz force felt by charged particles in the uniform magnetic field proportional to $\Omega$ [121]. The one well known example of this effect is a correspondence between arrays of vortices in superconductors placed in magnetic field and vortices appearing in the rotating superfluids. The drawback of this method is that additional harmonic potential have to be created to prevent shooting out particles due to centrifugal force, moreover the range of potentials which could be used is limited, as if the potential is not axisymmetric it becomes time dependent in the rotating frame and the theoretical description complicates significantly.

*Dressing atoms with laser light*

The simplest example of this class is a two-level atom in a laser beam. The internal eigenstates of the atom ($|g\rangle$ and $|e\rangle$) are coupled by the lightfield. The operator of coupling could be written as:

$$U(\mathbf{r}) = \frac{\hbar}{2} \begin{bmatrix} \Delta(\mathbf{r}) & e^{-i\phi(\mathbf{r})}\kappa(\mathbf{r}) \\ e^{i\phi(\mathbf{r})}\kappa(\mathbf{r}) & -\Delta(\mathbf{r}) \end{bmatrix}, \tag{112}$$

where $\Delta$ marks a detuning of laser frequency from the atomic transition, $\phi$ is a phase of the wave and $\kappa$ a strength of an atom-laser coupling (it linearly depends on the intensity of light), so called mixing angle $\theta$ could be defined as $\tan\theta = \kappa/\Delta$ (I have dropped the argument ($\mathbf{r}$) for clarity and for the extent of this section restored $\hbar$ for consistency with the descriptions in the literature). The eigenstates of the system are the dressed states:

$$|\chi_1\rangle = \begin{bmatrix} \cos(\theta/2) \\ e^{i\phi}\sin(\theta/2) \end{bmatrix}, \tag{113}$$

$$|\chi_1\rangle = \begin{bmatrix} \cos\left(-e^{-i\phi}\right)\sin(\theta/2) \\ \cos(\theta/2) \end{bmatrix}. \tag{114}$$

If we assume that the system is prepared in one of the dressed states ($|\chi_1\rangle$) and moved sufficiently slow, that it stays in it for the whole time, it is quite straightforward to check [122, 123] that due to a spatial variability of the dressed state new term appears in the Hamiltonian with the properties exactly as a vector potential[2]:

$$A(\mathbf{r}) = i\hbar\langle\chi_1|\nabla\chi_1\rangle = \frac{\hbar}{2}(\cos\theta - 1)\nabla\phi. \tag{115}$$

By taking the rotation of $A(\mathbf{r})$ we obtain an effective magnetic field:

$$B(\mathbf{r}) = \nabla \times A = \frac{\hbar}{2}\nabla(\cos\theta) \times \nabla\phi. \tag{116}$$

If only the phase ($\phi$) and the mixing angle ($\theta$) vary in the space in a noncollinear way the magnetic field is nonzero. If we consider running wave, the $\phi$ changes in the direction of the propagation, so the $\theta$ must change in the direction perpendicular to the beam. It could be done for example by changing intensity of light – effect which naturally occurs for narrow beams or by changing detuning of atoms by using another auxiliary laser beam. In reality, to realize described scenario one has to use atoms which have very long times of spontaneous emission as calcium or ytterbium. For other species, more complicated setups using several internal states have to be applied, although the general conception is similar.

---

2 Also the scalar potential appears, but it is not as interesting as there exist variety of techniques allowing creation a wide range of scalar potentials for ultracold atoms.

*Lattice systems*

Considering the lattice systems such as crystalline lattice or optical lattice, the artificial gauge fields could be introduced reversing the Peierls argument [124]. Originally, it was formed to simplify the description of behavior of the charged particles moving in the lattice. Upon applying the external magnetic field its effect could be written as a change of complex phases of tunnelings $t_{nm} \to t_{nm} \exp(i\phi_{nm})$, where the phase is calculated by integrating the vector potential on the line between the lattice sites:

$$\phi_{nm} = \frac{e}{\hbar} \int_{\mathbf{r_n}}^{\mathbf{r_m}} \mathbf{A}(\mathbf{r}) \cdot d\mathbf{r}. \tag{117}$$

If there is a nonzero flux of the magnetic field through some contour (in this context usually we consider smallest possible one – lattice plaquette) then the sum of the phases $\Phi$ will also be nonzero. Using the same 'reverse-engineering' as for continuous systems one can create complex tunnelings and make neutral particles show effects of orbital magnetism.

It could be done using laser-assisted tunneling and atoms with internal structure much like in the continuous case (for description see [122]), yet another option is to use fast periodic modulations to alter the tunneling phases.

The simplest scheme of modulation is periodic shaking of the lattice [125] which changes the amplitude of the tunnelings. It could be extended to create also the complex phases. The key point is the fact that the magnetic field breaks the time reversal symmetry, thus simple harmonic modulation could not simulate it as it is perfectly time symmetric. Usually, just using nonsymmetric modulation is sufficient to obtain complex phases of tunnelings. Such a simple scheme is proposed in [126], where driving is done with two dephased harmonic functions (with commensurate ratio). It indeed allows the creation of the complex tunnelings, although on the regular lattice the phases cancel and no artificial gauge field appears. However, it could give staggered fluxes on the triangular lattice. The more complicated systems have to be created to obtain field on the regular lattice or to get field with a nonzero mean, but it is possible [127].

CREATION OF THE RANDOM MAGNETIC FIELD.

The lattice system simulating the disordered magnetic field could be created using two species of the particles in the regular two-dimensional optical lattice undergoing fast periodic modulation of some parameters.

The sketch of the preparation procedure of such a system is following: First, on-site disorder is created using the second species of particles *frozen* in the lattice (as described in Sec. 1.6). In this case we use high density ($\rho_f = 2.5$) of quite weakly interacting ($t^f \gtrsim U^f$) *frozen* particles, so after a quench we obtain the Poisson distribution of the disorder. The Hamiltonian (with already immobilized fraction $t^f = 0$) reads:

$$H_D = t \sum_d \sum_{\langle ij \rangle_d} a_i^\dagger a_j + (V n_i^f) n_i, \tag{118}$$

where $\langle ij \rangle_d$ denotes nearest neighbors in direction $d$ which is either $x$ or $y$ and $(V n_i^f)$ is effectively the disordered on-site potential.

In the next step I want to add a gauge field to this picture. As having the gauge field in the lattice is equivalent to adding complex phases to the tunnelings I will proceed with creating those, using the fast periodic modulation of the lattice parameters. In the presented case I use the simultaneous modulation of the interspecies interaction $V \to V_0 + V_1 \sin(\omega \tau)$ and the tunneling rates $t \to t_0 + t_1^{(d)} f_\omega(\tau)$, where $\omega$ is the frequency of modulation and $f_\omega(\tau)$ is some periodic function. An important point is that I allow different modulations of the tunneling rates in the different lattice directions. The time dependent Hamiltonian reads:

$$\begin{aligned} H(\tau) = & \sum_d \left( t_0 + t_1^d f_\omega(\tau) \right) \sum_{\langle ij \rangle_d} a_i^\dagger a_j \\ & + \sum_i \left( (V_0 + V_1 \sin \omega\tau) n_i^f \right) n_i. \end{aligned} \tag{119}$$

As the sinus function is antisymmetric, as long as the $f_\omega$ is not antisymmetric the time reversal symmetry of the system will be broken and the gauge filed could appear.

The interspecies interactions could be driven in the same way as described in the section 2.2.2, while the tunneling rates could be modulated by change of the lattice height. This obviously will change also the interaction strength, but the tunneling amplitudes vary exponentially with the lattice height while the interactions vary approximately linearly, so we could neglect the latter effect.

As the (119) is time periodic we could use the results of the Floquet theory (App. A) to obtain the effective Hamiltonian for long times. Unfortunately in this case we could not use the 'standard' procedure of transforming system into rotating frame, as the time dependent parts of the Hamiltonian (119) do not commute. In this situation, if we just time average $H(\tau)$ (calculate zeroth order of the Magnus expansion) we will get errors of order $^1/_\omega$ and we will not obtain any magnetic field fluxes, as they are exactly of this order. To improve the accuracy and obtain desired effect, we could make a partial transformation:

$$\mathcal{U} = \exp \left( i \frac{V_1 \cos \omega\tau}{\omega} \sum_i n_i^f n_i \right). \tag{120}$$

This transformation removes the time dependence from the on-site part of the Hamiltonian, but more importantly it takes system to a frame in which the modulation is a symmetric function of time and therefore all odd elements of the Magnus expansion vanish [128], leaving errors of order $^1/\omega^2$:

$$H'(\tau) = \mathcal{U}H(\tau)\mathcal{U}^\dagger = \sum_i (V_0 n_i^f) n_i$$
$$+ \sum_d (t_0 + t_1^d f_\omega(\tau)) \sum_{\langle ij \rangle_d} e^{i\frac{V_1}{\omega}(n_j^f - n_i^f)\cos\omega t} a_i^\dagger a_j. \quad (121)$$

Now, upon the time averaging (denoted $\langle \ldots \rangle_T$ for $T = {}^{2\pi}/\omega$) we get effective Hamiltonian with errors of the order of $^1/\omega^2$:

$$H_{\text{eff}} = \langle H'(t) \rangle_T = \sum_d \sum_{\langle ij \rangle_d} J_{ij}^d[f_\omega] a_i^\dagger a_j + \sum_i (V_0 n_i^f) n_i, \quad (122)$$

where $J_{ij}^d[f_\omega]$ is an effective tunneling from site $i$ to $j$ (in direction $d$).

*Modulation scheme*

Depending on the scheme of the lattice modulation ($f_\omega$) we could get qualitatively different forms of the effective tunneling:

HARMONIC MODULATION    Modulation form: $f_\omega(\tau) = \cos(\omega\tau)$ results in an expression:

$$J_{ij}^d[\cos\omega\tau] = \langle (t_0 + t_1^d \cos\omega t) e^{i\frac{V_1}{\omega}(n_j^f - n_i^f)\cos\omega\tau} \rangle_T$$
$$= t_0 \mathcal{J}_0(\frac{V_1}{\omega}(n_j^f - n_i^f)) + it_1^d \mathcal{J}_1(\frac{V_1}{\omega}(n_j^f - n_i^f)), \quad (123)$$

where $\mathcal{J}_n(x)$ is $n$-th order Bessel function[3]. This form of effective tunneling is not the most convenient, as it creates not only random fluxes but also randomly changes the tunneling amplitudes, introducing yet another form of disorder.

If we set $t_1^d = \pm\sqrt{2}t_0$, the expression (123) could be approximated in the close vicinity of zero as:

$$J_{ij}^d[\cos\omega\tau] \approx \tilde{J}_{ij}^d[\cos\omega\tau] = t_0 \exp\left(\pm i\tan^{-1}\left(\frac{V_1}{\omega}\left(n_j^f - n_i^f\right)\right)\right). \quad (124)$$

The approximation diverges quickly from the exact result and I was unable to find a modulation scheme which better coincides with (124), however in principle it should exist and as the (124) have several favorable features I will present results for it alongside those for well defined modulation schemes. Among the properties of $\tilde{J}_{ij}^d[\cos\omega\tau]$ the most important are:

---

3 I used a property: $\langle \cos[n\omega t] \exp(ix\cos(\omega t)) \rangle_T = i^n \mathcal{J}_n(x)$.

- its amplitude is always one, so only the random fluxes are present (no random tunneling amplitudes),

- its phase depends nonlinearly on the argument, so in the case of the symmetric modulation in both directions $\left(t_1^x = t_1^y\right)$ we could still expect non-vanishing fluxes,

- $\tan^{-1}$ saturates on $\pm\pi/2$ and consequently the fluxes always takes values smaller than $2\pi$ ($2\pi$ is reached asymptotically for strong modulation), thus the mean flux amplitude is monotonic function of the modulation strength.

Although in principle any ratio of the modulations in different lattice directions could be chosen I restrict my analysis to two limiting cases of symmetric $t_1^x = t_1^y$ and antisymmetric $t_1^x = -t_1^y$ modulations. Those choices give smallest and biggest fluxes for the chosen modulation strength. It is also convenient choice in scope of used numerical method (MacKinnon method, see. B.3) as distinguishing one lattice direction over the other may pose problems due to different treatment of different directions in the method.

DELTA MODULATION     Another option of modulation of tunnelings is periodically repeated Dirac delta (so called Dirac comb):

$$f_\omega(\tau) = \text{III}(\omega\tau) = \sum_n \delta\left(\tau + \frac{2\pi}{\omega}n\right), \tag{125}$$

the resulting effective tunneling reads:

$$J_{ij}^d[\text{III}_\omega] = t_0 \exp\left(\pm i\frac{V_1}{\omega}\left(n_j^f - n_i^f\right)\right). \tag{126}$$

This form of modulation gives effective tunnelings with constant amplitude – which is desired property. However the phase is linear function of the occupation of frozen particles $\left(n_j^f - n_i^f\right)$ and thus to obtain nontrivial fluxes one has to use different shaking in different lattice directions[4] (I will use only the antisymmetric shaking in this case). Moreover, fluxes do not saturate but grow linearly with the growing argument of function, eventually they reach $2\pi$ and wind up, in the result it is possible to get smaller fluxes for bigger modulation amplitudes.

Delta modulation seem to be experimentally demanding but it could be obtained in at least two ways. Either by keeping the lattice very high to basically stop all transport and lowering it significantly (or simply turning off) only for short periods once every $T$ or by over-imposing several higher harmonic modulations with amplitude $t_1 = \pm 2t_0$ to approximate Dirac's comb:

$$\text{III}(\omega t) \approx 1 + 2\cos 2\omega\tau + 2\cos 3\omega\tau + 2\cos 4\omega\tau + \dots , \tag{127}$$

---

4 It is straightforward to check that $\Phi \propto \pm(n_1 - n_2) \pm (n_2 - n_3) \pm (n_3 - n_4) \pm (n_4 - n_1)$ and for symmetric modulation, when all signs are the same, flux is always zero.

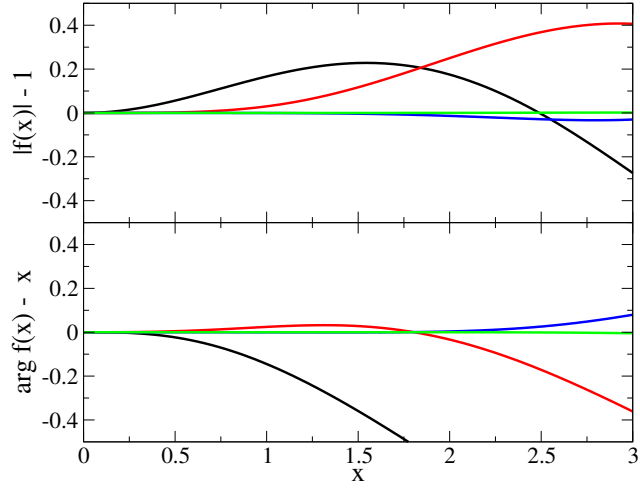

Figure 23: Convergence of the expansion (128). Curves shows deviations of the absolute value (upper panel) and the complex phase (lower panel) of approximated function $f(x)$ from $\exp(ix)$ ($|e^{ix}| = 1$, $\arg e^{ix} = x$). Black curves are results of 1st order of expansion, red of 2nd, blue of 4th and green of 6th. For 6th order there is no visible difference from approximated function in presented range.

after applying this approximated function to effective tunneling we get:

$$J_{ij}^{d}[\mathbb{M}_\omega] \approx \mathcal{J}_0\left(\pm\frac{V_1}{\omega}\left(n_j^f - n_i^f\right)\right) + 2i\mathcal{J}_1\left(\pm\frac{V_1}{\omega}\left(n_j^f - n_i^f\right)\right) + \ldots \ .$$
$$(128)$$

In contrary to the previous case the convergence is very good as it could be checked in the figure 23.

*Other investigated models*

Apart from the model described above I have studied several of its variations in order to better investigate various appearing phenomena.

UNCORRELATED DISORDER   In the model presented above the diagonal disorder is obviously correlated with the flux disorder as both came from the same distribution of the frozen particles. Only their relative amplitude could be changed by manipulating mean value of interactions $V_0$ and the modulation amplitude $V_1$.

As any correlations could affect localization properties (and indeed this is the case as we will see shortly) we want to investigate also uncorrelated version of the main model by simply picking the values of on-site and flux disorder independently. It is in principle possible to obtain such a system in ultracold atomic gases, although the setup is considerably more complicated. To create the uncorrelated disorder we need two different species of *frozen* particles which are independently placed randomly in the lattice. First one should interact with *mobile* atoms with constant amplitude $V_0$ generating the on-site disorder, while for the second species interaction strength should be modulated with amplitude $V_1$ around the zero so it will generate only the disordered complex tunnelings.

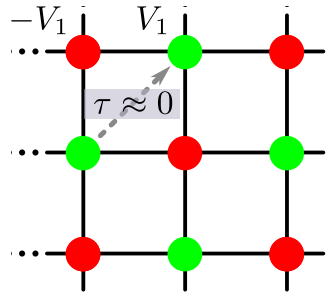

Figure 24: Setup in which staggered field could be created. Two dimensional regular lattice with auxiliary lattice with amplitude $2V_1$ rotated by $\pi/4$.

STAGGERED FIELD    Main model gives opportunity to crate disordered gauge field or no field at all. In order to check effect of breaking the time reversal symmetry without introducing new disorder I studied also the model with solely diagonal disorder placed in staggered field[5].

Such a field could be created for example using another lattice with lattice constant $2\sqrt{2}$ rotated by $\pi/4$ relative to the main lattice (as in Fig. 24). If only this auxiliary lattice is low enough it would not change tunnelings or introduce coupling across the plaquette and we could assume that it only changes on-site energies. If we now start to modulate the height of this new lattice $V_s(t) = \sin(\omega t)V_1$ (note that we do not have to change the interaction strength in this case) along with changing the tunnelings as in the main model, we will effectively get the staggered field.

CONTINUOUS DISORDER    In the main model as well as variations presented above, the disorder is taken from the random discrete distribution. It is possible that for some specific occupations and energy, the resonant transport will occur and significantly alter the results. To check if this is a case, one can redo the calculations for different densities $\rho_f$ (changing accordingly also the interaction strength) and observe if any qualitative changes appear. The other possibility, which

---

[5] I have chosen staggered field to preserve the mean flux $0$ appearing in the main model.

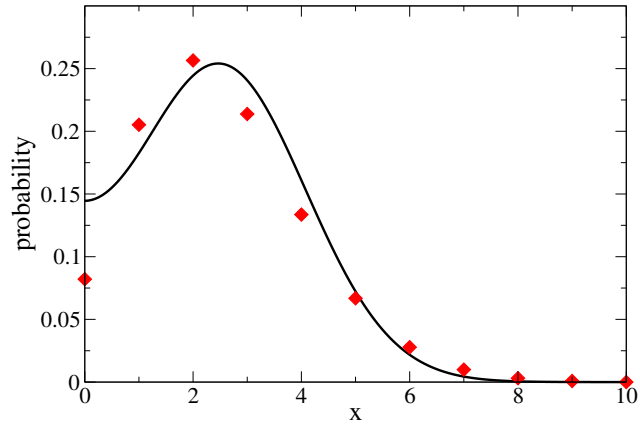

Figure 25: Probability density functions for Poisson distribution with mean 2.5 (red diamonds) and for folded normal distribution with mean 2.5 and standard deviation $\sqrt{2.5}$ (black solid line).

I have chosen, is to do calculations for the continuous disorder which does not suffer from this problem. I have used the folded normal distribution given by probability density function:

$$f(x|\mu, \sigma^2) = \frac{1}{\sqrt{2\sigma^2\pi}} \left( e^{-\frac{(x-\mu)^2}{2\sigma^2}} + e^{-\frac{(-x-\mu)^2}{2\sigma^2}} \right) \quad \text{for } x \geqslant 0, \quad (129)$$

which greatly resembles Poisson distribution for $\sigma = \sqrt{\mu}$ as shown in the figure 25. Simulations done for this disorder shows very little difference from ones for the discrete disorder so I could safely assume that no resonant transport happens. I do not try to make up an experimental setup in which this type of disorder could be created – and there is no need for that, as the transport properties are very similar to those of the systems with the discrete disorder.

RESULTS

The localization length for all studied cases has been calculated using the MacKinnon method (described in appendix B.3) for stripe widths in the range 16–128. Results presented below are calculated for mean occupation of *frozen* particles $\rho_f = 2.5$ and mostly for interspecies interaction strength $V_0 = 1.5$. For other values of interaction strength results are qualitatively similar but for lower $V_0$ they contain higher numerical errors due to rapidly growing localization length and for higher $V_0$ features become less distinctive as the localization length is only of order of several sites.

In Fig. 26 the typical results are shown. The localization lengths for the diagonal disorder and for strong delta modulation are plotted in the function of energy. Results for all investigated parameters (apart from ones for very strong disorder), reveal one peak in the localization length around the band center. I have not observed any signature of the mobility edge or isolated extended states so I could safely assume that maximal localization length is a good measure of the overall transport properties of the system for given parameters and restrict presentation of further results only to this one value.

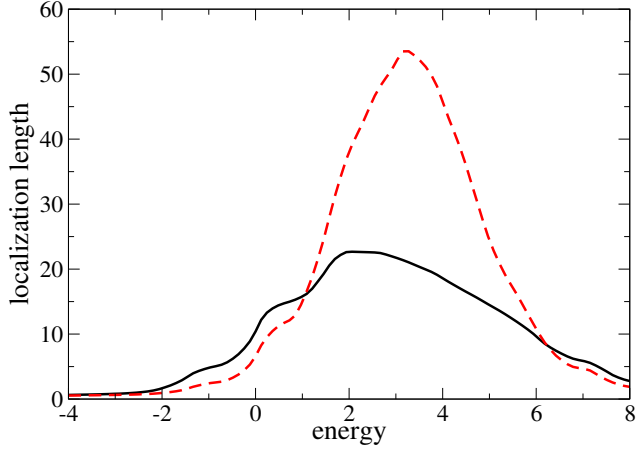

Figure 26: Localization length (in units of lattice spacing) as a function of energy (in units of tunneling amplitude), exemplary results for $V_0 = 1.5$ and $\rho_f = 2.5$. Black solid line is result for undriven system while red dashed stands for delta antisymmetric modulation with $V_1/\omega = 1.0$ (for correlated disorder).

The maximum localization length as a function of the modulation parameter $V_1/\omega$ is plotted for different models in Figs. 27-28. The results for following systems are presented:

FIG. 27(UPPER) Symmetric ($t_1^x = t_1^y$) harmonic (123) and approximated harmonic (124) modulation. For correlated and uncorrelated disorder.

FIG. 27(LOWER) Antisymmetric ($t_1^x = -t_1^y$) harmonic (123) and approximated harmonic (124) modulation. For correlated and uncorrelated disorder.

FIG. 28 Antisymmetric ($t_1^x = -t_1^y$) delta modulation. For correlated and uncorrelated disorder.

We could see that regardless the correlations in the disorder and the symmetry of the shaking, results for the exact harmonic modulation agrees quite well with the ones for the approximated form $\bar{J}_{ij}^d[\cos \omega t]$ up to modulation parameter around $V_1/\omega \approx 0.4$. The discrepancy for larger values $V_1/\omega \gtrsim 0.4$ is caused by the disorder in tunneling amplitudes, which for harmonic modulation becomes significant around this modulation amplitude. Difference of phase factors between exact and approximated expressions for the effective tunneling appears much earlier and do not seem to have large impact on the localization length.

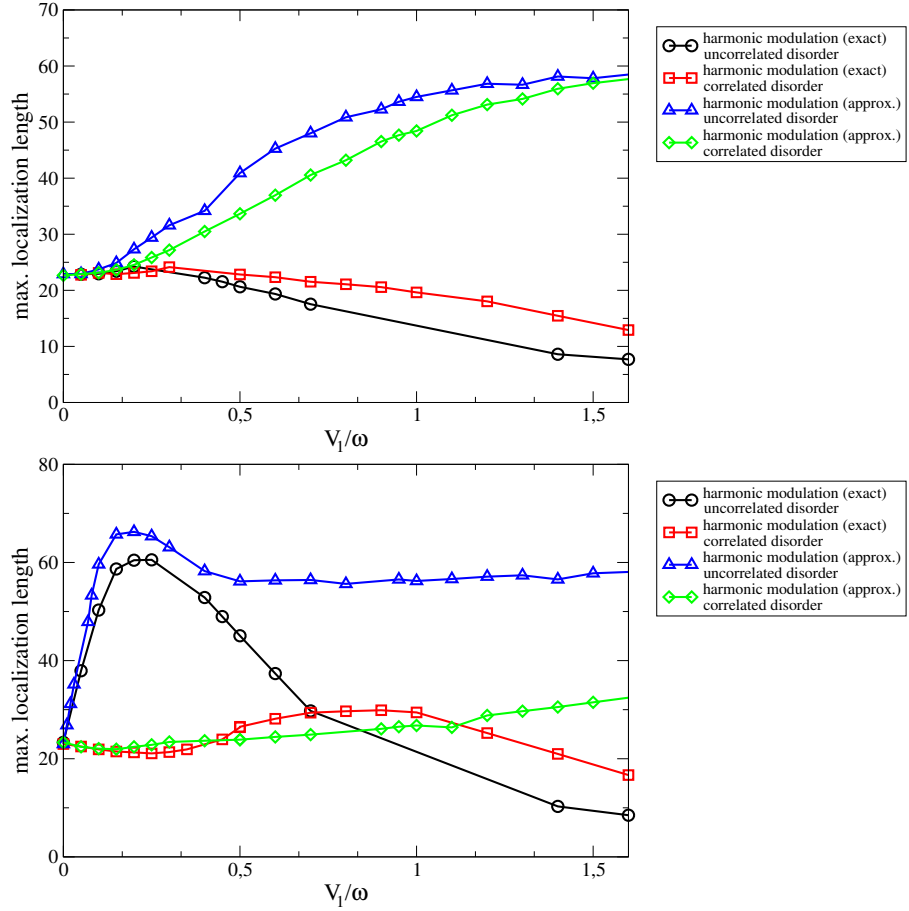

Figure 27: The maximum localization length (in units of the lattice constant) as a function of parameter of modulation $V_1/\omega$ for the cases of symmetric $t_1^x = t_1^y$ (upper panel) and antisymmetric lattice modulation $t_1^x = -t_1^y$ (lower panel). Black dots are results for the harmonic modulation and uncorrelated disorder; red squares stand for the harmonic modulation and correlated disorder; green diamonds are results for the approximated harmonic modulation and uncorrelated disorder and blue triangles for the approximated harmonic modulation and correlated disorder (lines are only guides for the eye).

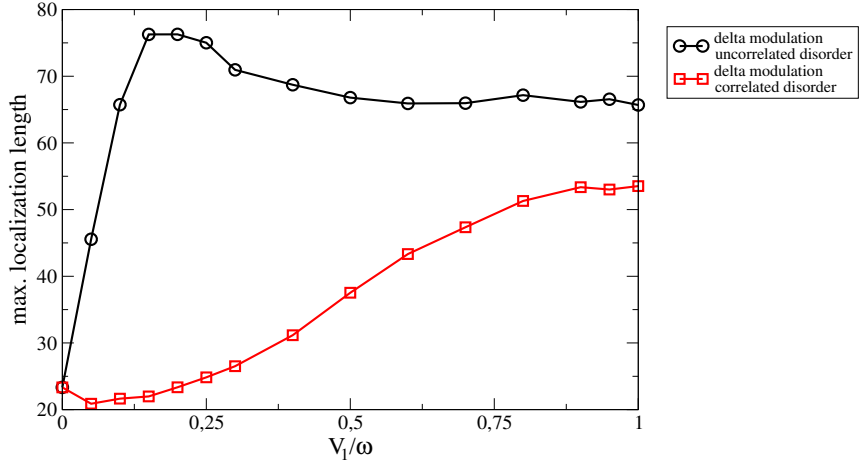

Figure 28: The maximum localization length (in units of the lattice constant) as a function of parameter of modulation $V_1/\omega$ for delta modulation. Black dots are results for uncorrelated disorder and red squares stand for correlated disorder (lines are only guides for the eye).

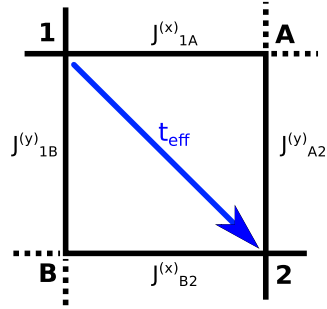

Figure 29: One plaquette disconnected from the lattice. Tunneling from site 1 to 2 is given by (130).

*Strengthening of the localization due to correlations*

Important feature clearly visible in the figures 27(lower panel) and 28 is the large discrepancy between the localization lengths for systems with correlated and uncorrelated disorder. For the uncorrelated case we see the rapid increase of the localization length which agrees with the intuition what should happen upon breaking of the time reversal symmetry. On the contrary, for the correlated disorder (the main model we discuss) after adding the disordered fluxes the localization length grows much slower, or even show some small decrease. With growing modulation parameter $V_1/\omega$ it starts to slowly approach the result for the uncorrelated case.

This unexpected behavior could be understood using a simple toy model of transport through one plaquette cut out of the whole lattice (presented in Fig. 29). The tunneling through such a structure

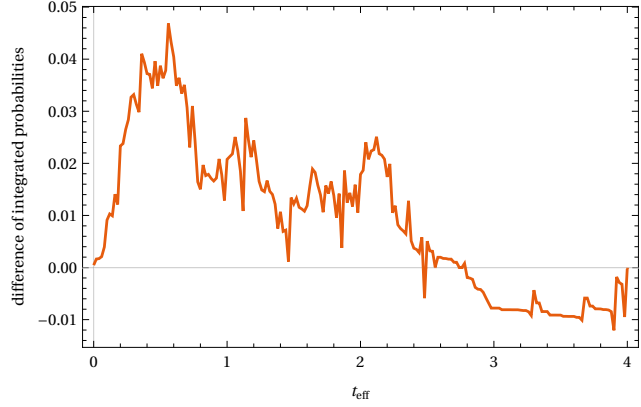

Figure 30: Plot of difference of integrated histograms of $t_{eff}$ for correlated and uncorrelated disorder for $E = 3.15$ and $V_1/\omega = 0.2$ (delta modulation).

could be easily calculated (for example using general method shown in section 2.4.2):

$$t_{eff} = \frac{1}{E - V_0 n_A^f} J_{2A}^y J_{A1}^x + \frac{1}{E - V_0 n_B^f} J_{2B}^x J_{B1}^y, \tag{130}$$

where the lattice sites are denoted as in Fig. 29 and $E$ is energy of the state. By calculating the large number of $t_{eff}$ for different disorder realizations for the cases with correlated as well as uncorrelated disorder (for delta modulations) we could check that indeed for the correlated case there are more low $t_{eff}$, as shown in Fig. 30 with the difference of integrated histograms for correlated and uncorrelated cases – thus the transport in this case is hindered.

To compare $t_{eff}$ for a wider range of parameters I use the average over logarithm $\langle \log t_{eff} \rangle$ much like in the case of the transmission (Sec. 1.5.4) to lower the impact of rare configurations giving very high $t_{eff}$. In Fig. 31 the difference of results for correlated and uncorrelated cases is presented ($\delta = \langle \log t_{eff} \rangle_{uncorrelated} - \langle \log t_{eff} \rangle_{correlated}$), showing that indeed for correlated case the transport is slower. Although, the effect is quite small and nearly vanish for $V_1/\omega > 0.7$, for which we still could observe significant discrepancy in the localization length, it appears also for tunneling through two plaquettes and may signal some general behavior.

*Interplay of localization and symmetry breaking*

As we have seen in preceding section, the correlations between on-site disorder and flux disorder could significantly alter the localization length. To avoid this effect and analyze how the appearance of the random fluxes affect the localization length I will use results for the uncorrelated case. In Fig. 32 the localization length is plotted as a function of the mean absolute flux through plaquette $\langle|\Phi|\rangle$[6]. The presented curves are results for the delta modulation and symmetric and antisymmetric harmonic modulation (I do not use results for exact harmonic modulation as it contains also the disorder in tunneling amplitudes which obscures discussed phenomena). We could see that all three curves scale similarly, which indicates that a particular distribution of disorder does not affect results very strongly. Moreover we can notice that for low mean absolute fluxes they coincide with (also plotted) results for staggered field. This suggests that in this regime the breaking of the time reversal symmetry is the most important effect. For higher $\langle|\Phi|\rangle$ we observe that results for disordered cases diverge from one for the staggered field – this marks the region in which flux disorder starts to significantly alter the localization length. Also, if we plot those three cases in a function of the flux variance they scale similarly as shown in figure 33.

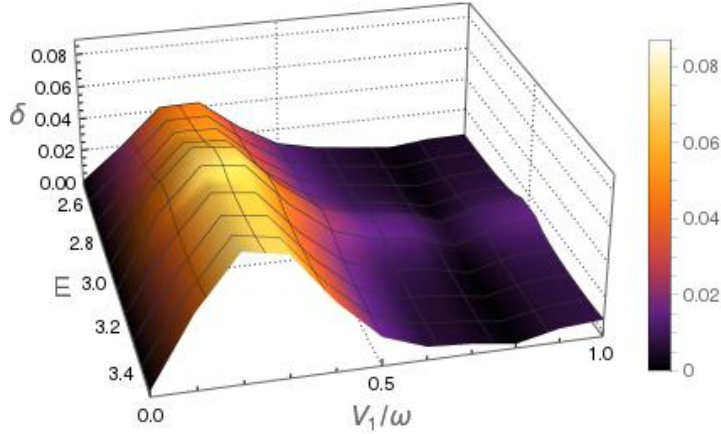

Figure 31: Plot of the difference of averaged logarithms of $t_{\text{eff}}$ for uncorrelated and correlated disorder as a function of state energy E (in units of tunneling amplitude) and modulation parameter $V_1/\omega$ (for delta modulation).

*Scaling of the mean free path*

At the end I will present preliminary results of comparison of the mean free path obtained from the localization length with one obtained analyzing the evolution of localized wavepacket.

Due to the scaling theory of the localization the localization length in the two dimensional systems in the *orthogonal* class is connected with the transport mean free path $\ell$ describing the distance the parti-

---

6  Absolute value have to be used as we are interested in amplitude of the disorder not the mean value of flux – which is zero in our case.

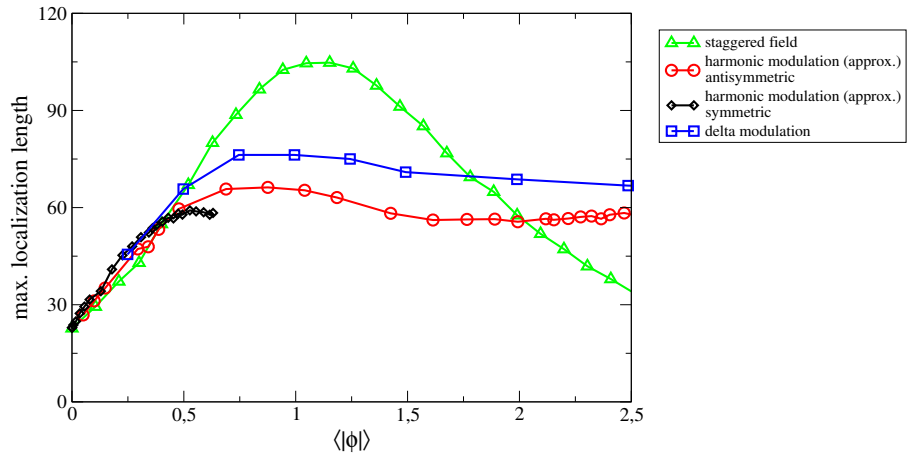

Figure 32: The maximum localization length (in units of the lattice constant) in function of mean absolute flux through plaquette. Green triangles are results for solely diagonal disorder and staggered gauge field; red squares and black diamonds for approximated harmonic modulation antisymmetric and symmetric respectively; blue squares for delta modulation (lines are only guides for the eye).

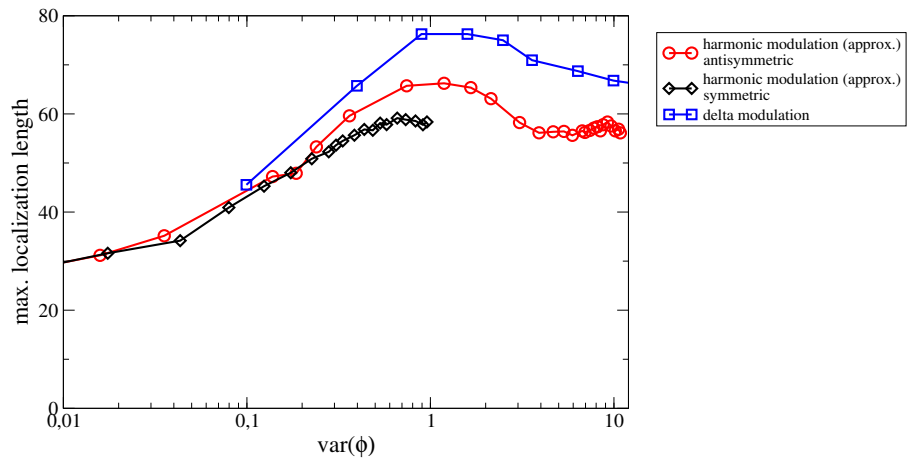

Figure 33: The maximum localization length (in units of the lattice constant) in function of variance of flux through plaquette. Red squares and black diamonds are results for approximated harmonic modulation antisymmetric and symmetric respectively; blue squares for delta modulation (lines are only guides for the eye).

cle can travel before its movement becomes randomized. The approximate relation is given by:

$$\lambda = \ell e^{\frac{\pi}{2} k \ell},\tag{131}$$

where $k$ is a momentum (in our case quasimomentum) and the factor $\pi/2$ comes from the self consistent theory of localization [55] and does not have to be exact for all cases, in particular for the strong disorder.

To check whether the formula (131) holds for the considered model, I have calculated the mean free path using two different methods. Both of them relies on a propagation of the initially localized wavepacket in the disordered system. The system size has been chosen to be $100 \times 100$ (and is sufficient, as the state never reaches the boundary) and the initial state has been Gaussian (with variance $\sigma^2 = 15$) multiplied by a plane wave with quasimomentum satisfying: $|k_x| + |k_y| = \pi$. The time evolution has been calculated with the help of QuTiP package [129] for times of order of 3 tunneling times (defined as an inverse of the tunneling rate). The mean free path could be extracted from the results in two ways:

1. By checking the mean distance at which expectation value of position operator becomes randomized (direct readout of the mean free path).

2. By calculating the slope of a mean squared displacement $\langle \mathbf{r}^2 \rangle$ of wave function. For intermediate times, when the slope is constant[7] it is connected with the mean free path by expression (valid in two-dimensions):

$$\ell = \frac{\partial_t \langle \mathbf{r}^2 \rangle(t)}{2v},\tag{132}$$

where $v$ is velocity of wave packet which could be obtained from the slope of expectation value of position operator for short times, before the mean free path is reached.

The mean free path obtained using the methods above and the one calculated from the localization length (for model with only diagonal disorder, $V_1/\omega = 0$) clearly coincide as shown in Fig. 34. The mean free path shows algebraic dependence on the disorder strength (the results of fitting the power-law are gathered in table 1). Although slopes differ, we observe that the expression (131) is approximately satisfied for my model, which is not trivial as it was derived by means of the perturbation theory in the regime of a weak disorder.

In the systems with broken time reversal symmetry the weak localization correction vanishes and the localization in two-dimensional systems is assured by the next order correction to scaling function $\beta(g)$. From this we could expect that relation between the mean free path and localization length should be given by:

$$\lambda = \ell e^{\zeta(k\ell)^2}. \tag{133}$$

---

7 For shorter times $\langle \mathbf{r}^2 \rangle$ grows quadratically due to ballistic expansion and for longer times wavefunction starts to localize and the expansion slow down and eventually stop

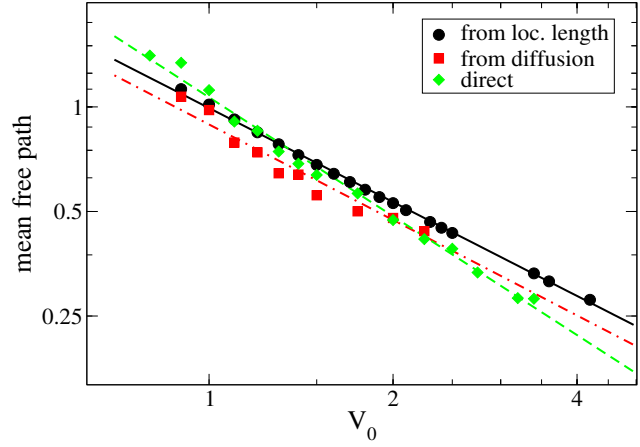

Figure 34: Mean free path (in units of lattice constant) as a function of parameter $V_0$ – disorder strength (in units of tunneling amplitude). Black dots are results calculated from the localization length using (131), red squares are results from diffusion coefficient and green diamonds directly from the mean distance traveled by wavepacket. Black solid, red dashed-dotted and green dashed lines are fitted to respective sets.

|  | slope($a$) | intercept($b$) |
|---|---|---|
| **no flux** | | |
| loc. length | $-0.9049(73)$ | $-0.049(54)$ |
| diffusion | $-0.983(75)$ | $-0.103(33)$ |
| direct | $-1.143(29)$ | $0.064(19)$ |
| $V_1/\omega = 0.1$ | | |
| loc. length | $-0.5964(78)$ | $-0.3720(78)$ |
| direct | $-1.083(17)$ | $0.987(10)$ |
| $V_1/\omega = 0.5$ | | |
| loc. length | $-0.5977(38)$ | $-0.3619(32)$ |
| direct | $-0.832(44)$ | $-0.742(25)$ |
| $V_1/\omega = 1.1$ | | |
| direct | $-0.450(33)$ | $0.298(20)$ |

Table 1: Results of fitting power law ($\ell = bV^a$) to the mean free path as a function of disorder strength.

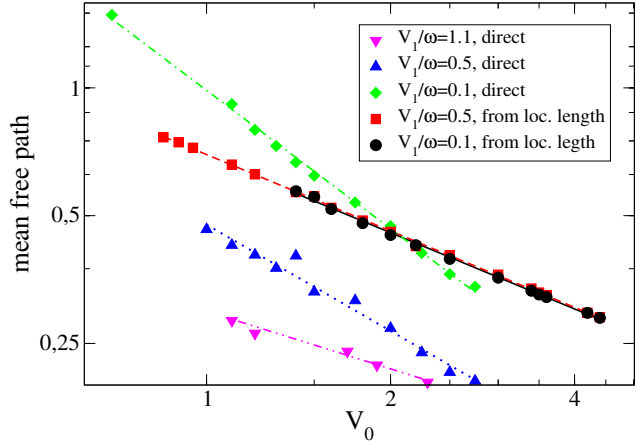

Figure 35: Mean free path (in units of lattice constant) as a function of parameter $V_0$ – disorder strength (in units of tunneling amplitude). Black dots and red squares are results calculated from the localization length using (133) for $V_1/\omega = 0.1$ and $0.5$ respectively, green diamonds, blue up-triangles and magenta down-triangles are results obtained directly from the mean distance traveled by wavepacket (for $V_1/\omega = 0.1$, $0.5$ and $1.1$). Lines are fits to respective sets.

As we can observe in the Fig. 35 and check in the table 1, for small and moderate modulation parameters $V_1/\omega$, the scaling of the mean free path obtained using expression (133) differs significantly form ones obtained directly form the time evolution. However, for larger fluxes $V_1/\omega \approx 1$. they start to scale similarly. It could suggest that only for the large amplitude of the flux disorder, those systems could be counted to the *unitary* class, while for lower fluxes the time-reversal symmetry is not completely broken. More probable explanation is that the approximate expression (133) holds only for sufficiently strong flux disorder. This topic could be continued, to check exact regimes in which the expression (133) gives good approximations. For sufficiently large set of results also the coefficient $\zeta$ could be calculated.

# CONCLUSIONS AND OUTLOOK

In the thesis I have presented results of the theoretical analysis of ultracold atomic gases in low-dimensional optical lattices with the disorder leading to the Anderson localization. I considered the disorder created using two species of particles: atoms of the first species are moving in the disordered potential created by interaction with the second fraction distributed randomly and immobilized.

The first part has been focused on the possibility of creation of the multi-mode, tunable filter for energies of atoms. I showed that such filters could be obtained using two one-dimensional optical lattices with perpendicular polarizations and different wavelength, holding *mobile* and *frozen* fractions respectively. This setup gives rise to a correlated disorder falling into a class of the generalized random N-mer model. For experimentally attainable parameters, those systems have several extended states providing 'windows of transport' in finite systems. By further extending presented setup with fast periodic modulation of the interactions between the species (which effectively works as a modulation of on-site energies) the off-diagonal disorder is created. It widens the range of possible setups, allowing changing the number of extended states as well as their relative positions. I have also presented general method of treating systems with generalized N-mer correlations, showing that energies of extended states can be calculated using commutation properties of transfer matrices for different blocks appearing in the system and that those states are indeed extended – in the infinite lattice transport do not vanish.

This studies can be continued in several directions. One, is usage of such filters to shape the wavefunction in a manner very similar to shaping of the pulses of light by adding several precisely chosen frequencies [71]. One may think also of extending presented formalism to larger number of dimensions (it is known that for random dimer correlations also in two-dimensional systems we could expect extended states [130]), or to check whether such correlations could have any impact on the many body localization. Another direction is further exploration of the properties of the transfer matrices, especially for systems with complicated functional dependence of transfer matrix on the state energy.

In the second part of the thesis, the setup allowing creation of the random artificial gauge field in two dimensional optical lattice has been presented. The random field could be created by transferring the disorder form on-site energies to complex phases of tunnelings by the means of the fast periodic modulation of the interspecies interactions and lattice height. In the described model both flux disorder and on-site disorder are present and their relative amplitude could be changed. It is also possible to create setups with both types of disorder strongly correlated as well as uncorrelated (however the latter is significantly more experimentally demanding). In principle those systems belong to the *unitary* class of disordered systems as the time-reversal symmetry is broken by the created magnetic field and the localization length should grow significantly comparing with the undriven model (which belongs to the *orthogonal* class). However, for the correlated case I have observed an unexpectedly small localization length. It appears that, for correlated disorder the probability of destructive interference when considering different paths between two points is increased, which have been shown on a toy model considering only one and two plaquettes. Results for the uncorrelated case show expected growth of the localization length but present other interesting effect: interplay between strengthening of localization due to increasing the amplitude of flux disorder and weakening it by breaking of the time-reversal symmetry caused also by the random fluxes. Finally, the theoretical expression connecting the localization length with mean free path is compared with results obtained by numerical solution of time dependent Schrödinger equation.

The maximum localization length grows for the rising flux amplitudes but the localization length as a function of energy of state becomes narrower (see Fig. 26), it means that possibly those systems could be used also as filters for energies (similarly to ones described in chapter 2). They do not provide so narrow peaks of delocalization, still they could be an interesting realization of the idea of the filter in two-dimensional systems. Moreover, the appearing peak in the localization length could be controlled solely by changing relative phases of modulation in different lattice directions. Another interesting question is whether it is possible to create also non-abelian gauge fields in similar manner, which will allow testing of the range of models described for example in [131].

APPENDIX

# EVOLUTION OF TIME PERIODIC SYSTEMS

In this appendix I am going to briefly describe methods, based on the Floquet theory, allowing to treat time periodic Hamiltonians. I will show method of calculating effective Hamiltonians governing behavior of periodically driven systems for times much longer than one modulation period. In the first section I will describe the properties of solutions of the Schrödinger equation with the time periodic Hamiltonian and present more 'intuitive' approach to deriving the effective time independent Hamiltonian[1]. In the second part I will provide a more rigorous description, starting from a precise separation of periodic and non-periodic dynamics and eventually approaching the Magnus expansion, which allows to find the effective Hamiltonian up to the desired accuracy.

## FLOQUET HAMILTONIAN AND ITS SPACE

For the time-periodic Hamiltonian $H(t)$ with period $T$, the Schrödinger equation reads:

$$i\partial_t|\psi_n(t)\rangle = H(t)|\psi_n(t)\rangle, \tag{134}$$

due to the Floquet theorem [40] we know that the solutions of this equation could be written as a product of a phase factor and a function with the same periodicity as the Hamiltonian ($|\phi_n(t+T)\rangle = |\phi_n(t)\rangle$):

$$|\psi_n(t)\rangle = e^{i\epsilon_n t}|\phi_n(t)\rangle, \tag{135}$$

where $\epsilon_n$ could be called quasi-energy in clear correspondence to a quasimomentum in the Bloch theorem for spatially periodic potentials [32]. We could also note that similarly as for the quasimomentum, adding $\omega$ to the quasienergy will not change a state but only take us to a next 'Brillouin zone' for the quasienergies. This marks a significant difference of the quasienergies from the energies of states – there is no natural ordering of energy, especially the quasienergy ground state could not be defined. Also, as the $|\psi_n(t)\rangle$ are time dependent they could not be considered as the eigenstates of $H(t)$. However, after imposing (135) into (134) we get:

$$\epsilon_n|\phi_n(t)\rangle = (H - i\partial_t)|\phi_n(t)\rangle, \tag{136}$$

---

1 This approach is in fact used in most of works concentrated on applying Floquet theorem, not on analyzing its mathematical properties.

which looks exactly as the eigenequation. To address this problem rigorously we could define a new operator: $\mathcal{H} = H - i\partial_t$, an Hamiltonian[2] acting in the extended space of T-periodic functions existing in the Hilbert space of the original Hamiltonian. To distinguish states in both spaces I apply convenient notation introduced in [125], and mark states in extended space by double brackets:

$$\mathcal{H}|\phi_n^m\rangle\rangle = (\epsilon_n + m\omega)|\phi_n^m\rangle\rangle. \tag{137}$$

In the definition of eigenstates in the extended space I have added new index $m$ denoting states differing by energy $m\omega$. We know that in the physical space those states are identical, but are they also in the extended space? To check this we have to define scalar product. We use scalar product from the physical space expanded by integration over one modulation period as in the extended space time is treated rather as a new coordinate than evolution variable:

$$\langle\langle\phi_n^m|\phi_n^{m'}\rangle\rangle \equiv \frac{1}{T}\int_0^T dt\langle\phi_n(t)e^{-i\omega m}|\phi_n(t)e^{i\omega m'}\rangle. \tag{138}$$

Indeed for $m \neq m'$ states $|\phi_n^m\rangle\rangle$ and $|\phi_n^{m'}\rangle\rangle$ are orthogonal. This means that one physical state $|\psi_n(t)\rangle$ represents whole class of orthogonal states $\{|\phi_n^m\rangle\rangle\}_{m\in\mathbb{N}}$ in the extended space. Two facts stem from this:

- We have to analyze properties of the full Hamiltonian spanning in $n$ and $m$. Especially the coupling between subspaces with different $m$-s is the reason of non trivial effects of periodic modulation.

- When going back to describe behavior of the system in physical space we need only one of whole class of solutions for different $m$-s as they correspond to only one physical state.

Very helpful in dealing with this problem is a representation of $\mathcal{H}$ as a block Hamiltonian with blocks ordered by the quantum number $m$:

$$\mathcal{H} = \begin{bmatrix} H_0 + \omega & H_1 & H_2 \\ H_{-1} & H_0 & H_1 \\ H_{-2} & H_{-1} & H_0 - \omega \end{bmatrix}, \tag{139}$$

where $H_m$ are Fourier components of the time dependent Hamiltonian:

$$H(t) = \sum_{k=-\infty}^{\infty} e^{i\omega k t} H_k. \tag{140}$$

---

2 In some works the name 'quasienergy operator' is used instead.

It is clear now that the diagonal blocks preserving number $m$ are just the time averaged $H(t)$ subsequently shifted by $\omega$, while the time dependent parts ($H_{j\neq0}$) provide the coupling between subspaces for different $m$. For now we could forget about in-block dynamics and concentrate on dealing properly with those couplings.

There are two main classes of the time dependent Hamiltonians considered in the literature: with the amplitude of modulation independent from the frequency and with linear dependence between the amplitude and the frequency of the modulation. In the first case it is obvious that in the infinite frequency limit ($\omega \to \infty$) separation between the subsequent diagonal blocks will grow and as the couplings stay constant we could just neglect them and rely only on $H_0$. This picture is quite intuitive – very fast modulations acts on such short time scales comparing to the other processes going on in a system that they have no effect at all (in fact we are making this silent assumption dealing with nearly any physical problem). Situation is more complicated when the couplings grow with the growing frequency, then we can not neglect off-diagonal blocks as values of their elements are always of the same order as the separation of blocks. In this case even in the infinite-frequency limit we could expect nontrivial effects.

Infinite frequency limit is clearly an idealization and usually we deal with systems modulated with high but finite frequencies. In fact in standard optical lattice setups the upper bound for modulation frequency is well stated, as when modulation frequency coincides with energy separation between bands, we start to excite our particles and one-band tight-binding could not be used anymore. Summarizing, we want a method giving us effective Hamiltonian valid for high but finite frequencies. Ideal operation will be transforming $H(t)$ into a block diagonal form. Then one block will give us exact behavior of the system for any frequency (although for too small frequencies results will be useless, as argued in the next section). However, it is nearly always impossible to write such a transformation. The most popular method, allowing to at least partially solve this problem, is making a transformation to a rotating frame. If we write $H(t)$ as a sum of a time independent Hamiltonian ($H_0$) and a perturbation ($H'$) multiplied by a time periodic function $f(t)$ (or in more general form, a sum of mutually commuting perturbations with different time dependencies: $\sum_i H_i' f_i(t)$):

$$H(t) = H_0 + H' f(t) \tag{141}$$

it is possible to make a transformation to a rotating frame $\mathcal{U}(t) = \exp(iH' \int dt f(t))$. If we apply it to the Schrödinger equation:

$$i\partial_t |\mathcal{U}(t)\psi(t)\rangle + \underline{f(t)H'|\mathcal{U}(t)\psi(t)\rangle}$$
$$= \mathcal{U}(t)H_0\mathcal{U}^\dagger(t)|\mathcal{U}(t)\psi(t)\rangle + \underline{f(t)H'|\mathcal{U}(t)\psi(t)\rangle}, \tag{142}$$

we could see that it effectively removes the term $H'f(t)$ from the Hamiltonian and instead dresses $H_0$ in the quickly rotating phase:

$$H_{rot}(t) = \mathcal{U}(t)H_0\mathcal{U}^\dagger(t). \tag{143}$$

As long as $H'$ is local it is usually easy to find a closed form of $H_{rot}$. As the transformation involves the integral of $f(t)$, it drops a coefficient of the order $1/\omega$. Effectively the off-diagonal terms (in sense of the block couplings) of $\mathcal{H}$ are now decaying with the growing $\omega$. For the case of the amplitude scaling with the frequency it allows us to find the infinite frequency limit so $\langle H_{rot}\rangle_T$ is the infinite frequency Hamiltonian for this case. If the amplitude is constant, then couplings decay with growing energy and in the effect the time averaged $H_{rot}(t)$ could be considered valid even for moderately fast modulations.

### THE MAGNUS EXPANSION

This section presents a slightly different approach to dealing with the time periodic Hamiltonian. In the last section we get an approximate expression for the effective Hamiltonian. The natural questions arise: what exactly is omitted to get the time-independent Hamiltonian and what is the accuracy of obtained approximation.

We will start with the first question. Let us consider the evolution operator in the system governed by the time dependent Hamiltonian $H(t)$:

$$U(t_2, t_1) = \mathcal{T}\exp\left(-i\int_{t_1}^{t_2} dt H(t)\right), \tag{144}$$

where $\mathcal{T}$ stands for a time ordering. For the time periodic $H(t)$ the evolution operator should be invariant under transformation $t_1 \to t_1 + nT$ and $t_2 \to t_2 + nT$. This feature along with the factorization property of $U(t_1, t_2)$ gives:

$$U(t_0 + 2T, t_0) = U(t_0 + 2T, t_0 + T)U(t_0 + T, t_0) = U(t_0 + T, t_0)^2, \tag{145}$$

which could be trivially generalized to any $n$. As the evolution operator is unitary it could be always written as an exponent of some other unitary operator so we get:

$$U(t_0 + nT, t_0) = U(t_0 + T, t_0)^n = \exp\left(-iH_{eff}[t_0]nT\right). \tag{146}$$

In this way we have obtained an important result: if we study only stroboscopic evolution of the the system (consider only times $t_0 + nT$), it is possible to find an unitary operator $H_{eff}[t_0]$ which is generating this evolution and is time independent[3] – the celebrated effective

---

3 It has only a parametric dependence on $t_0$, which is in fact a gauge choice (see Fig. 36).

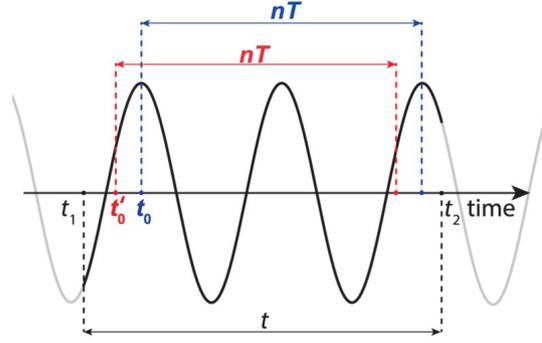

Figure 36: Construction of evolution operator $U(t_1, t_2)$ in time periodic system, with two different choices of starting points of stroboscopic evolution (choice of the Floquet gauge $t_0$), (source: [132]).

Hamiltonian. If we want to construct an evolution operator for arbitrary times we could write:

$$U(t_2, t_1) = U(t_2, t_0 + nT)e^{-iH_{eff}[t_0]nT}U(t_0, t_1)$$
$$= P(t_2 - nT - t_0)e^{-iH_{eff}[t_0](t_2 - t_1)}P^\dagger(t_1 - t_0), \quad (147)$$

where $P$ is defined as:

$$P(t) \equiv U(t_0 + t, t_0)\exp(iH_{eff}[t_0]t). \quad (148)$$

It is straightforward to check that $P(nT) = 1$. In this way we see that the general evolution operator (for arbitrary times) could be decomposed into two parts: $P(t)$ is describing the *micromotion* a periodic changes, which the system undergoes in the every turn of the modulation and time independent $H_{eff}$ governing the long term dynamics. Exact forms of the decomposition depends on choice of starting time $t_0$, the physical meaning of this choice is clearly shown in fig.36. We could see also that proper change of gauge allows us to left only one of micromotion operators in (147). The main concern for us is finding the $H_{eff}$ for fast modulations. However, we have to remember that even if we somehow obtain the exact $H_{eff}$ it will not give us full description of the system and, especially for smaller frequencies the micromotion has to be taken into account.

In general, finding the form of the evolution operator for the time dependent Hamiltonian is formidable task. A method allowing calculation of the approximate solution have been created by Wilhelm Magnus [133]. One of main concerns upon deriving the expansion has been to preserve group properties of the evolution operator i.e. if we are in framework of the quantum mechanics we want any order of the expansion to be an unitary operator. Such an approach allows one to safely use a desired order of expansion without worry that it contains unphysical effects which have to be neglected or countered with terms from higher orders etc.

I will not present derivation of the expansion as it is quite complicated and could be found in source work [133] or a good review [128], instead I will just summarize main results. If we have a system:

$$\partial_t Y(t) = A(t) Y(t), \tag{149}$$

we could assume that there exist solution of the form:

$$Y(t) = e^{\Omega(t)} Y(0) = e^{\sum_{n=1}^{\infty} \Omega_n(t)} Y(0) \tag{150}$$

The terms of the sum could be calculated using following procedure (unfortunately there is no closed form of expansion):

$$\Omega_1(t) = \int_0^t d\tau A(\tau) \tag{151}$$

$$\Omega_n(t) = \sum_{j=1}^{n-1} \frac{B_j}{j!} \int_0^t d\tau S_n^{(j)}(\tau), \tag{152}$$

where $B_j$ are Bernoulli numbers and $S_n^{(j)}(t)$ are defined by the recurrence relation:

$$S_n^{(1)}(t) = [\Omega_{n-1}(t), A(t)], \tag{153}$$

$$S_n^{(n-1)}(t) = \mathrm{ad}_{\Omega_1(t)}^{n-1}(A(t)), \tag{154}$$

with $\mathrm{ad}_X^n(Y)$ defined recursively (again!): $\mathrm{ad}_A^0 B = B$, $\mathrm{ad}_A^i B = [A, \mathrm{ad}_A^{i-1} B]$.

In the search for the effective Hamiltonian in time periodic case we could use the separation of the evolution operator:

$$|\psi(t)\rangle = P(t) \exp(iH_{\mathrm{eff}}) |\psi(t_0)\rangle \tag{155}$$

Using the fact that $P(nT) = \mathbb{I}$ we could calculate terms of Magnus expansion for stroboscopic times ($t = nT$) and in this way obtain the effective Hamiltonian:

$$|\psi(T)\rangle = \exp(iH_{\mathrm{eff}}) |\psi(t_0)\rangle \tag{156}$$

$$H_n = \frac{\Omega_n(T)}{T}. \tag{157}$$

It is all we have wanted, but it is worth knowing that with $H_{\mathrm{eff}}$ determined up to $n$-th term and using straightforward extension of the Magnus expansion it is possible to calculate also the *micromotion* evolution operator up to the same order (as presented in [128], section 3.2).

It is not simple to determine arbitrary terms in Magnus expansion, however in practice only three first terms are used (usually only the first one). If the modulation is symmetric around zero, first correction vanish and we need the second to calculate deviations from the time-average. Those first three terms reads:

$$H_F^{(0)} = \frac{1}{T} \int_0^T dt H(t), \tag{158}$$

$$H_F^{(1)} = \frac{1}{2!Ti} \int_{t_0}^{T+t_0} dt_1 \int_{t_0}^{t_1} dt_2 [H(t_1), H(t_2)], \tag{159}$$

$$H_F^{(2)} = -\frac{1}{3!T} \int_{t_0}^{T+t_0} dt_1 \int_{t_0}^{t_1} dt_2 \int_{t_0}^{t_2} dt_3 ([H(t_1), [H(t_2), H(t_3)]]$$
$$+ (1 \leftrightarrow 3)). \tag{160}$$

For our purposes usually more convenient is to use a Fourier transformed form of $H(t)$:

$$H_F^{(0)} = H_0, \tag{161}$$

$$H_F^{(1)} = \frac{1}{\omega} \sum_{l=1}^{\infty} \frac{1}{l} \left( [H_l, H_{-l}] - e^{il\omega t_0}[H_l, H_0] + e^{-il\omega t_0}[H_{-l}, H_0] \right). \tag{162}$$

The second order of the expansion could not be written in short form, only the general one:

$$H_F^{(2)} = -\frac{1}{3!T} \sum_{klm} \mathcal{I}_{klm}^{(\omega, t_0)} [H_k, [H_l, H_m]] \tag{163}$$

where:

$$\mathcal{I}_{klm}^{(\omega, t_0)} = \int_{t_0}^{T+t_0} \int_{t_0}^{t_1} \int_{t_0}^{t_2} dt_1 dt_2 dt_3 \left( e^{i\omega(kt_1 + lt_2 + mt_3)} + (1 \leftrightarrow 3) \right) \tag{164}$$

and the integral is nonzero only for $k$, $l$ or $m = 0$; $k = -l$; $k = -m$; $l = -m$ and $k + l + m = 0$. In this form also the connection with the Floquet Hamiltonian in the extended space is more clearly visible as the first term, is the diagonal block while further corrections are calculated by taking into account couplings $H_{i \neq 0}$

The Magnus expansion gives results in powers of $1/\omega$ as long as modulation amplitude is independent on frequency. If, on the other hand, we have linear dependence of amplitude on frequency the powers of expansion get mixed, and it is possible to check that for every level of the expansion one can find element which is of order 1 (independent on frequency) – thus whole expansion is useless in this scope. However by making transformation to the rotating frame frame we regain strict connection between order of the expansion and order of $1/\omega$. What is a reason of doing the same base change in case of systems with constant amplitude, one can ask? As then the off-diagonal blocks $H_{i \neq 0}$ drops terms of order $1/\omega$ the convergence of the series is enhanced to $1/\omega^{2n}$. In this way, if we make transformation and assure that modulation is symmetric we could neglect the first order of the expansion and the second is of the order $^1/_{\omega^4}$. Thus relying on just the time average in rotating frame is equivalent to calculating second term of Magnus expansion, and usually is much simpler.

# CALCULATION OF THE LOCALIZATION LENGTH

Set of analytical results in the Anderson localization theory is quite limited and usually restricted to asymptotic or very specific cases (weak disorder, special types of correlations), thus a wide variety of different numerical methods have been developed to calculate the localization length. In this appendix I have gathered some of them – especially those used by myself in course of my studies, but also several similar methods for comparison. I will begin with the model-specific methods, from simplest applicable only to one-dimensional systems with the nearest neighbor tunnelings, through ones usable in systems with long-ranged tunnelings or quasi one-dimensional stripes, to techniques applicable for higher dimensions. At the end I will discuss viability of using an exact diagonalization in investigating the localization and specific cases in which it could be especially useful.

## ONE DIMENSIONAL SYSTEMS

### Nearest neighbors tunneling

The standard method of calculating the localization length in one-dimensional systems emerges directly from a simple recursive calculation of a wavefunction, using the time independent Schrödinger equation to get the wavefunction in further parts of the system. We can expect that for the Anderson localized system the wavefunction amplitude will exponentially fall or rise and the slope of the exponent will give the localization length. Such a calculation could be made using the transfer matrices (which is harder albeit more general method). The transfer matrix for one dimensional system with only nearest neighbor tunneling reads (as defined in section 2.4):

$$T_i = \begin{bmatrix} \frac{\epsilon_i - E}{t_i} & -\frac{t_{i-1}}{t_i} \\ 1 & 0 \end{bmatrix}. \tag{165}$$

As transfer matrices could be simply chained:

$$\begin{bmatrix} \psi_{j+1} \\ \psi_j \end{bmatrix} = T_j \cdot \ldots \cdot T_i \begin{bmatrix} \psi_i \\ \psi_{i-1} \end{bmatrix} \equiv T_i^j \begin{bmatrix} \psi_i \\ \psi_{i-1} \end{bmatrix}, \tag{166}$$

we could find the transfer matrix describing the transport through arbitrary chosen part of the system – even the whole lattice. In the disordered system the transfer matrices are random matrices parametrized by state energy $E$. According to Fustenberg theorem [134] (variant of the Central Limit Theorem for non-commuting matrices) multiplying $n$ of such matrices in the limit of $n \to \infty$ give matrix with eigenvalues $e^{\pm \gamma(E)n}$ where parameter $\gamma(E, W)$ (Lapunov exponent) depends on the state energy $E$ and the disorder type $W$, but not on specific random realization of the disorder. An inverse of the Lapunov exponent is the localization length: $\lambda = 1/\gamma$. The procedure seems simple, we have to multiply the transfer matrices until we get the convergent value of $\gamma$. The problem is that the ratio of the matrix elements grows exponentially fast and usually reach machine precision way faster than the localization length converges. To overcome this problem one have to make a regularization of matrices in every step or at least once in several steps (depending on how fast they are rising). Details of the renormalization procedure are described in section B.2.3.

Fortunately, in the case of one-dimensional model with only nearest neighbor coupling, there exist easier and faster procedure. The Schrödinger equation could be rewritten into a form:

$$R_{i+1} = \frac{\epsilon_i - E}{t_i} - \frac{t_{i-1}}{t_i} \frac{1}{R_i}, \quad \text{where}: R_i = \frac{\psi_i}{\psi_{i-1}}. \tag{167}$$

It is a recursive equation for $R_i$, value which measures change of the wavefunction from site to site. Even for exponential growth of $\psi_i$, order of $R_i$ remains unchanged. From the assumption that:

$$\psi_n \xrightarrow[n \to \infty]{} A \exp\left(\frac{n}{\lambda}\right) \tag{168}$$

and the relation

$$\frac{\psi_i}{\psi_0} = R_i R_{i-1} \dots R_2 R_1, \tag{169}$$

it is easy to derive an equation for the localization length:

$$\lambda(n) = n \left( \sum_{i=1}^{n} \log R_i \right)^{-1} \xrightarrow[n \to \infty]{} \lambda. \tag{170}$$

Thus, it is sufficient to iterate (167) until value given by (170) converges.

*Long range tunnelings*

For long-range tunnelings, recurrence method presented above could not be easily used, especially if the long-range tunnelings are vanishingly small (which is usually the case). Of course one could write the transfer matrices for the systems with the tunnelings reaching $l$ sites (then the transfer matrix will have size $2l$), but this method can also pose problems. If we have inhomogeneous tunnelings which sometimes are zero, it would be hard or impossible to properly define the transfer matrices.

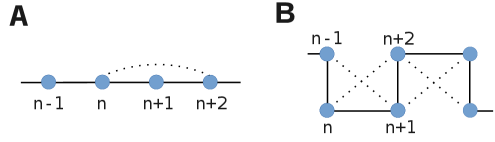

Figure 37: Method of winding a chain with long-range tunnelings to get quasi one-dimensional stripe with only nearest neighbor tunnelings (cross tunnelings between adjacent slides could also appear) [81].

Safe option in such a case is winding the lattice into a form of a quasi one-dimensional stripe to get only nearest neighbor tunnelings between slices (as on Fig. 37) and using method described in the subsequent section.

### QUASI ONE-DIMENSIONAL SYSTEMS

In this section I am going to describe a method which allows numerical calculation of the localization length in a stripe with width $M$ in one direction (denoted by the index $m$) and infinite in the other (denoted by the index $n$). We will start by properly defining change of Hamiltonian upon elongating the stripe, then we will obtain recurrence relation for a Green function and finally introduce renormalization procedure allowing us to avoid exponentially rising variables and obtain the localization length.

*Hamiltonian in stripe*

The system is described by a Hamiltonian

$$H = \sum_{n=1} H_n + \sum_{n=1} \left(V_n + \text{h.c.}\right), \tag{171}$$

where

$$H_n = \sum_{m=1}^{M} \epsilon_m^{(n)} |w_{n,m}\rangle \langle w_{n,m}|$$
$$- \sum_{m=1}^{M-1} \left( t_{m,m+1}^{(n,n)} |w_{n,m}\rangle \langle w_{n,m+1}| + \text{h.c.} \right) \tag{172}$$

is a tight binding Hamiltonian of the $n$-th slice and

$$V_n = -\sum_{m=1}^{M} \sum_{m'=1}^{M'} t_{m,m'}^{(n,n+1)} |w_{n,m}\rangle \langle w_{n+1,m'}| \tag{173}$$

is a coupling between slices $n$ and $n + 1$. Tunneling between site $(n, m)$ and $(n', m')$ is denoted $t_{m,m'}^{(n,n')}$, $\epsilon_m^{(n)}$ is energy on site $(n, m)$. Lets take the Hamiltonian of the stripe with length $N$:

$$H(N) = \sum_{n=1}^{N} H_n + \sum_{n=1}^{N-1} \left( V_n + \text{h.c.} \right).$$ (174)

Now if we add one extra slice to the system we get:

$$H(N + 1) = H(N) + H_{N+1} + V_N + V_N^\dagger,$$ (175)

which could be simplified as:

$$H(N + 1) = H^0(N + 1) + V(N),$$ (176)

by defining:

$$H^0(N + 1) \equiv H(N) + H_{N+1},$$ (177)

$$V(N) \equiv V_N + V_N^\dagger.$$ (178)

*A recursive equation for a Green's function*

To determine a transport through the system we have to find the Green's function $G_E^+(N + 1)$ for the Hamiltonian $H(N + 1)$. This can be done with the help of a resolvent of the Hamiltonian, defined as an operator satisfying an equation:

$$(z - H(N + 1)) \, G_z(N + 1) = 1.$$ (179)

The Green function could be found by taking the limit of the resolvent:

$$G_E^+(N + 1) = \lim_{\epsilon \to 0+} G_{E+i\epsilon}(N + 1).$$ (180)

Using (176) and an identity

$$\frac{1}{A} = \frac{1}{B} + \frac{1}{B}(A - B)\frac{1}{A},$$ (181)

with a substitution:

$$A = z - H(N + 1), \quad B = z - H^0(N + 1),$$ (182)

we get the equation for the resolvent of $H(N + 1)$:

$$G_z(N + 1) = G_z^0(N + 1) + G_z^0(N + 1)V(N)G_z(N + 1),$$ (183)

where $G_z^0$ is the resolvent of $H^0$. To get the transport properties of the system we do not need whole Green's function matrix but only its matrix elements between the first $|1\rangle$ and the last $|N+1\rangle$ slices: $\langle 1|G_z(N+1)|N+1\rangle$. In order to simplify equations we denote:

$$G_{n,m} \equiv \langle n|G_z(m)|m\rangle. \tag{184}$$

From the equation (183) using the relation:

$$G_z^0(N+1) = G_z(N) + \frac{1}{z - H_{N+1}}, \tag{185}$$

(which could be used due to a fact that $H(N)$ and $H_{N+1}$ act in orthogonal subspaces of the Hilbert space, see (177)) we get:

$$G_{1,N+1} = G_{1,N} V_N G_{N+1,N+1}. \tag{186}$$

To get a recursive equation for $G_{1,N}$, we need to calculate $G_{N+1,N+1}$ matrix. It could be done by taking equation (179) and multiplying it from the right side by a projection $P = |N+1\rangle\langle N+1|$, and by $P$ (or $Q = 1 - P$) from the left side:

$$\begin{aligned} P(z - H(N+1))(P+Q)G_z(N+1)P &= P, \\ Q(z - H(N+1))(P+Q)G_z(N+1)P &= 0. \end{aligned} \tag{187}$$

By solving the set of equations above we get:

$$G_{N+1,N+1} = \frac{1}{z - H_{N+1} - V_N^\dagger G_{N,N} V_N}. \tag{188}$$

Further, using both (186) and (188) we obtain:

$$G_{1,N+1} \left( z - H_N + 1 - V_N^\dagger G_{N,N} V_N \right) = G_{1,N} V_N \tag{189}$$

and again using (186) to extract $G_{N,N}$ we finally arrive to the recursive equation:

$$A_{N+2} = (z - H_{N+1}) V_N^{-1} A_{N+1} - V_N^\dagger V_{N-1}^{-1} A_N, \tag{190}$$

where $A_N = G_{1,N-2}^{-1}$. Similarly to recursive calculation in one dimension, choice of the initial values of $A_N$ do not affect the localization length, the convenient choice is for example:

$$A_0 = 0, \quad A_1 = V_0. \tag{191}$$

As there is no singularity in the equation (190), $z$ could be just replaced with state energy $E$. We have now all ingredients for:

*Calculation of the Anderson localization length*

The localization length in stripe of width $M$ ($\lambda_M$) is defined as:

$$\frac{2}{\lambda_M} = -\lim_{n \to \infty} \frac{1}{n} \ln \mathrm{Tr}|G_{1,n}|^2. \tag{192}$$

It can be calculated using the recursive equation (190). However when solving the equation iteratively, similar problem as for the transfer matrix method in one-dimensional systems appear – there is an exponential growth of the elements of $A_n$ for a large $n$. The regularization could be done by multiplying the both sides of the equation in each step by some matrix $R_n$. Starting from $n = 1$ and defining $A_k^{(1)} \equiv A_k R_1$ we get

$$\begin{aligned} A_3 &= (E - H_2) V_1^{-1} A_2 - V_1^\dagger V_0^{-1} A_1 \ \bigg| \times R_1 \\ A_3^{(1)} &= (E - H_2) V_1^{-1} A_2^{(1)} - V_1^\dagger V_0^{-1} A_1^{(1)}. \end{aligned} \tag{193}$$

In order to prevent matrix elements from growing exponentially we put $R_1 = A_2^{-1}$ and

$$A_3^{(1)} = A_3 A_2^{-1}, \quad A_2^{(1)} = 1, \quad A_1^{(1)} = A_1 A_2^{-1}. \tag{194}$$

By repeating the procedure in every step, we get: $A_k^{(n)} = A_k^{(n-1)} R_n$, with $R_n = \left[ A_{n+1}^{(n-1)} \right]^{-1}$, it satisfies:

$$A_{n+2}^{(n)} = (E - H_{n+1}) V_n^{-1} - V_n^\dagger V_{n-1}^{-1} A_n^{(n)}. \tag{195}$$

To 'store' information taken by $R_i$ lets define matrix:

$$B^{(n)} = B^{(n-1)} R_n / b_n, \quad b_n = \| B^{(n-1)} R_n \|, \tag{196}$$

with $\| \cdot \| = \sqrt{\mathrm{Tr} | \cdot |^2}$ a matrix norm. Using values of $\{b_n\}$ it is possible to get the localization length, as:

$$\begin{aligned} b_n &= \| B^{(n-1)} R_n \| = \frac{1}{b_{n-1}} \left\| B^{(n-2)} R_{n-1} R_n \right\| \\ &= \frac{1}{b_{n-1}} \left\| B^{(n-2)} \left[ A_{n+1}^{(n-2)} \right]^{-1} \right\| \\ &= \frac{1}{b_{n-1} b_{n-2}} \left\| B^{(n-3)} \left[ A_{n+1}^{(n-3)} \right]^{-1} \right\| = \dots \ . \end{aligned} \tag{197}$$

Making subsequent iterations we could notice that in fact:

$$\| A_{n+1}^{-1} \| = b_1 b_2 \dots b_n, \tag{198}$$

$$\ln \mathrm{Tr} | G_{1,n} |^2 = 2 \left( \ln b_{n+1} + \dots + \ln b_1 \right). \tag{199}$$

With the help of yet another variable $c_{n+1} = c_n + \ln b_{n+1}$ the Anderson localization length can be expressed as:

$$\lambda_M = -M \lim_{n \to \infty} \frac{n}{c_{n+1}}. \tag{200}$$

Same as in one-dimensional case we continue iterating (190) until $\lambda_M$ converges with prescribed accuracy. It is worth noting that the whole procedure of renormalization of the $A_i$ matrix (193–196) do not have to be done in every step. In fact, as it is most computationally costly part of the calculations (due to finding the inverses of the matrices), it should be done as rare as it is possible.

MACKINNON METHOD

Calculation of the localization length in one (or quasi one) dimension is somewhat straightforward as we have only one direction and limited (if any) loops. On the other hand two dimensional cases are much harder: the transfer matrix method could not be used, exact diagonalization is much slower as a size of space grows quadratically with linear size of system (which should be larger than the expected localization length). The localization length itself, even if it is finite (which is the case for most of the two dimensional systems), could grow fast with weakening disorder. A lot of works have been published, which mistakenly reported existence of the mobility edge because calculations have been made in too small systems. Presented method was specifically created for two dimensions but for above reason results showing the absence of the localization should be taken with care and askance. The method have been devised by MacKinnon and Kramer and described in [135, 136]. It is in some way based on a calculation for quasi one-dimensional stripes, as one could expect that calculating wider and wider stripes could eventually gave us the localization length for two-dimensional system. Using the method described in Sec. B.2 we carry out calculations for chosen disorder and state energy (denoted by composite variable $W$) and for several width of stripes $M$, obtaining set of the localization lengths $\{\lambda_{M_i}(W)\}$. More convenient for further analysis is the localization length divided by strip width $\Lambda(M, W) \equiv \lambda_M(W)/M$.

If we plot $\Lambda(M, W)$ versus $M$ we have two possible types of behavior: If $\Lambda(W, M)$ grows with growing $M$, we could expect that for $M \to \infty$ the localization length diverges and we have extended state. On the other hand, if $\Lambda(M, W)$ goes down, we could expect that the state is localized – $\Lambda$ falling to zero means $\lambda_M(W)$ reaching some constant value in the limit $M \to \infty$. In order to extract two dimensional localization length from the behavior of $\Lambda(M)$, we have to appeal to one parameter scaling theory (see Sec. 1.5.3). Upon assuming that under the change of a length scale $M' \to bM$ the change of $\Lambda$ depends on $\Lambda$ itself not on $M$ or $W$ separately [53, 137], we could readily write that:

$$\Lambda(M, W) = f\left(\frac{\xi(W)}{M}\right). \tag{201}$$

Where $f(x)$ is one parameter scaling function and $\xi(W)$ in localized cases could be considered the localization length for two dimensional system. As $f(x)$ is an injective function, if for two different sets of parameters $\Lambda(W_1, M_1) = \Lambda(W_2, M_2)$ also the arguments have to be the same. By taking logarithm we get:

$$\log \xi(W_1) - \log \xi(W_2) = \log M_2 - \log M_1. \tag{202}$$

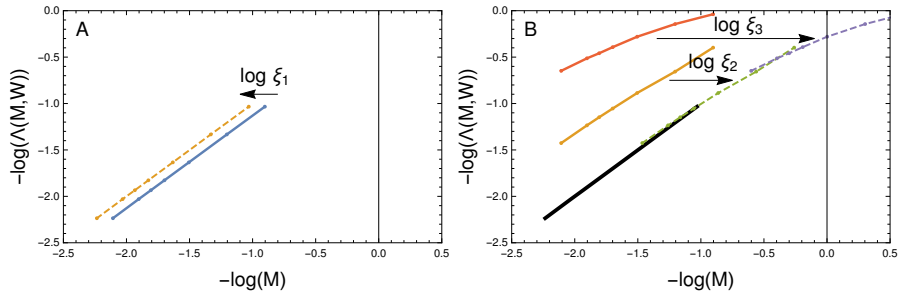

Figure 38: Procedure of finding the localization length. First (A) we calculate the localization length for the strongest disorder $\xi(W_1)$ to obtain the absolute position of scaling curve. Further (B) we shift subsequent curves for different disorders and from the shift we get their localization lengths. We could see that it is important that curves overlap – if the curve for $W_2$ (yellow) will be absent, we won't be able to fit curve for $W_3$ (red) as it do not overlap with already obtained part of scaling function (black).

In this way we could only find the relative values of localization length in comparison to the localization lengths for other $W_i$. To get the absolute values we have to pick at least one case for which we could find the localization length by other means. For strongly localized system we could expect that the localization length is significantly smaller than width of stripes used in calculations. We could then expect that $\lambda_M$ will converge fast to $\lambda_\infty = \xi$[1]. Using the value of localization length computed for the largest disorder (in case of constant disorder amplitude – for energy at the band edge, where the localization is the strongest), it is possible to establish absolute scale of the localization length. Then by the means of equation (202) we could find localization length for other $W_j$.

One could ask why we should bother with calculating whole set of $\lambda_M$ for each disorder $W$, as equation (202) allows us to calculate the localization length from just one data point? There are two motives. First reason is that such a procedure will be subject to large errors. If we have set of values for the same $W_i$ but different $M_j$-s, we know that for all of them $\xi(W)$ have to be the same and the fitting procedure is much more precise. Second reason is that, in most of the cases some of points for given $W_i$ is in the range of already fitted part of the scaling curve, while some are out of it. By using first part of them to find $\xi(W_i)$ and then using $\xi(W_i)$ to shift rest of points, we effectively find next part of $f(x)$. Such procedure let us to continue fitting for systems with even larger localization lengths.

Those arguments may become clearer, if I present the method in convenient graphical form (Fig. 38). We plot $\log \Lambda(W, M)$ in the func-tion of $\log M^{-1}$. Then we find the localization length ($\xi(W_1)$) for set with the smallest localization length (lowest curve) and then translate it by $\log \xi(w_i)$ establishing the first part of the scaling curve. Then by subsequently shifting further sets onto $f(x)$ we obtain $\xi(W_j)$. It can be done by minimizing a variance:

$$S_j = \sum_i \left( \frac{1}{N_{(i)}} \sum_{k=1}^{j} (\log M_{ik} - \log \xi(W_k))^2 \right.$$
$$\left. - \left( \frac{1}{N_{(i)}} \sum_{k=1}^{j} (\log M_{ik} - \log \xi(W_k)) \right)^2 \right) \qquad (203)$$

where $i$ goes over strip widths and $k$ over $W$-s, from strongest disorder to first not fitted ($W_j$) all $\xi(W_j)$ for $k < j$ are are known parameters while $\xi(W_j)$ is variable fitted to minimize $S_j$. The value $\log M_{ij}$ is a distance from scaling curve in Fig. 38. In order to find it, we could interpolate (but not extrapolate) known part of $f(x)$. As long as there are no gaps between sets for the different $W_j$ – upon projecting them on $\Lambda(M, W)$ axis they overlap – we are able to find the localization length for all disorder strengths. It is worth mentioning that $f(x)$ describes universal scaling law and therefore should be the same for wide classes of the systems.

---

[1] In other words we could say that we have asymptotic form of scaling function $f(x)$: $f(x) \xrightarrow[x \to 0]{} x$

### EXACT DIAGONALIZATION

Method applicable to any system, regardless its dimensionality, range of tunnelings etc. is an exact diagonalization of the Hamiltonian. However, in most of the cases it is less efficient than other methods described above, still there are some situations when it is worth considering.

Of course one have to remember that, if the chosen system size is too small no localization properties will be found. Unluckily, diagonalization gives best results for one dimensional systems with nearest neighbor tunneling[2], but those systems could be very easily studied using the recurrence method described in B.1.1.

Two cases, known for me, for which exact diagonalization is really worth using are fractal lattices and Aubry-Andre model. Fractals due to their non-integer dimensionality which is hard to implement into schemes presented above and the fact that usually finite systems are considered, thus we do not need methods giving the localization length for infinite systems. The Aubry-Andre, model as it have quite specific spectrum: several vary flat bands separated by the wide gaps. All the methods presented before used the energy of state as a parameter. In this way we will either lose a lot of computational time for nothing calculating out of the band or have to do diagonalization nevertheless to determine for which energies calculate the localization length.

---

[2] Then the size of the Hamiltonian scales linearly with system size and for tridiagonal matrices diagonalization is faster and generates significantly smaller numerical errors than for Hamiltonian matrices having more nonzero elements.

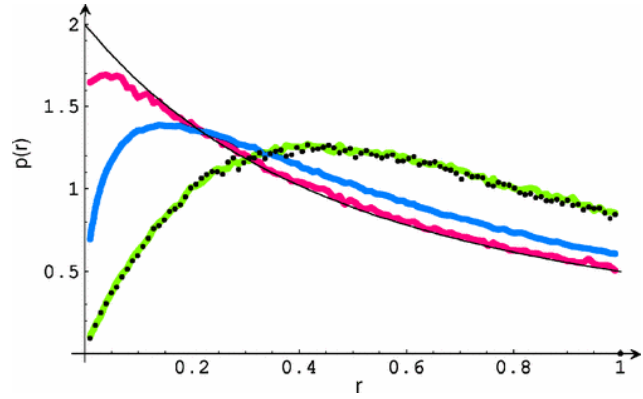

Figure 39: Example of usage of statistics of $r_i$ to determine localization of states. Black solid and black dotted lines are theoretical Poisson and GOE distributions. Green, blue and purple lines are results for increasing disorder strengths (source: [138]).

I will present three methods allowing investigation of the transport properties using the exact diagonalization:

*Level statistics*

In this method we need to know only eigenvalues which is computationally cheaper than obtaining eigenstates. Instead of using eigenvalues $\{E_i\}$ itself we concentrate on ratio of consecutive gaps:

$$r_i = \frac{\min(\delta_i, \delta_{i-1})}{\max(\delta_i, \delta_{i-1})}. \tag{204}$$

where $\delta_i = E_i - E_{i-1}$. Further we average $r_i$ close to chosen energy E over many realizations of disorder $r(E) = \langle r_i \rangle$. How the value $r(E)$ could provide us information about the localization properties? In the case of localized eigenstates, their overlap is vanishingly small, as they occupy different parts of the configuration space thus their positions should be uncorrelated and consequently distribution of gaps should be given by Poisson distribution for which $r(E) \approx 0.3863$. On the other hand extended states overlap and 'see' each other which induces level repulsion (lower probability of two states lying close to each other) leading to a Wigner-Dyson distribution of gaps which is characterized by $r(E) \approx 0.5359$. Those statements could be further confirmed by plotting $\{r_i\}$ on histogram and comparing with mentioned distributions (Fig. 39).

*Density of states*

This method, created by Thouless [139], may seem quite similar to one above, as again the distribution of energy levels will be connected with the localization. Whole method is based on the fact that element $G_{1N}$ of Green function of one dimensional tight-binding Hamiltonian (7) is:

$$G_{1N}(E) = \frac{\prod_{i=1}^{N-1} t_i}{\det(E\mathbb{I} - H)} = \frac{\prod_{i=1}^{N-1} t_i}{\prod_{\alpha=1}^{N}(E - E_\alpha)}. \tag{205}$$

The Green function has poles for all eigenvalues of Hamiltonian, moreover residues of those poles contain informations about eigenstate associated with given eigenenergy. Residue of $G_{1N}$ for energy $E_\beta$ reads:

$$\psi_1^\beta \psi_N^\beta = \frac{\prod_{i=1}^{N-1} t_i}{\prod_{\alpha \neq \beta}(E - E_\alpha)}. \tag{206}$$

Where $\psi_i^\beta$ is value of eigenstate $\beta$ on site i. As we have

$$\frac{\psi_N^\beta}{\psi_1^\beta} = \exp\left(-\frac{N-1}{\lambda_\beta}\right), \tag{207}$$

by putting (207) into (206) and taking the logarithm we obtain:

$$\lambda_\beta^{-1} = (N-1)^{-1} \sum_{\alpha \neq \beta} \log|E_\beta - E_\alpha| - (N-1)^{-1} \sum_{i=1}^{N-1} \log|t_i|. \tag{208}$$

This relation could be also replaced by the integral, if the length of the chain is sufficient and density of states is well-behaved:

$$\lambda_\beta^{-1} = \int \rho(x) \log|E_\beta - x| dx - \log|t|, \tag{209}$$

where $\rho(x)$ is the density of states and $t = \sqrt[N-1]{t_1 \ldots t_{N-1}}$ is the geometric mean of tunnelings.

*Properties of the eigenstates*

In this method, the eigenstates of Hamiltonian have to be found. As it is more computationally expensive it should be used, if for example we are not sure if the localization is exponential and we want to check precisely form of the eigenstates. If we have eigenstates found, instead of the localization length we could use another measure of the localization degree: inversed participation ratio (IPR)[3] defined as:

$$IPR(\psi) = \left(\sum_i |\psi_i|^4\right)^{-1} \tag{210}$$

---

3 Comparing with other works one have to be cautious as IPR could be sometimes called participation ratio (PR) as well as the Participation ratio (PR = $IPR^{-1}$) called IPR.

it is straightforward to check that for particle localized on only one site $\text{IPR}(\psi_{\text{loc}}) = 1$ (the localization length for this case is 0) whereas for wave evenly distributed over whole system $\text{IPR}(\psi_{\text{ext}}) = N$ (the localization length is $\infty$). Usually for extended states IPR takes large but smaller than $N$ values, quite good method of checking whether state is truly extended is calculation of IPR for growing system size – for localized functions it should be constant, whereas for extended it will scale with the growing system size. Of course for too small system sizes, the states which should be localized will exhibit all properties of extended ones. Notable distinction from the localization length is that IPR could pinpoint systems in which the particle is spread across the system but still occupies small fraction of the sites.

Assuming that the wavefunction have exponential profile: $\psi_i = \exp(-|i|/\lambda)$ it is possible to calculate the localization length from IPR

$$\lambda = \text{arctanh}\left(x^{1/3} - \frac{1}{3}x^{-1/3}\right)^{-1}, \quad x = \frac{1 + \sqrt{1 + \text{IPR}(\psi)^2/3^3}}{\text{IPR}(\psi)}. \tag{211}$$

However, such an approach generates significant errors as IPR depends strongly on the distribution of the wavefunction around maximum occupation. Thus IPR is sensitive to the disorder realization, while the localization length is a slope of the wavefunction for large distances. If there is a need to compare localization length with the results of exact diagonalization much safer method is calculating the localization length directly by fitting exponent to the slope of the wavefunction.